\documentclass[11pt]{article}
\pdfoutput=1 

\usepackage{jheppub} 

\usepackage[T1]{fontenc} 

\usepackage{graphics}
\usepackage{graphicx}
\usepackage{epsf}
\usepackage{epsfig}
\usepackage{float}
\usepackage{makecell}
\usepackage{multirow}
\usepackage{color}
\usepackage{xcolor}
\usepackage{natbib}
\usepackage[utf8]{inputenc}
\usepackage{amsmath}
\usepackage{amssymb}
\usepackage[export]{adjustbox}
\usepackage{subcaption}
\usepackage{soul}

\date{\today}

\newcommand{\beq}{\begin {equation}}  
\newcommand{\eeq}{\end   {equation}} 
\newcommand{\bea}{\begin {eqnarray}} 
\newcommand{\eea}{\end   {eqnarray}}  
\newcommand{\baa}{\begin {array}   } 
\newcommand{\eaa}{\end   {array}   }     
\newcommand{\bit}{\begin {itemize} }
\newcommand{\eit}{\end   {itemize} }
\newcommand{\be }{\begin {equation}} 
\newcommand{\ee }{\end   {equation}}
\newcommand{\nn }{\nonumber        }

\newcommand{\MeV}{\ensuremath{\mathrm{MeV}}}
\newcommand{\GeV}{\ensuremath{\mathrm{GeV}}}
\newcommand{\TeV}{\ensuremath{\mathrm{TeV}}}

\def\m{\mathcal}
\def\brc#1{\left(#1\right)}

\def\MeV{\mathrm{MeV}}
\def\GeV{\mathrm{GeV}}
\def\TeV{\mathrm{TeV}}

\newcommand\braket[2]{\left\langle{#1}\right|\left.{#2}\right\rangle}

\def\SUtwo{\mathrm{SU\brc{2}_L}}
\def\Uone{\mathrm{U\brc{1}_Y}}
\def\Ztwo{\mathcal{Z}_2}
\def\bphi{{\Phi}}
\def\bh{{H}}
\def\rt2{\sqrt{2}}
\def\Re{\mathrm{Re}}
\def\MSbar{\overline{\mathrm{MS}}}

\def\MHn{M_{H^0}}
\def\MAn{M_{A^0}}
\def\Mn{M_\Delta}
\def\Mp{M_{H^\pm}}
\def\Mpp{M_{H^{\pm\pm}}}

\title{\boldmath Probing Extended Scalar Sectors with Precision $e^+e^-\to Zh$ and Higgs Diphoton Studies
}

\author[a, b, c, d]{Michael J. Ramsey-Musolf }
\author[e, f, g, h, i]{Jiang-Hao Yu,}
\author[c]{Jia Zhou.}

\affiliation[a]{Tsung-Dao Lee Institute and School of Physics and Astronomy, Shanghai Jiao Tong University, 800 Dongchuan Road, Shanghai, 200240 China}
\affiliation[b]{Shanghai Key Laboratory for Particle Physics and Cosmology, Key Laboratory for Particle Astrophysics and Cosmology (MOE), Shanghai Jiao Tong University, Shanghai 200240, China}
\affiliation[c]{Amherst Center for Fundamental Interactions, Department of Physics, University of Massachusetts, Amherst, MA 01003}
\affiliation[d]{Kellogg Radiation Laboratory, California Institute of Technology, Pasadena, CA 91125 USA}
\affiliation[e]{CAS Key Laboratory of Theoretical Physics, Institute of Theoretical Physics, Chinese Academy of Sciences, Beijing 100190, P. R. China}
\affiliation[f]{School of Physical Sciences, University of Chinese Academy of Sciences, Beijing 100049, P.R. China}
\affiliation[g]{Center for High Energy Physics, Peking University, Beijing 100871, China}
\affiliation[h]{School of Fundamental Physics and Mathematical Sciences, Hangzhou Institute for Advanced Study, UCAS, Hangzhou 310024, China}
\affiliation[i]{International Centre for Theoretical Physics Asia-Pacific, Beijing/Hangzhou, China}

\emailAdd{mjrm@sjtu.edu.cn}
\emailAdd{jhyu@itp.ac.cn}
\emailAdd{jia@umass.edu}

\abstract{  

   We compute the one-loop corrections to  $\sigma(e^+ e^- \to Z h)$ arising from representative
  extended Standard Model scalar sector scenarios. According to the new scalar 
  $\SUtwo$ representations, we consider the inert doublet, real and complex triplet, quintuplet, and
  septuplet models. With the sub-percent level precision expected for prospective future $e+e-$ collider
  measurements of $\sigma(e^+ e^- \to Z h)$, studies of the Higgsstrahlung process will probe 
extended scalar sector particle spectrum and interactions in a manner complementary to direct searches
at the Large Hadron Collider and possible future $pp$ colliders. We also compare with the sensitivity of
future Higgs diphoton decay rate measurements. We find that the   $\sigma(e^+ e^- \to Z h)$ and $\Gamma(h\to\gamma\gamma)$ complementarity
is particularly pronounced for the complex triplet model.  

}

\begin{document} 
\preprint{ACFI-T21-01}
\maketitle
\flushbottom

\newpage

\section{Introduction}
\label{sec:intro}


%
%
%

The discovery of the Higgs boson at the Large Hadron Collider (LHC) in 2012 was an important milestone in high energy physics, with completing the particle spectra and validating the Higgs mechanism in the Standard Model (SM). It is known that there must be new physics beyond the SM (BSM), for example, to account for dark matter and matter-antimatter asymmetry in the universe. 
Given that the Higgs boson plays essential roles in various BSM scenarios, it is natural to expect BSM will influence properties of the Higgs boson, and thus lead to deviations of the Higgs properties from the SM predictions.

After the second run of the LHC, several of the Higgs couplings have been measured and found to agree with the SM predictions within about 15\% accuracy \cite{ATLAS:2019slw,Sirunyan:2018koj}. 
At the high luminosity LHC, measurements of the Higgs couplings are expected to be improved to around 5 - 10\% precision level \cite{Cepeda:2019klc,deBlas:2019rxi,Abada:2019lih}, which can yield BSM sensitivity comparable to that of direct searches for BSM at the LHC. 
Thus to probe BSM significantly beyond the LHC direct searches one would require the precision measurements of the Higgs boson couplings at percent-level accuracy. 
Achieving such precision would require new facilities like the lepton collider, since the lepton collider has the advantage of clean signatures and high statistics samples of the Higgs boson, i.e., a Higgs factory. 
Several proposals of such Higgs factories have been made, including the Circular Electron Positron Collider (CEPC) in China, the International Linear Collider (ILC) in Japan, and the Future Circular Collider with $e^+ e^-$ (FCC-ee) and the Compact Linear Collider (CLIC) at CERN.

The CEPC, ILC, and FCC-ee plan to operate their (first) runs at a center of mass energy of around 240 - 250 GeV. With this energy, the dominant Higgs production channel is the Higgs-strahlung process $e^+ e^- \to Z h$. 
These Higgs factories (with $\sqrt{s} = $240 GeV for the CEPC and FCC-ee $\sqrt{s}=$250 GeV for the  ILC) 
expect to obtain  similar integrated luminosities and should provide similar physics sensitivities, assuming adoption of optimized detectors and analyses \cite{CEPCStudyGroup:2018ghi,Abada:2019zxq,Barklow:2015tja,Fujii:2017vwa,Fujii:2019zll}. At the CEPC, with the expected integrated luminosity of 5.6 ab$^{-1}$, over one million Higgs boson are expected to be produced. With such number of events, the CEPC will be able to measure the Higgs boson couplings to the $Z$ boson pair with an accuracy of $0.25$\% \cite{TheATLAScollaboration:2014ewu,CMS:2013xfa,CEPCStudyGroup:2018ghi}, more than 10 times better than the expected accuracy at the HL-LHC. Such a precision measurement offers unprecedented discovery potential for BSM connected with the Higgs boson.

Extended scalar sectors constitute a well-motivated class of such BSM scenarios. 
According to the $\SUtwo$ quantum number of a given representation, the extended scalar could be classified as real or complex singlet, doublet, real or complex triplet, quadruplet, and so on. These new scalars usually affect the electroweak symmetry breaking pattern and in principle may catalyze a first order electroweak phase transition (EWPT) \cite{Profumo:2007wc,Espinosa:2011ax,Niemi:2018asa,Niemi:2020hto,Barger:2007im,Barger:2008jx,Papaefstathiou:2020iag}. Furthermore, in some cases these scalars could provide the dark matter candidates \cite{Cirelli:2005uq,McDonald:1993ex,Burgess:2000yq,Diaz:2015pyv,Ma:2006km,Barbieri:2006dq,Banerjee:2016vrp,FileviezPerez:2008bj,Araki:2011hm,Hambye:2009pw,AbdusSalam:2013eya,Chao:2018xwz} or mechanism for neutrino mass generation~\cite{Konetschny:1977bn,Magg:1980ut,Schechter:1980gr,Cheng:1980qt}. 
In this work, we consider the following specific cases: 
inert doublet, real triplet, complex triplet, quintuplet and septuplet. These new scalars are expected to be discovered at the future HL-LHC runs, which has been extensively studied in various literature \cite{Diaz:2015pyv,Banerjee:2016vrp,FileviezPerez:2008bj,Blank:1997qa,Chen:2008jg,Konetschny:1977bn,Magg:1980ut,Schechter:1980gr,Cheng:1980qt,Du:2018eaw,Chao:2018xwz}.

In these extended scalar models, the Higgs portal term of the generic form $|\Phi|^2 |H|^2$ usually exists. Its presence could allow for a first order EWPT. In this case the scalar potential in the extended scalar sector is directly relevant to the EWPT. We expect the $e^+ e^- \to Z h$ process could help
determine important aspects of the extended scalar potential, 
which is in general quite difficult to measure at the HL-LHC. These new scalars modify the Higgs couplings to the Z pair through radiative correction with new scalars running in the loop. These radiative corrections
depend on the new scalar mass spectra, the gauge interactions of the new scalars, and the scalar Higgs couplings in the scalar potential. Given the fact that gauge interactions are determined by gauge invariance, one could extract information on the scalar Higgs couplings by measuring the total rate and distributions of the $e^+ e^- \to Z h$ process, once the new scalar bosons are discovered and the scalar mass spectra are determined. 


On the other hand, it is possible the HL-LHC may yield no direct evidence for new scalars. 
Nevertheless, it is possible that deviations of the Higgs boson properties from the SM predictions could well provide the first evidence for BSM. Even in this case, the $e^+ e^- \to Z h$ process could help us indirectly probe the scalar sector, and thus tell us the information on the scalar mass spectrum.

In this context, it is important to emphasize that the Higgs diphoton process $h\gamma\gamma$ provides
a theoretically clean, complementary, indirect probe of an extended scalar sector at colliders. 
The target of the future lepton collider - CEPC \& FCC-ee - precision at this channel is a few
percent (refer to Table \ref{tbl:zh:hgg:precision})~\cite{CEPCStudyGroup:2018ghi,Abada:2019zxq}.
Unlike the $e^+ e^- \to Z h$ process, 
only charged scalars from the extended scalar sector contribute to $h\gamma\gamma$ via triangle loops. 
In this context one expects complementarity on the parameter space of the Higgs portal couplings between
the $e^+ e^- \to Z h$ and Higgs diphoton processes. 
We explore this complementarity for each model scenario under consideration in this paper.
We find that this complementarity is most pronounced for the complex triplet model,
whereas it is less evident for the other scenarios considered here.

In what follows, we systematically classify the scalar sector according to their $\SUtwo$ quantum number,
and obtain the form of the Higgs potential. Then we perform one-loop calculation of 
the $e^+ e^- \to Z h$ process and separate the BSM contribution from the SM one
in the $\MSbar$ and the On-shell schemes.
Calculations of the complete SM one-loop corrections are reported in Refs.~\cite{Fleischer:1982af,Kniehl:1991hk,Denner:1992bc}~\footnote{The complete SM one-loop corrections consist of the weak and QED corrections, which are individually gauge invariant. 
In Refs.~\cite{Fleischer:1982af,Kniehl:1991hk,Denner:1992bc}, the real photon emission - photon initial-state radiation (ISR) - in 
the QED correction was approximated with soft photon bremsstrahlung formalism. 
The emitted photon is deemed as soft if its energy $E_\gamma$ in the rest frame satisfies
$0<E_\gamma<\delta_s\frac{\sqrt{s}}{2}$, where $\delta_s$ is the soft cutoff parameter.
$\delta_s\sim0.1$ was used in~\cite{Fleischer:1982af,Denner:1992bc}. 
Ref.~\cite{Kniehl:1991hk} also approximated the hard photon contribution
as a convolution of the reduced cross section with a photon radiator in the hard/collinear region,
so that the sum of the soft and hard photon contribution was free of the choice of $\delta_s$.
In principle, the hard/non-collinear region should also be included even though its contribution
may not be numerically significant.}. 
At LEP200 energies the weak correction in the $\alpha$-parameterization~\footnote{The weak relative correction in the $G_\mu$-parameterization $\delta_\text{weak}^{G_\mu}$ differs from the one in the $\alpha$-parameterization $\delta_\text{weak}^{\alpha}$ by a constant shift $\Delta r$ that $\delta_\text{weak}^{G_\mu}=\delta_\text{weak}^{\alpha}-\Delta r$, where $\Delta r$ summarizes the one-loop electroweak corrections to muon decay~\cite{Sirlin:1980nh}.} has a positive sign and a magnitude of a few percent. The sign of the correction becomes negative at higher energies~\cite{Denner:1992bc}. 
The full one-loop radiative corrections to $e^+e^-\to Zh/H^0A^0$ processes in the inert doublet
model has been recently calculated~\cite{Abouabid:2020eik}.
The Higgs diphoton decay rate including the BSM contribution can be readily obtained by means of
the well known loop functions. 
We investigate the projected sensitivities to the scalar mass splitting and scalar Higgs couplings given the expected precision levels at the lepton colliders. 




Our study is organized as follows.
In Section \ref{sec:model} we discuss the generic form of the multiplet $\bphi_n$ in representation $j$
and the corresponding scalar potentials  
(explicit expressions are given in Appendix \ref{sec:potentials}).
The NLO corrections from the scalar multiplets to the process
$e^+e^-\rightarrow Zh$ are calculated in $\MSbar$ scheme in Section \ref{sec:nlo}.
Numerical results are presented in Section \ref{sec:result}, where the constraints on the scalar
potential parameters are obtained from the comparison of the $Zh$ cross section measurement to that of
the Higgs diphoton decay rate assuming the expected CEPC integrated luminosity.
Finally we summarize the significance of the study in Section \ref{sec:conclusion}.
We also add Feynman rules for the models studied in Appendix \ref{sec:feyn},
the expressions for self energy and vertex corrections in Appendix \ref{sec:cor},
and calculation in on-shell renormalization scheme used to compare with the that in $\MSbar$
in Appendix \ref{sec:onshell}.

\section{Extended Scalar Sector}
\label{sec:model}


In an extended scalar sector, an additional scalar multiplet,
either charged or singlet under
$\SUtwo\times\Uone$, is introduced. 
We could systematically classify the scalar multiplet based on its transformation under the $\SUtwo$ group. Symbolically one can write the scalar multiplet as
$\bphi$ as opposed to SM Higgs doublet written as $\bh$.
For a $\SUtwo$ $n = 2j+1$ multiplet $\Phi$ following the representation in
Ref.~\cite{Chao:2018xwz, Pilkington:2016erq}, using the notation $(j, j_3)$ as the subscript of the field it is written as
\begin{equation}
\Phi=\left[\begin{array}{c}
\phi_{j,j}\\
\phi_{j,j-1}\\
.\\
.\\
\phi_{j,-j}\\
\end{array}
\right]\,, \qquad {\textrm{with electric charge}} \quad Q_i = j_3 + \frac{Y}{2},
\end{equation}
where it has $2j+1$ component fields, each having electric charge $Q_i$. 
It is convenient to consider the associated conjugate $\overline{\Phi}$ instead of $\Phi^\ast$
\begin{equation}
{\overline\Phi}_{j,m} = (-1)^{j-m} \Phi^\ast_{j,-m}\ \ \ .
\end{equation}
Since $\Phi$ and ${\overline\Phi}$ transform in the same way under $\SUtwo$, it is convenient to build the $\SUtwo$ invariant using  $\Phi$ and ${\overline\Phi}$.
For integer isospin $j$, the scalar multiplet $\Phi$ can be real or complex, while for half integer isospin $\Phi$ is always complex for any value of hypercharge $Y$. 
If $\Phi$ is a real multiplet (integer isospin), there is a redundancy $\Phi=\overline{\Phi}$ such that the realness condition $\phi_{j,m}=(-1)^{j-m}\phi^{*}_{j,-m}$ is satisfied. 
If $\Phi$ is a complex multiplet (all kinds of isospin), each component represents one degree of freedom field, and it can be decomposed into two real multiplets as follows
\begin{equation}
\Phi=\frac{1}{\sqrt{2}}\left(S + i A\right),\qquad \overline{\Phi} =\frac{1}{\sqrt{2}}\left(S - i A\right)\,,
\end{equation}
where both $S$ and $A$ satisfy the realness condition $S=\overline{S}$ and $A=\overline{A}$. 
Thus the scalar extension with a complex multiplet $\Phi$ is equivalent to a model of two real multiplets $S$ and $A$.

For a general $2j+1$ multiplet $\Phi$ with the hypercharge $Y$, the kinetic term  reads
\begin{equation}
  \m{L}_{\mathrm{kin}} = \brc{D_{\mu}\bh}^{\dagger}\brc{D_{\mu}\bh}
  + \brc{D_{\mu}\bphi}^{\dagger}\brc{D^{\mu}\bphi}. 
\end{equation}
The covariant derivative for the multiplet can be written in a general form
\begin{equation}
  D_\mu = \partial_\mu  + ig_2W_\mu^a T^a + ig_1\frac{Y}{2}B_\mu, \qquad \brc{a=1,2,3}
  \label{eq:cov_deriv}
\end{equation}
where $T^a$ are the $\SUtwo$ generators. It is convenient to define the raising and lowering generators and the charged $W$ bosons accordingly 
\[
T^\pm=T^1\pm iT^2, \qquad W_\mu^\pm=W_\mu^1\mp iW_\mu^2.
\]
The other two physical neutral bosons are obtained via the diagonalization in such a way that
\[
\left[
  \begin{tabular}{c}
    $Z_\mu$ \\
    $A_\mu$
  \end{tabular}
  \right]=
  \left[
    \begin{tabular}{cc}
      $c_W$ &$-s_W$ \\
      $s_W$ &$c_W$
    \end{tabular}
    \right]
  \left[
    \begin{tabular}{c}
    $W_\mu^3$ \\
      $B_\mu$
    \end{tabular}
    \right], 
\]
with $s_W,~c_W$ the sine and cosine of the SM weak mixing angle $\theta_W$.   
Since in the representation $j$ the generator $T^a~\brc{a=\pm,3}$ is a $\brc{2j+1}\times\brc{2j+1}$
matrix\footnote{See the generic form of the generators $T^a$ in Appendix. \ref{sec:feyn},
which adopts the representation in Ref. \cite{Chao:2018xwz}.},
the $\brc{n,n'}$ element of the last two terms in Eq. (\ref{eq:cov_deriv}) reads 
\bea
\brc{ g_2 W_\mu^a T^a + g_1\frac{Y}{2}B_\mu }_{n,n'}  &=& \brc{g_2 W_\mu^+ T^+ + g_2 W_\mu^- T^- + g_2 W_\mu^3 T^3 + g_1\frac{Y}{2}B_\mu }_{n,n'} \nn\\
&=& \brc{g_1\frac{Y}{2}B_\mu - g_2 W_\mu^3 m}\delta_{n',n}
+g_2 W_\mu^+\sqrt{\brc{j-m}\brc{j+m+1}}\delta_{n',n+1} \nn \\
&&+ g_2 W_\mu^-\sqrt{\brc{j+m}\brc{j-m+1}}\delta_{n',n-1}, 
\eea
where $n=j+m+1$ and $m=-j,\ldots,j$. 
In particular, the hypercharge $Y$ takes $1$ for  doublet, $2$ for complex triplet,
and $0$ for real triplet, quintuplet and septuplet, respectively. 
Therefore, the gauge Feynman rule are quite generic for scalar multiplets, see Appendix \ref{sec:feyn}.
%


Although the kinetic Lagrangian is quite generic for a scalar multiplet, the scalar potential is representation dependent. 
To construct the general scalar potential, one could proceed~\cite{Chao:2018xwz, Pilkington:2016erq} to build $\SUtwo$ invariants by first pairing the field ingredients into irreducible representations and finally into $\SUtwo$ invariants. 
The pairing of the fields has the possibility of
\bea
	{\textrm{Pairing: }}\ \overline{H} H, \quad \Phi \Phi, \quad \overline{\Phi} \Phi , \quad \overline{\Phi}~\overline{\Phi}.
\eea
For half integer isospin multiplet, additional pairing $HH, H\Phi, \overline{H}~\overline{\Phi}$ appears with $Y =  1, -3$, and similarly
$HH, \overline{H} \Phi,  H \overline{\Phi} $ for $Y =  -1, +3$.   
From these building blocks, the  $\SUtwo$ invariants can be constructed.  
For any multiplet, it must contain the following universal terms
\bea
\left[\left(\Phi{\Phi}\right)_J \left({\overline\Phi}\,{\overline\Phi}\right)_J\right]_0, \quad  \left({\overline H} H\right)_0 \left({\overline \Phi} \Phi\right)_0, \quad \left[\left({\overline H} H\right)_1 \left({\overline \Phi} \Phi\right)_1 \right]_0.
\eea
In the case of $Y = 0$, there are additional terms
\bea
 \left[\left({\overline H} H\right)_1 \left({\Phi} \Phi\right)_1 \right]_0, \ \ {\textrm{for $j$ = half integer}}; \quad \left({\overline H} H\right)_0 \left({\Phi} \Phi\right)_0, \ \ {\textrm{for $j$ = integer}};
\eea
and the scalar self interactions
\bea
\left[\left(\Phi{\Phi}\right)_J \left({ \Phi}\,{ \Phi}\right)_J\right]_0, \quad
\left[\left({\overline \Phi}{\Phi}\right)_J \left({\Phi}\,{ \Phi}\right)_J\right]_0.
\eea
For other special multiplet, there could have additional invariant terms.
The potentials in the specific models are given in Appendix \ref{sec:potentials}. 
\section{NLO Calculation in $\MSbar$ Scheme}
\label{sec:nlo}

\begin{figure}[thpb]
  \centering
  \includegraphics[scale=0.6]{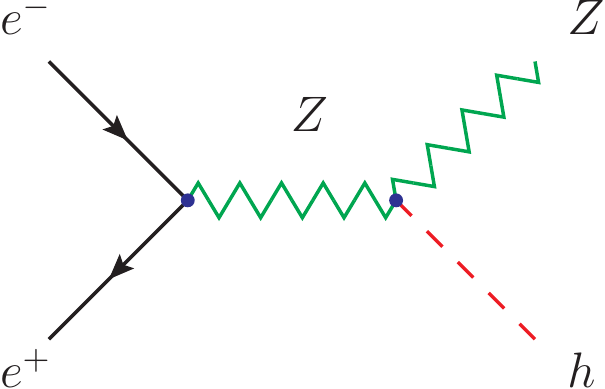}
  \caption{LO process: $e^-(p_1)~+~e^+(p_2)\rightarrow Z(k_1)~+~h(k_2)$}
  \label{fig:lo}
\end{figure}

The LO process
$e^+e^-\rightarrow Zh$
is depicted in Fig. \ref{fig:lo}. 
The expression for the matrix element squared has the following form:
\def\MM{{\bf M}}
\begin{equation}
  \left|\MM_{e^+e^-\rightarrow Zh}^{\mathrm{LO}}\right|^2 = \frac{2e^4\brc{g_v^2+g_a^2}}{s_W^2 c_W^2}\cdot\frac{t u + 2s M_Z^2 - M_h^2 M_Z^2}{\brc{s-M_Z^2}^2+M_Z^2\Gamma_Z^2},
  \label{eq:msq:lo}
\end{equation}
where $s_W=\sin\theta_W,~c_W=\cos\theta_W$ with $\theta_W$ the weak mixing angle, 
and
\[
g_v=\frac{-1/2+2s_W^2}{2s_Wc_W}, ~ g_a=-\frac{1}{4s_Wc_W}
\]
are the vector and axial-vector couplings to to the $e^+e^-Z$ vertex, respectively. The Mandelstam variables are defined as
\[
\brc{p_1+p_2}^2 = \brc{k_1+k_2}^2 = s, \, \brc{p_1-k_1}^2=\brc{p_2-k_2}^2=t, \, \brc{p_1-k_2}^2=\brc{p_2-k_1}^2 = u, 
\]
and the relation \(s+t+u=M_Z^2+M_h^2\).
The total cross section reads 
\begin{equation}
  \sigma = 
  \frac{1}{4}\frac{1}{48\pi s^2}\frac{e^4\brc{g_v^2+g_a^2}}{s_W^2c_W^2}\cdot
  \frac{\left[\kappa\brc{s,M_Z^2,M_h^2}+12 s M_Z^2\right]\kappa\brc{s,M_Z^2,M_h^2}}{\brc{s-M_Z^2}^2+M_Z^2\Gamma_Z^2},
  \label{eq:msqtot:lo}
\end{equation}
where 
the front $\frac{1}{4}$
is the spin average over the initial states, and the function $\kappa$ takes
the form:
\[
\kappa\brc{x,y,z}=\sqrt{x^2+y^2+z^2-2xy-2xz-2yz}. 
\]

\begin{figure}[htpb]
  \centering
  \includegraphics[scale=0.75]{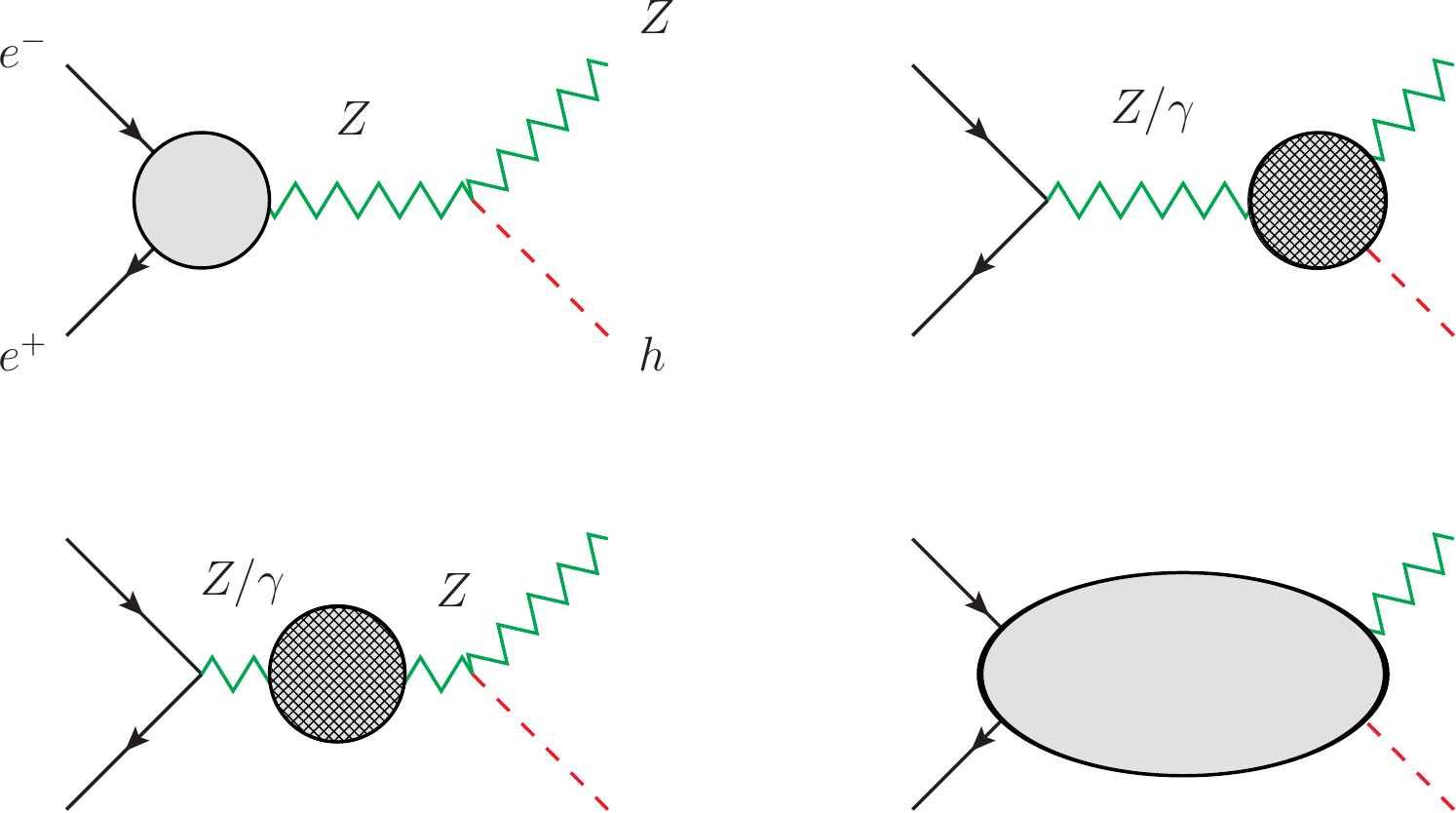}
  \caption{General one loop contributions to $e^+e^-\rightarrow Zh$ amplitude.
    The blobs denote one loop irreducible diagrams,
    and these hatched indicate where the extended scalar loop corrections can take place.}
  \label{fig:nlo}
\end{figure}

In general one loop corrections to the $e^+e^-\rightarrow Zh$ process shown in Fig. \ref{fig:nlo}
consist of vertex corrections to the initial and finial vertex,
self energy corrections to the mediated propagator,
and box corrections with virtual particles in loop in contact with both initial and final states,
respectively.
The interaction between the extended scalars and initial fermions is negligible due to its suppressed
Yukawa couplings of $\m{O}\brc{\frac{m_f}{M_W}}$. 
Therefore, in what follows, the NLO contribution originated from extended scalar loop corrections to
the $Zh$ production consists of self energy and vertex corrections to the vector and scalar bosons,
as indicated in Fig. \ref{fig:nlo} for the diagrams with hatching blobs.

Loop corrections alone are often UV divergent and need be renormalized
in certain renormalization scheme.
We use dimensional regularization in $d=4-2\epsilon$ spacetime dimensions and
the modified minimal subtraction ($\MSbar$) scheme to perform the calculation.
Here, we use the terms with caret $~\hat{}~$ denote the renormalized quantities,
which are the bare quantities subtracting the part proportional to
$\frac{1}{\bar{\epsilon}}=\frac{1}{\epsilon}-\gamma_E+\log4\pi$ by definition. 
In this scheme, the one loop amplitude can be written as below 
\begin{equation}
  \begin{aligned}
    i\MM_{e^+e^-\rightarrow Zh}^{\mathrm{NLO}}
    &=i\MM_{e^+e^-\rightarrow Zh}^{\mathrm{tree}}
    +i\MM_{e^+e^-\rightarrow Zh}^{\mathrm{self}}
    +i\MM_{e^+e^-\rightarrow Zh}^{\mathrm{vert}}, \\
    &=-i\frac{\hat{e}^2\hat{M}_Z}{\hat{s}\hat{c}}
    \hat{\rho}_{NC}\brc{s}
    \bar{v}\brc{p_2}\gamma^\mu\brc{g_v^{eff}-g_a^{eff}\gamma_5}u\brc{p_1}\epsilon_\mu\brc{k_1}
    + i\MM_{Z^*\rightarrow Zh}^{\mathrm{vert}}
    + i\MM_{\gamma^*\rightarrow Zh}^{\mathrm{vert}},
  \end{aligned}
  \label{eq:amp:msbar}
\end{equation}
where the last two terms on the right hand side
denote the contribution of $ZZh$ and $\gamma Zh$ vertex corrections, and 
$\hat{s},\hat{c}$ are sine and cosine of the weak mixing angle at 1-loop level in $\MSbar$ scheme
and can be expressed in terms of
the $\MSbar$ masses such that $\hat{c}^2=1-\hat{s}^2=\hat{M}_W^2/\hat{M}_Z^2$.
The effective vector and axial-vector coupling $g_v^{eff},g_a^{eff}$ have the following form
\begin{equation}
  g_v^{eff} = \frac{I_{W,e}^3-2\hat{\kappa}\brc{s}\hat{s}^2Q_e}{2\hat{s}\hat{c}},
  \quad
  g_a^{eff} = \frac{I_{W,e}^3}{2\hat{s}\hat{c}},
  \label{eq:gv:ga:eff}
\end{equation}
where the factor $\hat{\kappa}\brc{s}$ modifying $\hat{s}^2$ 
originates from the $\gamma-Z$ mixing tensor
and
\[
\hat{\kappa}\brc{s}=1-\frac{\hat{c}}{\hat{s}}\frac{\hat{\Sigma}_T^{\gamma Z}\brc{s}}{s}.
\]
The factor $\hat{\rho}_{NC}\brc{s}$ takes the form
\begin{equation}
  \begin{aligned}
    \hat{\rho}_{NC}\brc{s}=&\frac{1}{s-\hat{M}_Z^2+\hat{\Sigma}_T^{ZZ}\brc{s}}
    \brc{1+\frac{1}{2}\delta\hat{Z}_{ZZ}+\frac{1}{2}\delta\hat{Z}_h}, \\
    =&\frac{1}{s-M_Z^2+i\Gamma_Z M_Z}
    \brc{1+\frac{\Re\hat{\Sigma}_T^{ZZ}\brc{M_Z^2}-\Re\hat{\Sigma}_T^{ZZ}\brc{s}}{s-M_Z^2}
      +\frac{1}{2}\delta\hat{Z}_{ZZ}+\frac{1}{2}\delta\hat{Z}_h} \\
      & +\m{O}\brc{\alpha^2,\frac{\Gamma_Z^2}{M_Z^2}}
  \end{aligned}
  \label{eq:rhoNC}
\end{equation}
where we expand the expression in the second(third) line and omit terms of $\m{O}\brc{\alpha^2}$
or $\m{O}\brc{\frac{\Gamma_Z^2}{M_Z^2}}$ using
\[
\hat{M}_Z^2=M_Z^2+\Re\hat{\Sigma}_T^{ZZ}\brc{M_Z^2}, \quad
\Gamma_ZM_Z=\mathrm{Im}\hat{\Sigma}_T^{ZZ}\brc{s}, 
\]
and
\[
\delta\hat{Z}_{ZZ}=\hat{Z}_{ZZ}-1=
-\Re\left.\frac{\partial\hat{\Sigma}_T^{ZZ}\brc{k^2}}{\partial k^2}\right|_{k^2=M_Z^2},
\quad
\delta\hat{Z}_h=\hat{Z}_h-1=
-\Re\left.\frac{\partial\hat{\Sigma}^h\brc{k^2}}{\partial k^2}\right|_{k^2=M_h^2}, 
\]
with $\hat{Z}_{ZZ},~\hat{Z}_h$ the finite residues of $Z$ and $h$ propagators.
The self energy functions in all scalar multiplet models that are relevant in our calculation are listed
in Appendix \ref{append:NLO:self}.

We separate the vertex corrections in Eq. (\ref{eq:amp:msbar})
because they cannot fully factorize to Born
(extra Lorentz structures other than that of the LO amplitude).
It should be noted that the vertex corrections alone are free of UV divergences,
and thus can be written individually apart from the Born-like structure. 
Therefore, the total NLO corrections at matrix element square level can be obtained as below  
  \begin{equation}
    \left|\MM\right|_{\mathrm{tot,corr}}^2=\left|\MM_{e^+e^-\rightarrow Zh}^{\mathrm{NLO}}\right|^2
    - \left|\MM_{e^+e^-\rightarrow Zh}^{\mathrm{LO}}\right|^2,
    \label{eq:msq:corr}
  \end{equation}
  where $\MM_{e^+e^-\rightarrow Zh}^{\mathrm{LO}}$ is the amplitude at the leading order. 
  And the amplitude square of $\MM_{e^+e^-\rightarrow Zh}^{\mathrm{NLO}}$ is 
  \begin{equation}
    \left|\MM_{e^+e^-\rightarrow Zh}^{\mathrm{NLO}}\right|^2=
    \frac{\hat{e}^4\hat{M}_Z^2\brc{g_{v,eff}^2+g_{a,eff}^2}}{M_Z^2\hat{s}^2\hat{c}^2}
    \left|\hat{\rho}_{NC}\right|^2\brc{tu+2sM_Z^2-M_h^2M_Z^2} + \ldots, 
  \end{equation}
  where the ellipsis denotes other contributions that are at NLO and NNLO level.
  For the NLO contribution it is
  contained in the interference between the vertex term and the first term in the last line of
  Eq. (\ref{eq:amp:msbar}) 
and in particular the exact NLO level contribution reads
\begin{equation}
  \begin{aligned}
    &2~\Re\left[\brc{\MM_{Z^*\rightarrow Zh}^{\mathrm{vert}}
        +\MM_{\gamma^*\rightarrow Zh}^{\mathrm{vert}}}
      \cdot
      \brc{\MM_{e^+e^-\rightarrow Zh}^{\mathrm{tree}}}^*\right] \\
    = &-\frac{1}{16\pi^2}\frac{4~e^3}{M_Z s_W c_W}
    \sum_{s,\brc{M_1,M_2}}\brc{\frac{g_{ssh}g_{ssZ}^2\brc{g_v^2+g_a^2}}{\brc{s-M_Z^2}^2+M_Z^2\Gamma_Z^2} + \frac{g_{ssh}g_{ss\gamma}g_{ssZ}~g_v\brc{s-M_Z^2}}{s\left[\brc{s-M_Z^2}^2+M_Z^2\Gamma_Z^2\right]}}\times \\
    &2\Re\Bigg\{\brc{t u + 2s M_Z^2 - M_h^2 M_Z^2}\brc{2C_{24s}\brc{M_1,M_2} - \frac{1}{2}B_0\brc{k_2,M_2,M_2}} \\
    &+ \brc{s+M_Z^2-M_h^2}\brc{M_h^2 M_Z^2 - t u}\brc{C_{22s}\brc{M_1,M_2} - C_{23s}\brc{M_1,M_2}}\Bigg\}, 
  \end{aligned}
  \label{eq:vert:m1m2}
\end{equation}
where the expressions of $C_{22s}\brc{M_1,M_2}$, $C_{23s}\brc{M_1,M_2}$ and $C_{24s}\brc{M_1,M_2}$
can be found in Appendix \ref{append:NLO:vert}, and 
$\sum_{s,\brc{M_1,M_2}}$ denotes the sum of all possible scalar couplings 
and corresponding pair of scalar masses $\brc{M_1,M_2}$.
$g_{ssh},~g_{ssZ},~g_{ss\gamma}$ are the couplings of scalar-scalar-Higgs, scalar-scalar-$Z$ and
scalar-scalar-photon, respectively. The Feynman rules of the multiplets in the gauge coupling sector
can be found in Appendix \ref{sec:feyn}. 
The masses in the pair $\brc{M_1,M_2}$ are degenerate ($M_1=M_2$) in all scalar models
for each induced scalar triangle loop
except for that in the inert doublet model which has the two masses assigned differently. 


\section{Numerical Results in $\MSbar$ Scheme}
\label{sec:result}

In this section, we present the contour comparison between relative corrections to $Zh$ production
and the Higgs diphoton ($h\gamma\gamma$) decay rate 
induced by scalar multiplet
loop contributions,
with an effort
to delineate the prospective BSM parameter space constraints implied by each process. 
The relative correction by the scalar-induced loop contribution to the $e^+e^-\rightarrow Zh$ process
is defined as follows,
\begin{equation}
  \delta\sigma_{Zh}=\frac{\sigma_{\mathrm{BSM}}^{\mathrm{1-loop}}}{\sigma_{\mathrm{SM}}^{\mathrm{LO}}}, 
\end{equation}
where $\sigma_{\mathrm{BSM}}^{\mathrm{1-loop}}$ is the total cross section based on
the matrix element square term in Eq. (\ref{eq:msq:corr}),
and $\sigma_{\mathrm{SM}}^{\mathrm{LO}}$ is the LO total cross section given in Eq. (\ref{eq:msqtot:lo}).
There are three commonly-used input-parameter schemes for the calculation of EW radiative corrections:
the $\alpha\brc{0}$, $\alpha\brc{M_Z}$ and $G_\mu$ schemes, respectively. All are
suited for both the $\MSbar$ and OS renormalization schemes.
Each scheme has its advantages and disadvantages.
The most suitable choice depends on the nature of process under consideration. 
In general, the $\alpha\brc{M_Z}$ and $G_\mu$ schemes render the SM EW corrections 
free of explicit large logarithms involving light fermion masses and are, therefore, preferable over the $\alpha\brc{0}$ scheme
if no external photons are involved in the process.
A detailed discussion of these input-parameter schemes as well as the relation between one another can be found
in Ref.~\cite{Denner:2019vbn}.
In the present context, the one-loop contribution from the extended scalar sector particles in our study can be performed independently of the
overall SM EW contribution, and the result is manifestly free from light fermion mass logarithms.
Moreover, the impact on $\sigma(e^+e^-\to Zh)$ of the renormalized extended scalar sector loop contribution is less sensitive to the choice of input parameter scheme
than is the SM one-loop contribution. 
Hence, for simplicity we use the $\alpha\brc{0}$ input-parameter scheme
and set the corresponding SM input parameters as follows: 
\begin{equation}
\begin{aligned}
  &\alpha^{-1}=\brc{\frac{e^2}{4\pi}}^{-1}=137.036, \\
  &M_W=80.385~\GeV, ~ M_Z=91.1876~\GeV, ~\Gamma_Z=2.4952~\GeV, ~ M_h=125.1~\GeV.
\end{aligned}
\label{eq:para}
\end{equation}
At one-loop level in $\MSbar$ scheme the vector boson masses ($M_V,~\brc{V=W,Z}$), 
mixing angle ($\hat{s},\hat{c}$) and the electric charge ($\hat{e}$) are 
\begin{equation}
  \begin{aligned}
    &\hat{M}_V^2=M_V^2+\Re\hat{\Sigma}_T^{VV}\brc{M_V^2}, \\
    &\hat{c}^2=1-\hat{s}^2=\frac{\hat{M}_W^2}{\hat{M}_Z^2}, \\
    &\hat{e}=e\brc{1-\frac{1}{2}\delta\hat{Z}_{\gamma\gamma}
      -\frac{1}{2}\frac{\hat{s}}{\hat{c}}\delta\hat{Z}_{Z\gamma}},  
  \end{aligned}
  \label{eq:c:s:e:msbar}
\end{equation}
where the expressions of $\delta\hat{Z}_{\gamma\gamma},~\delta\hat{Z}_{Z\gamma}$ can be found in
Eq. (\ref{eq:reno:wavefunc}) and the self energy functions are given in Appendix \ref{append:NLO:self}.

Turning now to the $h\to\gamma\gamma$ process, the relative correction by scalar-induced loop contributions to the Higgs diphoton decay rate is defined below
\begin{equation}
  \delta R_{h\gamma\gamma}=
  \frac{\Gamma^{\mathrm{BSM}+\mathrm{SM}}_{h\rightarrow\gamma\gamma}
      -\Gamma^{\mathrm{SM}}_{h\rightarrow\gamma\gamma}}
       {\Gamma^{\mathrm{SM}}_{h\rightarrow\gamma\gamma}},
       \label{eq:dr:hgg}
\end{equation}
where the expression of Higgs diphoton decay width is given by,
\begin{equation}
  \begin{aligned}
    \Gamma^{\mathrm{BSM}+\mathrm{SM}}_{h\rightarrow\gamma\gamma} = &\frac{G_F\alpha^2M_h^3}{128\sqrt{2}\pi^3}
    \left|\sum_f N_c Q_f^2 g_{hff}A_{1/2}^h\brc{\tau_f}
    + g_{hWW}A_f^h\brc{\tau_W}
    - \sum_s \frac{M_W}{g_2M_s^2}g_{ss\gamma}^2 g_{ssh} A_0^h\brc{\tau_s}
    \right|^2, 
  \end{aligned}
  \label{eq:hgg}
\end{equation}
with the loop functions $A_{1/2}^h,~A_1^h,~A_0^h$ defined as \cite{Djouadi:2005gj}
\begin{equation}
  \begin{aligned}
    A_{1/2}^h\brc{\tau_i} &=-2 \tau_i\left[1+\brc{1-\tau_i} \mathcal{F}\brc{\tau_i}\right] \\
    A_{1}^h\brc{\tau_i} &=2+3 \tau_i+3 \tau_i\brc{2-\tau_i} \mathcal{F}\brc{\tau_i} \\
    A_{0}^h\brc{\tau_i} &=-\tau_i\left[1-\tau_i \mathcal{F}\brc{\tau_i}\right], 
  \end{aligned}
  \label{eq:hgg:loopfunc}
\end{equation}
and the function $\mathcal{F}\brc{\tau_i}$ reads
\[
  \mathcal{F}\brc{\tau_i} =\left\{
  \begin{array}{ll}
   \left[\sin^{-1}\brc{\sqrt{\frac{1}{\tau_i}}}\right]^{2}, &\tau_i \geq 1; \\
    -\frac{1}{4}\left[\ln \brc{\frac{1+\sqrt{1-\tau_i}}{1-\sqrt{1-\tau_i}}} -i \pi\right]^{2}, &\tau_i < 1;
  \end{array}
  \right.
\]
and $\tau_i=4M_i^2/M_h^2~\brc{i=f,W,s}$.
$\Gamma^{\mathrm{SM}}_{h\rightarrow\gamma\gamma}$ shares the same expression with
$\Gamma^{\mathrm{BSM}+\mathrm{SM}}_{h\rightarrow\gamma\gamma}$ but the last term in the module square
in Eq. (\ref{eq:hgg}).
One can see that only charged scalars contribute to the last term in Eq. (\ref{eq:hgg})
since $g_{ss\gamma}$ vanishes for neutral scalars.

Table~\ref{tbl:zh:hgg:precision} shows 
the estimated precision for $\sigma(Zh)$ production and $h\gamma\gamma$ decay signal rate for the CEPC and FCC-ee at $\sqrt{s}=240$ GeV
as well as the ILC at $\sqrt{s}=250$ GeV. Note that all three exhibit roughly the same level of precision for $\sigma(Zh)$ measurements.
Besides, the uncertainty on the measurement of the Higgs production via gluon and $b\bar{b}$ fusion
in $h\to\gamma\gamma$ decay channel at the HL-LHC is also listed in the right end column in the table,
whose precision is a few percent and somewhat better than those estimated in the future Higgs factories
even though the $\sigma\brc{Zh}$ measurement is much less precise. 
In what follows, provide
numerical results assuming the CEPC projections~\cite{CEPCStudyGroup:2018ghi} for purposes of concrete
illustration~\footnote{The precision for the Higgs production via the gluon and $b\bar{b}$
fusion in $h\to\gamma\gamma$ channel is applied in addition in the complex triplet model to illustrate
a straightforward improvement for the constraints in the parameter space.  
}.
In the following, we categorize our result into two cases where the constraints on the BSM parameter
space by the $Zh$ production are complementary and degenerate to that by the Higgs diphoton decay rate,
respectively.
We assume the constraints on the quartic Higgs (portal) couplings from the perturbativity in each model
are subject to the relation 
\begin{equation}
\lambda_i\brc{\mu}\lesssim\frac{\lambda_\text{FP}}{3}, 
\label{eq:lam:pert}
\end{equation}
where $\lambda_\text{FP}=12.1\ldots$ is the value at a fixed point
where $\beta_\lambda=0$ .
The renormalization scale $\mu\in\left[M_Z,\Lambda\right]$, and $\Lambda$ is the cutoff scale of the theory. 
This relation is an extension to the approximate perturbativity constraint on the couplings in
the complex singlet model in Ref. \cite{Gonderinger:2012rd}
which is based on the work in Ref. \cite{Riesselmann:1996is}.


\begin{table}[thpb]
\centering
\caption{
Estimated precision of CEPC, FCC-ee and ILC for  $\sigma(Zh)$  and $h\to\gamma\gamma$ signal rate.
The projected precision for the CEPC and FCC-ee are taken from
Refs.~\cite{CEPCStudyGroup:2018ghi,Abada:2019lih}
and those for the ILC are from Table 3 in Ref.~\cite{Durieux:2017rsg}, which are rescaled from the 
luminosity samples of 250 fb$^{-1}$ in the ILC operating
Scenarios~\cite{Barklow:2015tja,Bambade:2019fyw,Fujii:2019zll}.
All quoted values are are for statistical uncertainties at the 68\% confidence level.
The total uncertainty (stat + syst) on the measurement for $h\to\gamma\gamma$
with Higgs production via gluon and $b\bar{b}$ fusion at the HL-LHC~\cite{Cepeda:2019klc}
is listed in the last column on the right in comparison of that at the CEPC, FCC-ee and ILC. 
} 
\label{tbl:zh:hgg:precision}
\small
\begin{tabular}{|c|c|c|c|c|}
\hline\hline
 Measurement &CEPC  &FCC-ee  &ILC &HL-LHC\\
 &(240 GeV, 5.6 ab$^{-1}$)& (240 GeV, 5 ab$^{-1}$) &(250 GeV, 2 ab$^{-1}$) &(14 TeV, 3 ab$^{-1}$) \\
 \hline
 $\sigma\brc{Zh}$ &$0.50\%$ &$0.50\%$ &$0.71\%$ &-- \\
 $\sigma\times\mathrm{BR}\brc{h\rightarrow\gamma\gamma}$ &$6.8\%$ &$9.0\%$ &$12\%$ &$4\%$ (ggF+bbH) \\
 \hline\hline
\end{tabular}
\end{table}

It is worth mentioning that in order to verify that our result is independent of scheme choice, 
we also perform the one-loop calculation in an alternative renormalization
scheme - on-shell renormalization scheme and obtain the consistent result with that in $\MSbar$ scheme.
One can refer to Appendix \ref{sec:onshell} for a detailed procedure of on-shell renormalization scheme.

\subsection{Complementary with Higgs Diphoton Decays}
\label{subsec:res:comp}
\paragraph{Complex Triplet}
The complex scalar triplet model with the hypercharge of the extended
scalar triplet $Y=2$ is a component of
the type-II seesaw model~\cite{Konetschny:1977bn,Magg:1980ut,Schechter:1980gr,Cheng:1980qt}.
The neutrino masses are obtained through the Yukawa interaction term in the type-II seesaw model below
\begin{equation}
\m{L}_\text{Yuk}=h_{ij}\overline{L^{Ci}}i\tau_2\Delta L^j + \text{h.c.},
\label{eq:Lyuk}
\end{equation}
where $h_{ij}$ is the neutrino Yukawa coupling that is complex and symmetric,
$L_i=\brc{\nu_i,e_i}_L^T$ $\brc{i=e,\mu,\tau}$ is a left-handed lepton doublet,
and $\Delta$ is a $2\times2$ representation of the complex triplet fields
(see expression below (\ref{eq:V:ct})).
After electroweak symmetry breaking (EWSB), the neutrinos acquire masses due to non-vanishing
triplet vacuum expectation value (VEV) $v_\Delta\ne0$, and the neutrino mass matrix reads
\begin{equation}
m_{\nu,ij}=\sqrt{2}h_{ij}v_\Delta.
\label{eq:mnu}
\end{equation}
Meanwhile the minimization of the potential (\ref{eq:V:ct}) gives rise to the triplet VEV
\begin{equation}
v_\Delta\approx\frac{\mu v_\phi^2}{\sqrt{2}\left[M^2+\frac{\lambda_4+\lambda_5}{2}v_\phi^2\right]}, 
\end{equation}
when $v_\Delta/v_\phi\ll1$.
One can see that for fixed $\lambda_4,\lambda_5$, $v_\Delta\rightarrow0$ with $M/v_\phi\gg1$.
Being constrained by the $\rho$ parameter the upper bound of the triplet VEV
$v_\Delta\lesssim3~\GeV$~\cite{Du:2018eaw}, 
so that the sine of the mixing angles~\footnote{There are three mixing angles
$\brc{\alpha,~\beta_0,~\beta_\pm}$
respectively for the neutral and singly charged components of the triplet; the doubly charged component $H^{\pm\pm}$, it is already in a mass eigenstate.
See details in Ref.~\cite{Du:2018eaw}.}
are highly suppressed since $\propto v_\Delta/v_\phi\ll1$. 
Therefore, in what follows, we consider the scalar triplet unmixed with the SM Higgs doublet after EWSB,
as being treated in the other models with zero VEV in the subsequent subsection.

\begin{figure}[htpb]
  \centering
  \includegraphics[scale=0.75]{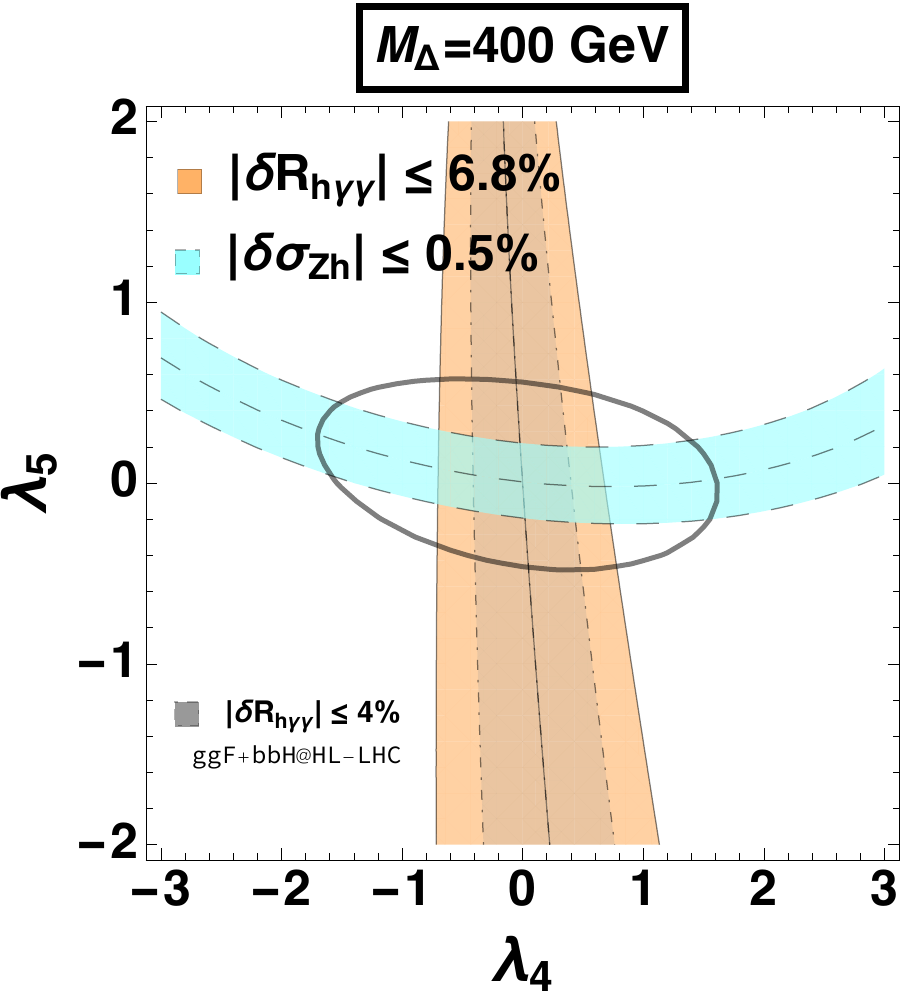}
   \hfill
   \includegraphics[scale=0.75]{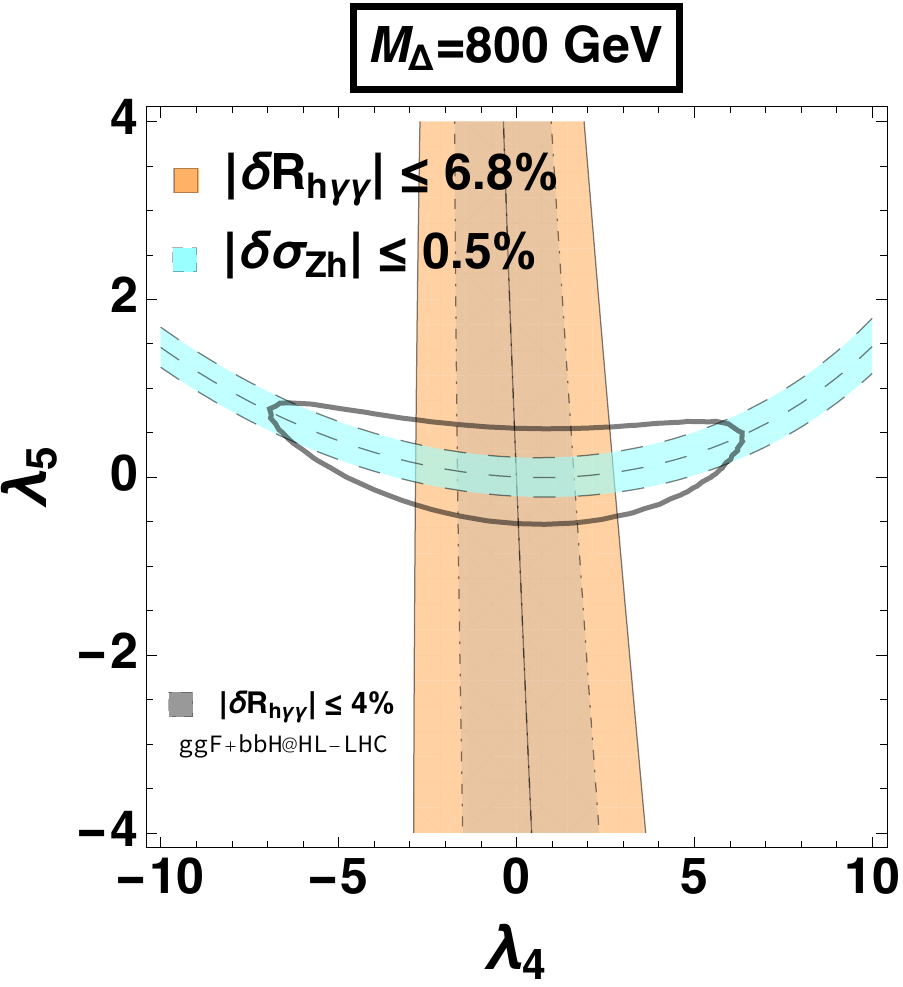}
   \caption{Contour comparison of $\left|\delta\sigma_{Zh}\right|\le0.5\%$
   ($1\sigma$ cyan band with dashed central and outer lines)
  and $\left|\delta R_{h\gamma\gamma}\right|\le6.8\%$
  ($1\sigma$ orange band with solid central and outer lines)
    in complex triplet model with $\Mn=400~\GeV$ (left) and $\Mn=800~\GeV$ (right), respectively.
    The ellipse on each plot denotes the $95\%$ confidence level for the parameter
    $\lambda_4$ and $\lambda_5$ with corresponding fixed $\Mn$.
    In addition, the contour of $\left|\delta R_{h\gamma\gamma}\right|\le4\%$
    ($1\sigma$ grey band with dashdotted central and outer lines) corresponding
    to the accuracry at the HL-LHC via gluon and $b\bar{b}$ fusion is added for the comparison,
    where one can see that with higher precision on $h\to\gamma\gamma$ decay rate a better constraint
    on the parameter $\lambda_4$ can be archieved.}
  \label{fig:ct}
\end{figure}

We observe that
the complex triplet makes the most interesting case in our study, as it exhibits the complementary
sensitivity between the $Zh$ and $h\gamma\gamma$ decay rate precision measurements.
In this model, there are three free parameters that enter the one-loop calculation, which are
$\lambda_4,~\lambda_5$ and $M_\Delta$, respectively. The scalar triplet masses are given by
\begin{equation}
  \begin{aligned}
    \Mpp^2 &= \Mn^2 - \frac{\lambda_5 v_\phi^2}{2}, \\
    \Mp^2 &= \Mn^2 - \frac{\lambda_5 v_\phi^2}{4}, \\
    M_H^2 &= M_A^2 = \Mn^2,
  \end{aligned}
  \label{eq:mass:ct}
\end{equation}
with $\Mpp$, $\Mp$, $M_H$ and $M_A$ the masses of the doubly and singly charged, 
CP-even and CP-odd neutral scalar, respectively.
In Fig. \ref{fig:ct} the contours of $\left|\delta\sigma_{Zh}\right|\le0.5\%$
($1\sigma$ cyan band with dashed central and outer lines)
and $\left|\delta R_{h\gamma\gamma}\right|\le6.8\%$
($1\sigma$ orange band with solid central and outer lines)
are plotted in
$\brc{\lambda_4,\lambda_5}$ plane with $\Mn=400,~800~\GeV$.
To illustrate a straightforward improvement for probing region of parameter space
in Higgs diphoton decay channel when increasing the precision for the measurement in this channel,  
a contour of $\left|\delta R_{h\gamma\gamma}\right|\le4\%$ ($1\sigma$ grey band with
dashdotted central and outer lines) corresponding to the precision at the HL-LHC is also plotted.
A shrinking of the band width constraining parameter $\lambda_4$ is apparent with smaller uncertainties
on the Higgs diphoton decay measurement. 

The ellipse encompassing the crossed range between the two bands on each plot
outlines the constraints of the parameter $\lambda_4$ and $\lambda_5$ at $95\%$ confidence level
when combining $\left|\delta\sigma_{Zh}\right|\le0.5\%$ and $\left|\delta R_{h\gamma\gamma}\right|\le6.8\%$
and using $\chi^2$ test for 2 degree-of-freedom. 
One can see a cross-like
shape for the two contours, indicating complementary results between the two processes and
thereby putting 
the most stringent constraints one another.
We find that as the mass of the neutral component
($\Mn$) increases the cyan band (with dashed lines)
as the contour from $\left|\delta\sigma_{Zh}\right|\le0.5\%$
gets flatter and becomes less dependent of the parameter $\lambda_4$
\footnote{One should compare the flatness of the cyan band (with dashed lines) in the same area of
$\brc{\lambda_4,\lambda_5}$ plane.
Compared to the left-hand subfigure in Fig. \ref{fig:ct}, the cyan band in the right-hand subfigure
in a zoom-in range of $\lambda_4\in\brc{-3,3}$ appears fairly flat. 
}.
This can be explained as follows.
As one can see in Eq. (\ref{eq:mass:ct}) the mass difference of differently charged components is
proportional to $\lambda_5$.
Performing the large mass expansion in the range of
$s/\Mn^2\ll1$
to the self energy and vertex loop functions,
we find that the dominant contribution is from the $WW$
self energy contribution that enters $\hat{s}$ and $\hat{c}$
(see first and second lines in Eq. (\ref{eq:c:s:e:msbar}))
due to the mass splitting between
scalars with different number of charges and the mass difference is solely proportional to
$\lambda_5$ (see expression of $\Sigma_T^{WW}\brc{k^2}$ in (\ref{eq:self:ctriplet}) and
apply large mass expansion to $B_0$ functions).
On the other hand, as for the orange band, 
when neglecting the mass difference in the loop function $A_0^h\brc{\tau_i}$
(which is valid in the large mass limit as the case we discuss above using large mass expansion),
the minimization of the $\delta R_{h\gamma\gamma}$ (i.e., the third term in the modular square 
in Eq. (\ref{eq:hgg}) vanishes) gives rise to a simple linear relation between $\lambda_4$ and
$\lambda_5$ that $10\lambda_4+\lambda_5=0$, which roughly reflects the central position of the
orange band in Fig. \ref{fig:ct}.
In other words, the complementary parameter space probes by the precision measurement for $\sigma\brc{Zh}$ and $h\to\gamma\gamma$ decay rate are realized in two aspects: 
\begin{enumerate}
\item[1.] The BSM contribution to $\sigma\brc{Zh}$ is dominated by the $WW$ self energy via differently charged scalar in loops which is susceptible to the variation of the mass splitting parameter ($\lambda_5$), compared to other types of corrections.
\item[2.] The BSM contribution to $h\to\gamma\gamma$ decay rate involves triple Higgs couplings with two charged Higgs that have a stronger dependence on the parameter $\lambda_4$ than on the other couplings (see Feynman rules in Table~\ref{tbl:feyn:potential}),  making it more susceptible to variation in $\lambda_4$.
\end{enumerate}
Therefore, the complementarity is observed when exploring the parameter space in the plane $\brc{\lambda_4,\lambda_5}$. 
Furthermore, from the ellipses constraining the parameter $\lambda_4$ and $\lambda_5$ at $95\%$ confidence level
one also can see that the constraint on $\lambda_5$ is stronger than that on $\lambda_4$ with increase of
$\Mn$. 

Instead of fixing the neutral scalar mass, one may fix one of the two coupling parameters and
investigate the contours of $\delta\sigma_{Zh}$ and $\delta R_{h\gamma\gamma}$ at $1\sigma$ level 
in the neutral scalar mass and coupling parameter plane.
First we can make slice of the contours in the $\brc{\Mn,\lambda_5}$ plane with fixed $\lambda_4$.
Judging from Fig. \ref{fig:ct}, the boundary of $\lambda_4$ with $\Mn=400~\GeV$ is
$\left|\lambda_4\right|\sim1$ and that with $\Mn=800~\GeV$ is $\left|\lambda_4\right|\sim3$
which are still within the allowed perturbativity range (\ref{eq:lam:pert}); 
while the constraint on parameter $\lambda_5$ keeps it in a small range
$\left|\Delta\lambda_5\right|\lesssim0.4$ ($\left|\lambda_5\right|\lesssim0.2$ for small $\lambda_4$).

\begin{figure}[thpb]
  \centering
  \includegraphics[scale=0.75]{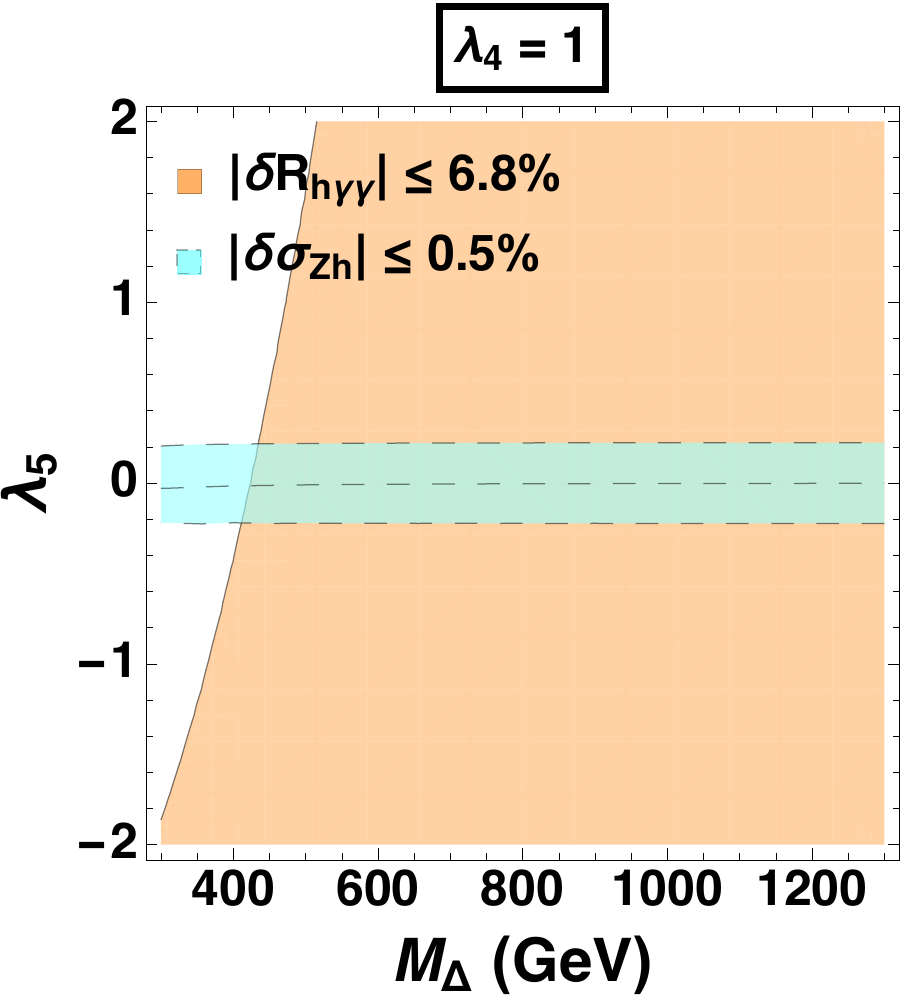}
  \hfill
  \includegraphics[scale=0.75]{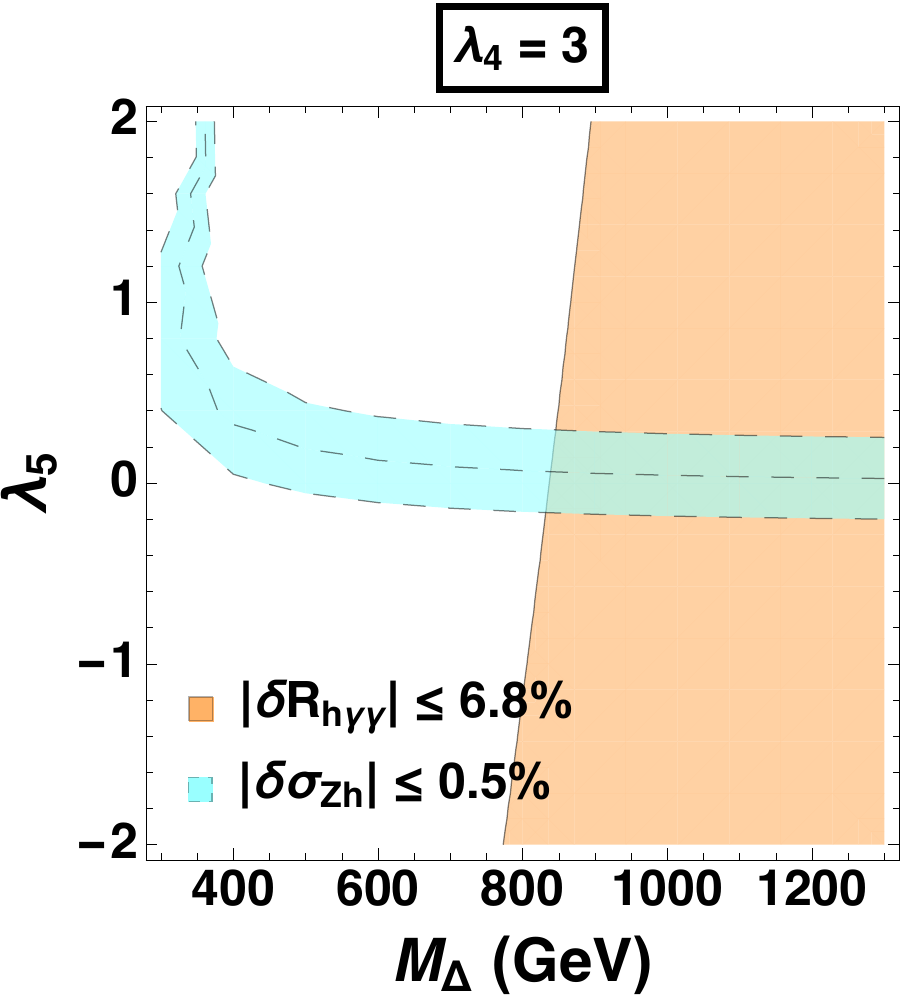}
  \caption{Contour comparison of $\left|\delta\sigma_{Zh}\right|\le0.5\%$
  (cyan band outlined with dashed lines)
    and $\left|\delta R_{h\gamma\gamma}\right|\le6.8\%$
    (orange area outlined with solid line)
    in complex triplet model in plane
    $\brc{\Mn,\lambda_5}$ with $\lambda_4=1$ (left)
    and $\lambda_4=3$ (right).}
  \label{fig:ct:l4}
\end{figure}

Fig. \ref{fig:ct:l4} shows the two contours of $\left|\delta\sigma_{Zh}\right|\le0.5\%$
and $\left|\delta R_{h\gamma\gamma}\right|\le6.8\%$ with the fixed $\lambda_4=1,~3$,
respectively. The contour of $\left|\delta\sigma_{Zh}\right|\le0.5\%$ constantly restricts $\lambda_5$
within a thin strip region, while the contour of $\left|\delta R_{h\gamma\gamma}\right|\le6.8\%$ puts
a lower bound $\Mn$ region, so that in the overlapping area of the two contours the lower bound
of the mass $\Mn$ is approximately $400~\GeV$ for $\lambda_4=1$ and $800~\GeV$ for $\lambda_4=3$.
Similar lower bound of $\Mn$ can be obtained for negative $\lambda_4$, i.e., $\lambda_4=-1,-3$, respectively. 
This is consistent with the result in Fig. \ref{fig:ct}, where the left figure ($\Mn=400~\GeV$)
has the boundary of $\left|\lambda_4\right|\lesssim1$ and the right figure ($\Mn=800~\GeV$) has
the boundary of $\left|\lambda_4\right|\lesssim3$. 

In addition to the constraints from perturbativity (\ref{eq:lam:pert}),
vacuum stability and perturbative unitarity also impose boundaries in the parameter space.
In general any of the above three sources of constraints 
impacts all the couplings in the potential.
For instance, in Ref.~\cite{Du:2018eaw} it was shown that 
vacuum stability and perturbative unitarity at tree-level with fixed values of $\lambda_2=0.2$ and
$\lambda_3=0$ require the parameter space in the $\brc{\lambda_4,\lambda_5}$ plane
to be mostly in the 4th quadrant, 
and initial condition ($\lambda_i$ fixed at the starting scale) should be set first 
for the renormalization group equations (RGEs) to test the perturbativity with running scales,
which are both illustrated in Fig. 1 therein. 
On the other hand, the constraints by the precision measurement on $Zh$ and $h\gamma\gamma$ decay rate
only rely on the Higgs portal couplings $\brc{\lambda_4,\lambda_5}$
in addition to the triplet mass spectrum, which therefore, complement that by perturbativity,
vacuum stability and perturbative unitarity.

As discussed above that the upper bound of the triplet VEV $v_\Delta\lesssim3$ GeV
is strictly constrained by the $\rho$ parameter and the sine of the mixing angles are highly suppressed
so that one can consider the scalar triplet and the SM doublet fields do not mix together and
treat the virtual scalar triplet loop corrections to the $Zh$ production separately.
On the other hand, 
as can be seen in (\ref{eq:mnu}) that the neutrino masses derived from the type-II seesaw model
depend on the neutrino Yukawa couplings $h_{ij}$ and the triplet VEV $v_\Delta$. 
The sum of neutrino masses has an upper bound, and in the context of a minimal 7-parameter model 
($\mathrm{\Lambda CDM}+\sum m_\nu$)
one obtains  $\sum m_\nu < 0.12$ eV (95\%CL) from the combination of {\it Planck} 2018 
and BAO~\cite{Aghanim:2018eyx,Zyla:2020zbs} data sets. 
The parameters $h_{ij}$ and $v_\Delta$, however, are not individually rigorously constrained
({\em e.g.}, $v_\Delta$ could be in the KeV or GeV regime), 
because their product yields the light neutrino mass matrix.
For a fixed $v_\Delta$, a larger (smaller) $h_{ij}$ corresponds to a larger (smaller) value for 
neutrino masses, and vice versa.

As discussed in Ref.~\cite{Du:2018eaw}, this interplay of $h_{ij}$ and $v_\Delta$ affects the sensitivity of 
collider probes of the complex triplet model. At the LHC and a prospective 100 TeV $pp$ collider, 
the dominant discovery channels could be
$H^{++}H^{--}$ and $H^{\pm\pm}H^\mp$ depending on the value of $v_\Delta$ \cite{Du:2018eaw}. 
The doubly charged Higgs $H^{\pm\pm}$ could decay into same-sign di-leptons (di-bosons)
and the branching ratio Br$\brc{H^{\pm\pm}\to l^\pm l^\pm}$ (Br$\brc{H^{\pm\pm}\to W^\pm W^\pm}$)
depends on the parameters $\Mn$, $\lambda_5$, and $h_{ij}$
($\Mn$, $\lambda_5$, and $v_\Delta$).
The singly charged Higgs $H^\pm$ has the decay vertex $H^\pm h W^\mp$
and Br$\brc{H^{\pm}\to h W^\pm}$ depends on the parameters
$\Mn$, $\lambda_4$, $\lambda_5$ and $v_\Delta$.
In Ref.~\cite{Du:2018eaw} the regions of $> 5\sigma$ significance~\footnote{The significance is defined as
$\frac{\text{Sig}}{\sqrt{\text{Sig+Bkg}}}$ with Sig and Bkg the total signal and background event numbers.}
 are obtained using the discovery channels
$H^{++}H^{--}$ and $H^{\pm\pm}H^\mp$ with $H^{\pm\pm}$ decaying into same-sign di-leptons and/or di-bosons
in the $\brc{\Mn,v_\Delta}$ plane with fixed neutrino masses ($m_{\nu_l}=0.01$ eV for $l=e,\mu,\tau$)
and Higgs portal couplings ($\lambda_4=0,\lambda_5=-0.1$) (see Fig. 7 therein). 

In our study, we find that the parameter $\lambda_5$ could be tightly constrained
($\left|\Delta\lambda_5\right|\lesssim0.4$) by the precision measurement for the $Zh$ cross section
since $\sigma\brc{Zh}$  is sensitive to the mass splitting between the scalar components. 
Moreover,  measurements of $\sigma\brc{Zh}$ and $\Gamma(H\to\gamma\gamma)$ provide complementary information, as illustrated
in Fig.~\ref{fig:ct:l4}.
The Higgs di-photon decay rate is relatively insensitive to $\lambda_5$ but strongly-dependent on $\lambda_4$. 
Therefore, with additional information about the Higgs portal parameters one might further delineate  the discovery regions in the $\brc{\Mn,v_\Delta(h_{ij})}$ plane for given values of the neutrino masses. 

To illustrate, we take two benchmark points from Ref.~\cite{Du:2018eaw} to show the consistency
on the Higgs portal parameter space settings between the findings in our study
and in probing the discovery channels with neutrino masses involved, which read
\begin{equation}
\nonumber
\begin{tabular}{|*9{c|}}
\hline\hline
$\Mn$ &$M_Z$ &$M_h$ &$m_\nu$ &$v_\Delta$ &$\lambda_2$ &$\lambda_3$ &$\lambda_4$ &$\lambda_5$ \\
\hline
400~\&~800 GeV &91.1876 GeV &125 GeV &0.01 eV &$10^{-4}$ GeV &0.2 &0 &0 &-0.1 \\
\hline\hline
\end{tabular}
\label{eq:ct:bm}
\end{equation}
We stress that the value of $M_h$ that is off 0.1 GeV compared to our settings in (\ref{eq:para})
($125.1$ GeV for loop correction vs. $125$ GeV for the benchmark points)
does not noticeably affect other settings. 
The mass of the neutrino is equal in the three generations.  
These two benchmark points represent two mass scales (low and high $\Mn$ region) and are covered
by various overlapping discovery regions in the $\brc{\Mn,v_\Delta}$ plane
(see Fig. 7 in Ref.~\cite{Du:2018eaw}).
In the analyses of Fig. \ref{fig:ct} and Fig. \ref{fig:ct:l4} we already observe that
with the choice of $\Mn=400~(800)~\GeV$, the regions of $\lambda_4$ and $\lambda_5$
constrained by the precision measurement of $Zh$ and $h\gamma\gamma$ decay rate approximate
$\left|\lambda_4\right|\lesssim1~(3)$ and $\left|\lambda_5\right|\lesssim0.2$,
which encompass $\brc{\lambda_4,\lambda_5}=\brc{0,-0.1}$ for the two benchmark points
in (\ref{eq:ct:bm}).


\subsection{Degenerate with Higgs Diphoton Decay Sensitivity}
\label{subsec:res:degen}
In this subsection the models of interest including the inert doublet, real triplet, quintuplet and septuplet
are studied,
and likewise the constraints on the parameter space in each model are outlined in terms of the respective
precision measurements on the $Zh$ production and Higgs diphoton decay rate. 
We find that in these models it gives less complementary sensitivity between the $Zh$ and Higgs diphoton 
decay rate precision measurement, showing (partial) degeneracy in constraining the parameter spaces between
the $Zh$ and $h\gamma\gamma$ processes.
In particular, there is a large degree of degeneracy on constraints of the parameter space
between the two processes in the real triplet model; while in the inert doublet, quintuplet and septuplet
models, since there are more free parameters it depends on which sliced parameter plane that
one looks into to determine the degree of degeneracy. 

\paragraph{Inert Doublet}
The inert doublet model is the simplest two Higgs doublet model (THDM) that has an imposition of $\Ztwo$
symmetry and therefore the lightest neutral component 
in the extended scalar sector could be a potential
weakly interacting massive particle (WIMP) dark matter candidate
\footnote{There are two neutral components: 
CP-even $H^0$ and CP-odd $A^0$. One can choose either one as the dark matter candidate
as long as it's the lightest particle. The convention in the studies is take $H^0$ the lightest particle
by requiring the parameter $\lambda_5$ non-positive (see Eq. (\ref{eq:idm:mass}) for the mass definition).}.
The extended scalar sector also decouple to
the SM fermions due to its vanishing VEV.
The interplay between the dark matter phenomenology and the EWPT
in the context of the inert double model has been studied in a variety of spectra
\cite{Borah:2012pu,Gil:2012ya,Blinov:2015vma,Chowdhury:2011ga,Cline:2013bln}. 
It shows that a strongly first order electroweak phase transition (SFOEWPT) requires
a large mass splitting between the dark matter candidate particle and the other extended scalars,
and the Higgs funnel regime ($\MHn\sim M_h/2$) is the only region of parameter space that can successfully
saturate the dark matter abundance and provide a SFOEWPT \cite{Borah:2012pu,Gil:2012ya,Blinov:2015vma}.

\begin{figure}[thpb]
\begin{center}
\includegraphics[scale=0.75]{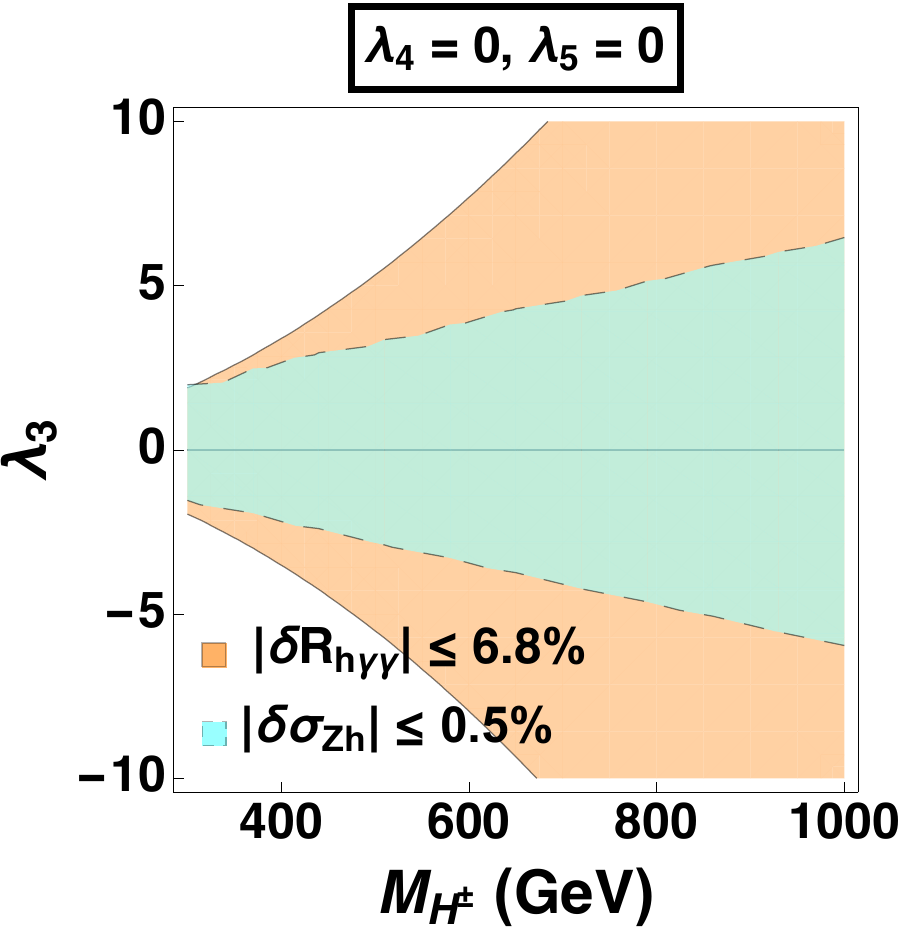} \\
\includegraphics[scale=0.75]{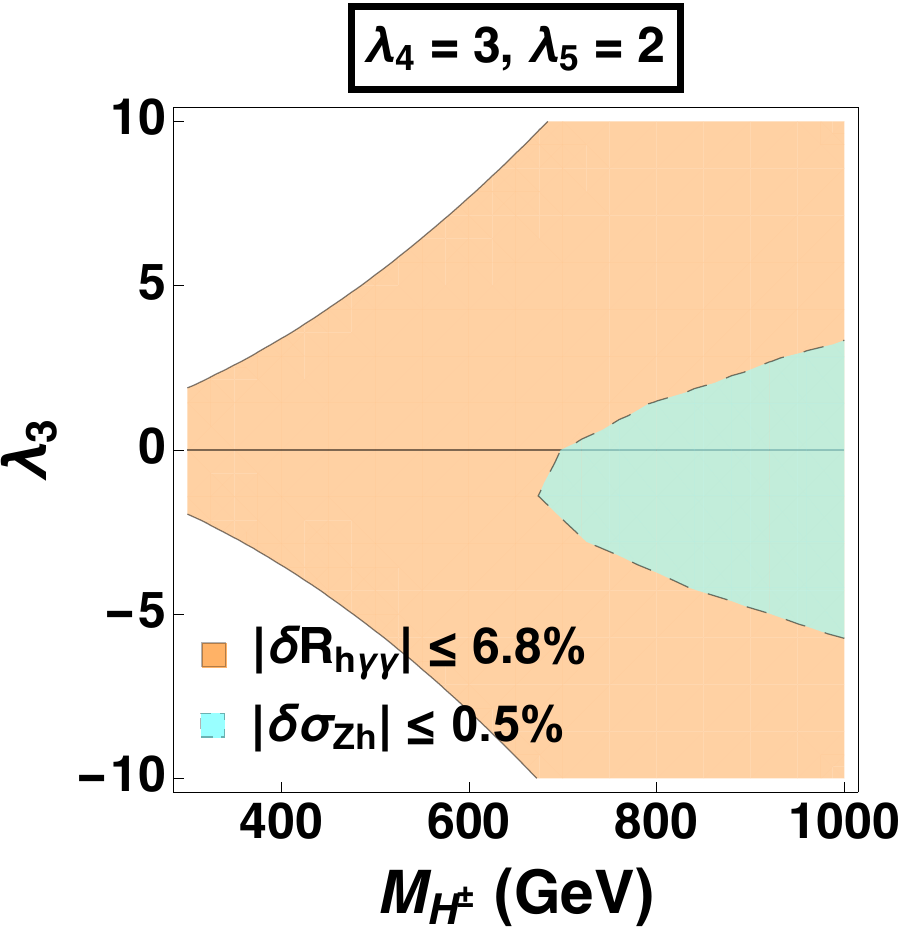}
\hfill
\includegraphics[scale=0.75]{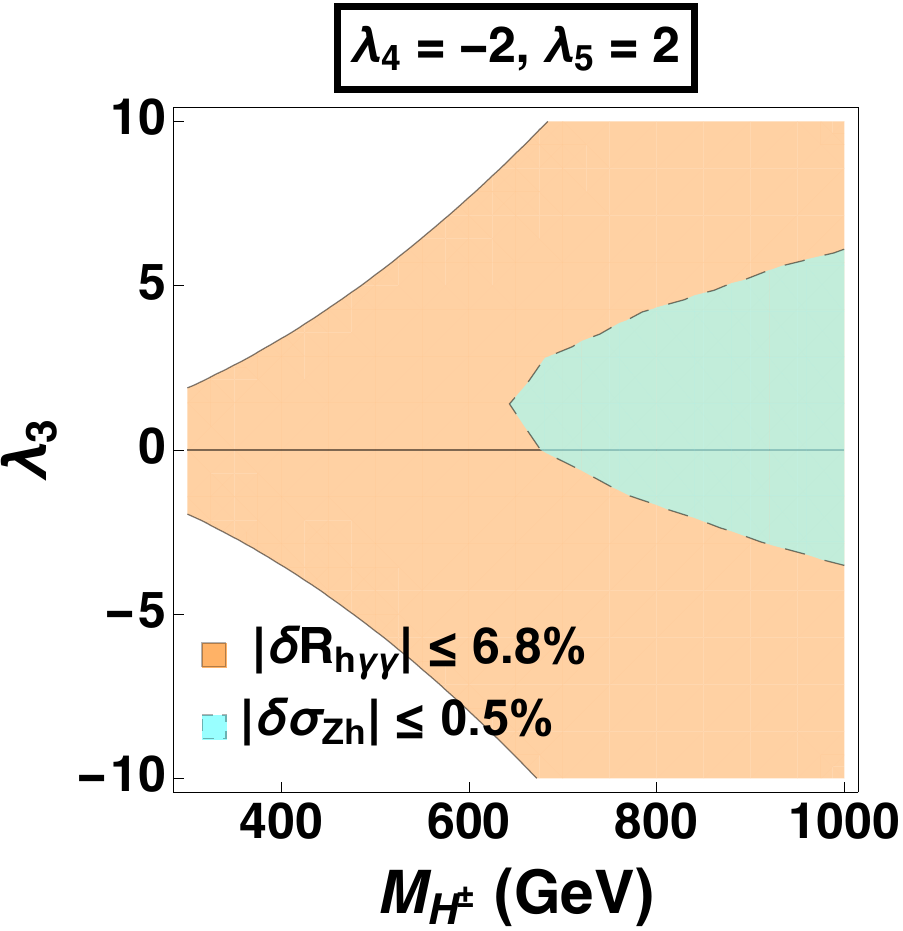}
\end{center}
\caption{Contours in $\brc{\lambda_3,\Mp}$ plane for $\left|\delta\sigma_{Zh}\right|\le0.5\%$
(cyan area outlined with dashed lines)
and
$\left|\delta R_{h\gamma\gamma}\right|\le6.8\%$
(orange area outlined with solid lines)
in inert double model.
From left to right panel, $\lambda_4$ and $\lambda_5$ are set at different values.
The positive and negative $\lambda_4$ shifts the contour by $Zh$ precision measurement downward
and upward accordingly. The non-zero $\lambda_5$ makes greater lower bound of $\Mp$ compared to
vanishing $\lambda_5$.}
\label{fig:idm:l3mu2}
\end{figure}

In the inert double model there are four free parameters that enter the one-loop calculation,
which are 
\(
  \{\mu_2^2,~\lambda_3,~\lambda_4,~\lambda_5\}. 
\)
The inert scalar masses are given by
\begin{equation}
  \begin{aligned}
    \Mp^2 &= \mu_2^2 + \frac{1}{2}\lambda_3 v_\phi^2, \\
    \MHn^2 &= \mu_2^2 + \frac{1}{2}\lambda_L v_\phi^2, \\
    \MAn^2 &= \mu_2^2 + \frac{1}{2}\lambda_A v_\phi^2, 
  \end{aligned}
  \label{eq:idm:mass}
\end{equation}
with $\lambda_{L,A}=\brc{\lambda_3+\lambda_4\pm\lambda_5}$.
Since only the charged component ($H^\pm$) contributes to the correction to the Higgs diphoton decay
(refer to the last term on the right hand side of Eq. (\ref{eq:hgg}) where $g_{ss\gamma}\equiv0$
for neutral scalars), 
it leaves us two parameters to plot the two contours of $\left|\delta\sigma_{Zh}\right|\le0.5\%$ and
$\left|\delta R_{h\gamma\gamma}\right|\le6.8\%$ in the same plane, which are $\{\mu_2,\lambda_3\}$.
By fixing $\lambda_4$ and $\lambda_5$, Fig. \ref{fig:idm:l3mu2} shows the two contours
by $Zh$ and Higgs diphoton precision measurements in the $\brc{\lambda_3,\Mp}$ plane,
where $\Mp$ is rescaled from $\mu_2$ according the first relation in Eq. \ref{eq:idm:mass}.
One can see that the contour by Higgs diphoton precision measurement is unchanged because it depends
on no other parameters but $\lambda_3$ and $\Mp\brc{\mu_2}$.
The contour by $Zh$ precision measurement is however dependent on the choice of parameters $\lambda_4$ and
$\lambda_5$, as illustrated in Fig. \ref{fig:idm:l3mu2}.
We find that compared to both vanishing $\lambda_4$ and $\lambda_5$ which makes a somewhat
central-sit contour,
the positive $\lambda_4$ pushes the contour downward and the negative upward,
and non-vanishing $\lambda_5$ gives rise to higher lower boundary of $\Mp$. 
Note that the sign of $\lambda_5$ alone does not change the shape of contour due to the fact
that the two neutral components interchange the mass definition in Eq. (\ref{eq:idm:mass})
while the Feynman rules involving these two particles are identical
(refer to the Feynman rule in Appendix \ref{sec:feyn}). 

\begin{figure}[thpb]
\begin{center}
\includegraphics[scale=0.8]{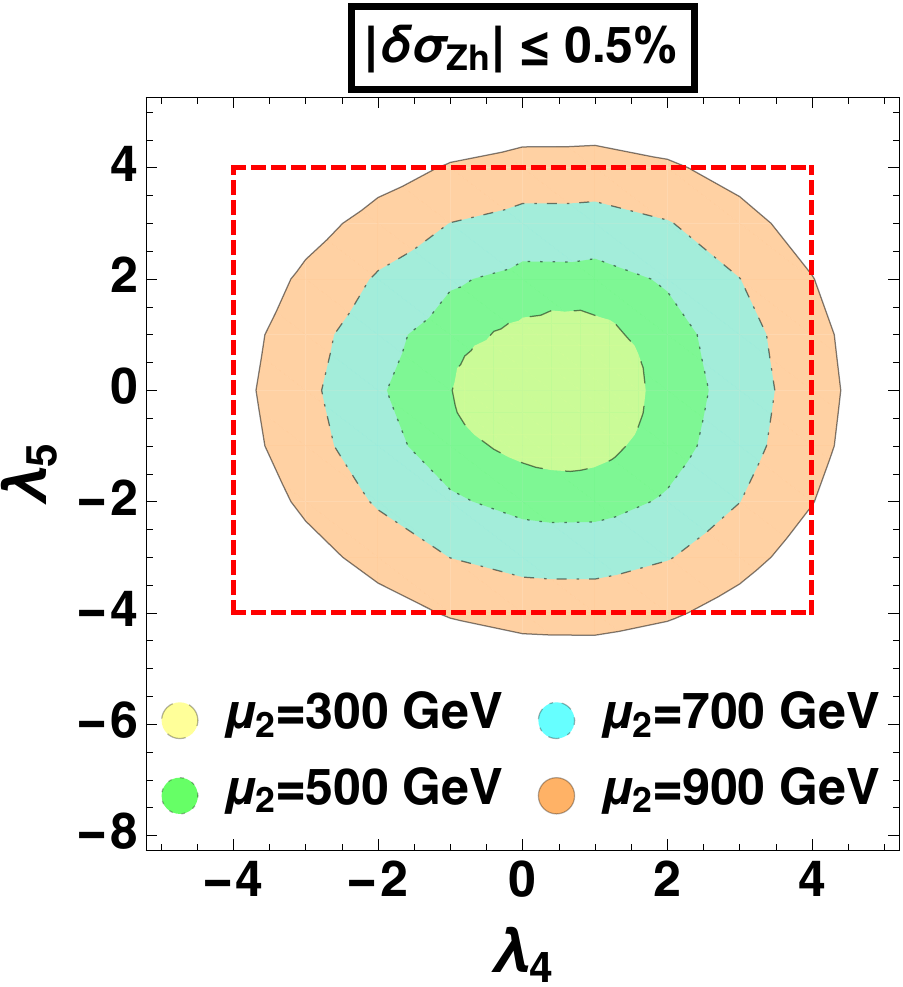}
\end{center}
\caption{Contours in $\brc{\lambda_4,\lambda_5}$ plane for $\left|\delta\sigma_{Zh}\right|\le0.5\%$
in inert double model with $\mu_2=300,~500,~700,~900~\GeV$,
which correspond to concentric areas in yellow, green, cyan and orange, and are respectively outlined with dashed, dotted, dash-dotted and solid lines.
With the increase of $\mu_2$ the contour area gets larger and eventually excesses the region of
perturbativity of the $\lambda$ parameters in the scalar potential.
The allowed region of perturbativity is dictated within the red dashed box according to (\ref{eq:lam:pert}).}
\label{fig:idm:l4l5}
\end{figure}

On the other hand, 
one observes that the inert scalars decouple from the Higgs diphoton process when setting $\lambda_3=0$
because $g_{H^+H^-h}\propto\lambda_3$. In the meantime, $\mu_2$ represents the mass of
the physical charged Higgs $H^\pm$.
In this case we consider the constraints of $\lambda_4$ and $\lambda_5$ by $Zh$ future precision
are utterly complementary with that by Higgs diphoton future precision. 
Hence, one can plot the $Zh$ contours in $\{\lambda_4,\lambda_5\}$ parameter space 
by fixing $\lambda_3=0$ and varying $\Mp\brc{\mu_2})$.
Fig. \ref{fig:idm:l4l5} shows contours of $\left|\delta\sigma_{Zh}\right|\le0.5\%$
with $\Mp=300,~500,~700~,900~\GeV$
in the inert double model, where the area within the red dashed square denotes
the perturbativity area of $\brc{\lambda_4,~\lambda_5}$ \cite{Hambye:1996wb}.   
The contours of $\delta R_{h\gamma\gamma}$ is missing in this case
due to $\delta R_{h\gamma\gamma}\equiv 0$
when $\lambda_3=0$.

\begin{figure}[thpb]
\begin{center}
\includegraphics[scale=0.8]{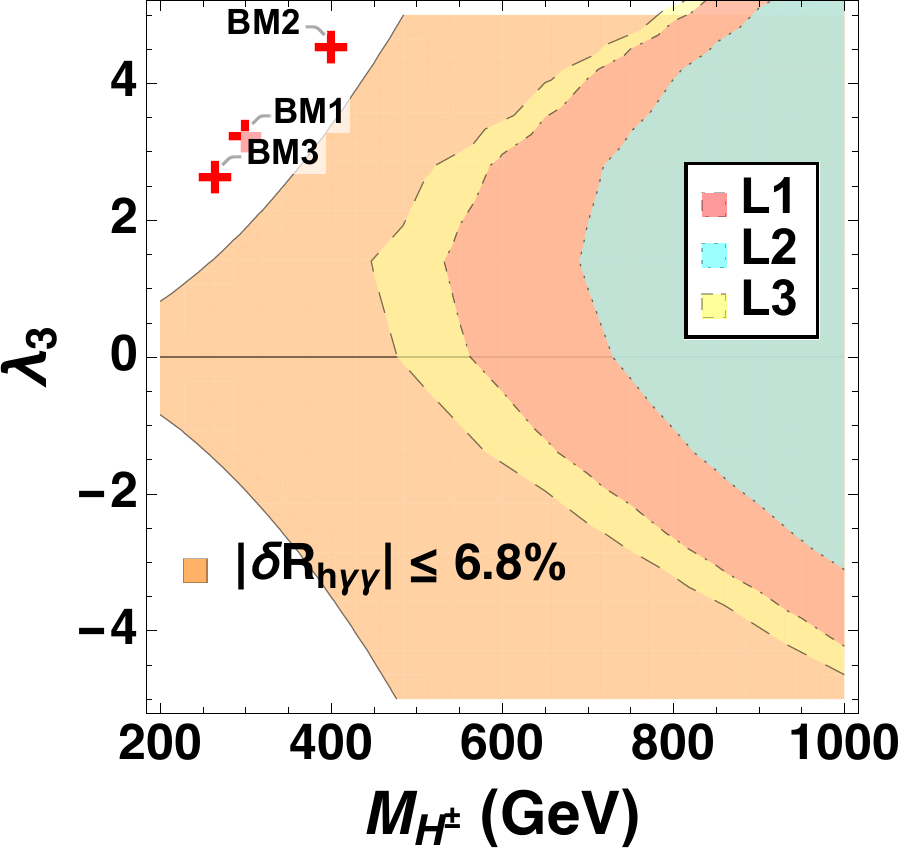}
\end{center}
\caption{Contours in $\brc{\lambda_3,\Mp}$ plane for $\left|\delta\sigma_{Zh}\right|\le0.5\%$
(red, cyan and yellow areas outlined with dotdashed, dotted and dashed lines and labelled with L1, L2 and L3)
and $\left|\delta R_{h\gamma\gamma}\right|\le6.8\%$ (orange area outlined with solid lines).
L1, L2 and L3 respectively denote the choice of $\lambda_4$ and $\lambda_5$ in accordance with
that in BM1, BM2 and BM3, which are listed in Table \ref{tbl:idm:ewpt}. The three benchmark points BM1, BM2 and BM3 are marked with red crosses. 
}
\label{fig:idm:ewpt}
\end{figure}

In order to see if the constraints on the parameters in our study give region of interest where the SFOEWPT
occurs, we exploit the benchmark models (BMs) specified in Ref. \cite{Blinov:2015vma} and mark the 
three benchmark points "BM1, BM2, BM3" corresponding to the three benchmark scenarios in the
$\brc{\lambda_3,\Mp}$ plane along with the contours by $h\gamma\gamma$ decay and $Zh$ 
precision measurement
in Fig. \ref{fig:idm:ewpt}. 
The legends "L1, L2, L3" on the right hand side of the figure denote the contours by $Zh$
precision measurement (i.e., $\left|\delta\sigma_{Zh}\right|\le0.5\%$)
with the fixed $\lambda_4$ and $\lambda_5$ in accordance with that in "BM1, BM2, BM3", respectively.
One can find the input parameters for the three benchmark scenarios in Table \ref{tbl:idm:ewpt}. 
As can be seen from Fig. \ref{fig:idm:ewpt}, all three benchmark points are excluded from the contours by
both the $h\gamma\gamma$ decay and $Zh$ precision measurement at 1-$\sigma$ confidence level.
This tells us that with the expected CEPC precision the constrained parameter space may further
exclude some region that has SFOEWPT and is permitted phenomenologically elsewhere
like the benchmark scenarios used above. 

\begin{table}[thpb]
\centering
\caption{Input parameters for the three benchmark scenarios given in Ref.\cite{Blinov:2015vma}.
Masses are in unit $\GeV$.
Couplings ($\lambda_3,~\lambda_4,~\lambda_5$) are valued at scale $Q=246~\GeV$
with finite temperature correction.}
\label{tbl:idm:ewpt}
\begin{tabular}{|c|c|c|c|c|c|c|}
\hline\hline
  &$\MHn$ &$\MAn$ &$\Mp$ &$\lambda_3$ &$\lambda_4$ &$\lambda_5$ \\
 \hline
 BM1 &66 &300 &300 &3.3 &-1.7 &-1.5 \\
 BM2 &200 &400 &400 &4.6 &-2.3 &-2.0 \\
 BM3 &5 &265 &265 &2.7 &-1.4 &-1.2 \\
 \hline\hline
\end{tabular}
\end{table}

\paragraph{Real Triplet}
The real triplet model has the extended scalar sector a real triplet with the neutral component a
potential WIMP dark matter candidate if the triplet VEV is vanishing~\cite{FileviezPerez:2008bj}.
Due to small mass splitting between the charged and neutral triplet scalars ($\Delta M=166~\MeV$)
\cite{Cirelli:2005uq,Cirelli:2007xd} in the large scalar mass regime $M_\Sigma\gg M_W$,
the direct search for the dark matter involves the process of charged scalar pair production
accompanying the decay to neutral scalar and a soft $\pi^\pm$, the so-called disappearing charge tracks
(DCTs).
Experimentally, the DCT signature is profiled in search for the charginos at the LHC
\cite{Aad:2013yna,CMS:2014gxa,Aaboud:2017mpt,Sirunyan:2018ldc}.
The reach of the real triplet with DCT signature at the 13 TeV LHC and a possible future 100 TeV $pp$
collider is studied in \cite{Chiang:2020rcv}. 
On the other hand, the Higgs portal parameter $\lambda_3$ may play a role in EWPT.
The EWPT in the real triplet has been thoroughly studied in \cite{Niemi:2018asa,Niemi:2020hto}
using dimensional reduction - a three dimensional effective field theory (DR3EFT)
that allows non-perturbative lattice simulation. 

\begin{figure}[thpb]
\begin{center}
\includegraphics[scale=0.75]{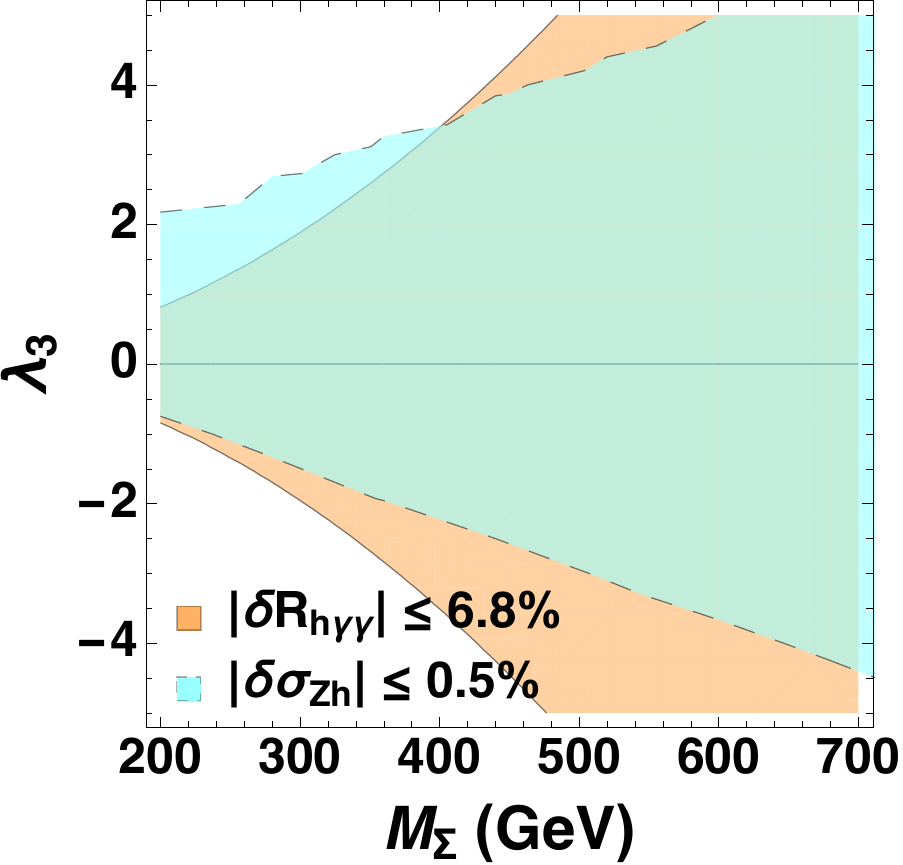}
\end{center}
\caption{Contours in $\brc{\lambda_3,M_\Sigma}$ plane for $\left|\delta\sigma_{Zh}\right|\le0.5\%$
(cyan area outlined with dashed lines)
and
$\left|\delta R_{h\gamma\gamma}\right|\le6.8\%$
(orange area outlined with solid lines)
in real triplet model.}
\label{fig:rt}
\end{figure}

\begin{figure}[thpb]
\begin{center}
\includegraphics[scale=0.75]{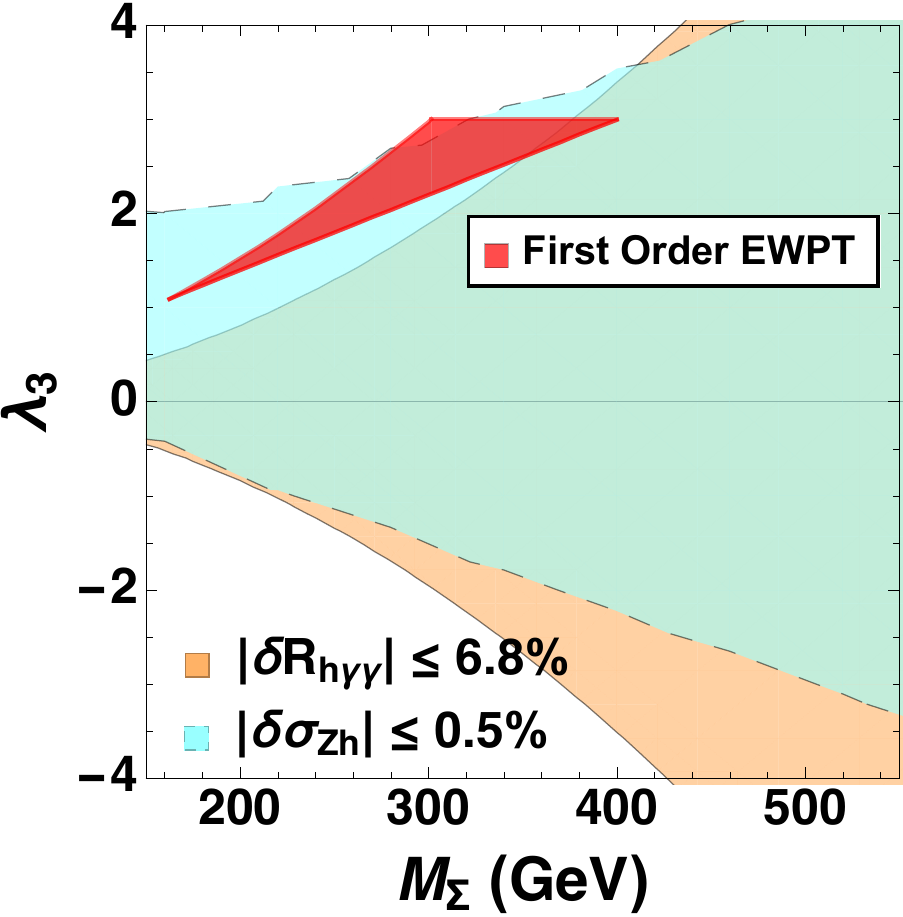}
\end{center}
\caption{Contours same as that in Fig. \ref{fig:rt} with superposition of the red region denoting 
the first order phase transition presented in Ref.~\cite{Niemi:2020hto}.}
\label{fig:rt:new}
\end{figure}

In the real triplet model the free parameters from the potential that enter our calculation are
the mass of the scalar triplet $M_{\Sigma}$ and $\lambda_3$, 
where we have omitted the mass splitting difference between
the neutral and charged components at loop level since the splitting is small
($\Delta M=166~\MeV$ \cite{Cirelli:2005uq,Cirelli:2007xd})
compared to the range of the mass investigated ($M_{\Sigma}\gtrsim 200~\GeV$). 
In this case, one has the contour plots in the parameter space $M_{\Sigma}$ versus $\lambda_3$
for both $\left|\delta\sigma_{Zh}\right|\le0.5\%$ and $\left|\delta R_{h\gamma\gamma}\right|\le6.8\%$
shown in Fig. \ref{fig:rt}, 
and can see that
the contour by the $Zh$ measurement largely overlaps with that by the Higgs diphoton decay
rate measurement. When $M_\Sigma\lesssim400~\GeV$ the Higgs diphoton measurement reduces the
positive bound of $\lambda_3$ while for the rest the $Zh$ measurement gives slight better constraints
upon these two parameters. 
It is interesting to mention that the triplet dark matter direct search gives rise to much narrower
window than the Higgs diphoton decay for $\lambda_3$ when $M_\Sigma$ is in $\TeV$ regime
(see Figure 7 in Ref. \cite{Chiang:2020rcv}). 
Given the first order EWPT region provided in \cite{Niemi:2018asa,Niemi:2020hto},
one can compare this region with 
the constraints on the triplet mass $M_\Sigma$ and the Higgs portal coupling $\lambda_3$
by the precision measurement for $Zh$ production and $h\gamma\gamma$ decay rate
at the CEPC to see in which area of the parameter
space does it occur the first order EWPT, which is illustrated in Fig. \ref{fig:rt:new}, 
where
the red strip represents the up-to-date first order EWPT region 
\footnote{The first order EWPT region obtained in Ref.~\cite{Niemi:2018asa} is a subset of that
calculated in Ref.~\cite{Niemi:2020hto}. It has the same lower bound of the region aligned
with the larger region given in the latter paper.
Both papers performed 
dimensional reduction to an effective three-dimensional theory (DR3EFT) to identify the phase transition regions.
The former paper integrates out the triplet field, assuming it is (super)heavy so that 
there is an area in which the dimensional reduction (DR) no longer holds;
while this area can be accessible with the treatment in the latter paper
and the result is validated with lattice simulation.}
($M_\Sigma\in[100,400]~\GeV$) calculated in~\cite{Niemi:2020hto}
\footnote{The red region represents the one-step first order phase transition $O\rightarrow\phi$. 
In \cite{Niemi:2020hto} there are three thin strips adjacent and above the red region leading to two-step
transitions $O\rightarrow\Sigma\rightarrow\phi$ with $\Sigma\rightarrow\phi$ a first order and
$O\rightarrow\Sigma$ a crossover, a first order or a second order EWPT.
See Figure 1 in \cite{Niemi:2020hto} for the details.}.
As can be seen that with the CEPC required accuracy 
the first order EWPT could be excluded by precision measurement of the $h\gamma\gamma$ decay rate
with $M_\Sigma\lesssim350~\GeV$ while it is permitted in reference to the $Zh$ measurement solely.

\paragraph{Quintuplet and Septuplet}
The quintuplet and septuplet are the high dimensional EW scalar multiplets ($n=5,7$)
carrying zero hypercharge ($Y=0$) so as to have the neutral components which could be potential WIMP
dark matter candidates. The EW scalar multiplets of dimension $n$ are widely studied and discussed in
\cite{Hambye:2009pw,AbdusSalam:2013eya,Chao:2018xwz,Pilkington:2016erq}.
Ref. \cite{Chao:2018xwz} updated the most general renormalizable potentials reported in previous studies
\cite{Hambye:2009pw,AbdusSalam:2013eya}, and focused on $n=5,7$ cases to illustrate the dark matter
phenomenology.
In what follows, we follow the representation of the $\SUtwo$ EW scalar multiplet
and the convention of the formulism adopted in Ref. \cite{Chao:2018xwz}.

The quintuplet and septuplet have the same parameters that appear in the one-loop calculation and the
difference is the fold of the scalar components respectively being $n=5$ and $n=7$.
The scalar masses are given by
\begin{equation}
  \begin{aligned}
    & M_{S_A}^2 =
    M_A^2 + \frac{1}{2}\lambda_1 v^2 + \frac{2}{\sqrt{n}}M_B^2 + \frac{1}{\sqrt{n}}\lambda_3 v^2, \\
    & M_{S_B}^2 =
    M_A^2 + \frac{1}{2}\lambda_1 v^2 - \frac{2}{\sqrt{n}}M_B^2 - \frac{1}{\sqrt{n}}\lambda_3 v^2,
  \end{aligned}
\end{equation}
where the subscripts $A,~B$ denote the type of the scalars, and the mass of each type is equal
regardless of the number of charges carried at tree level.
Analogous to that in the real triplet model the radiative correction gives rise to small mass splitting
between the charged and neutral component when $M_{S_{A,B}}\gg M_W$
that $M_{S_{A,B}^m}-M_{S_{A,B}^0}=m^2\Delta M$
($m=-\frac{n-1}{2},\ldots,\frac{n-1}{2}$, $\Delta M=166~\MeV$ \cite{Cirelli:2005uq,Cirelli:2007xd}).
There are four free parameters from the potential that enter our calculation.
They are the couplings $\lambda_1,~\lambda_3$ and the masses either the unphysical $M_A,~M_B$ or
the physical $M_{S_A},~M_{S_B}$.

\begin{figure}[htp]
  \centering
  \begin{subfigure}{0.45\textwidth}
    \centering
    \includegraphics[scale=0.65]{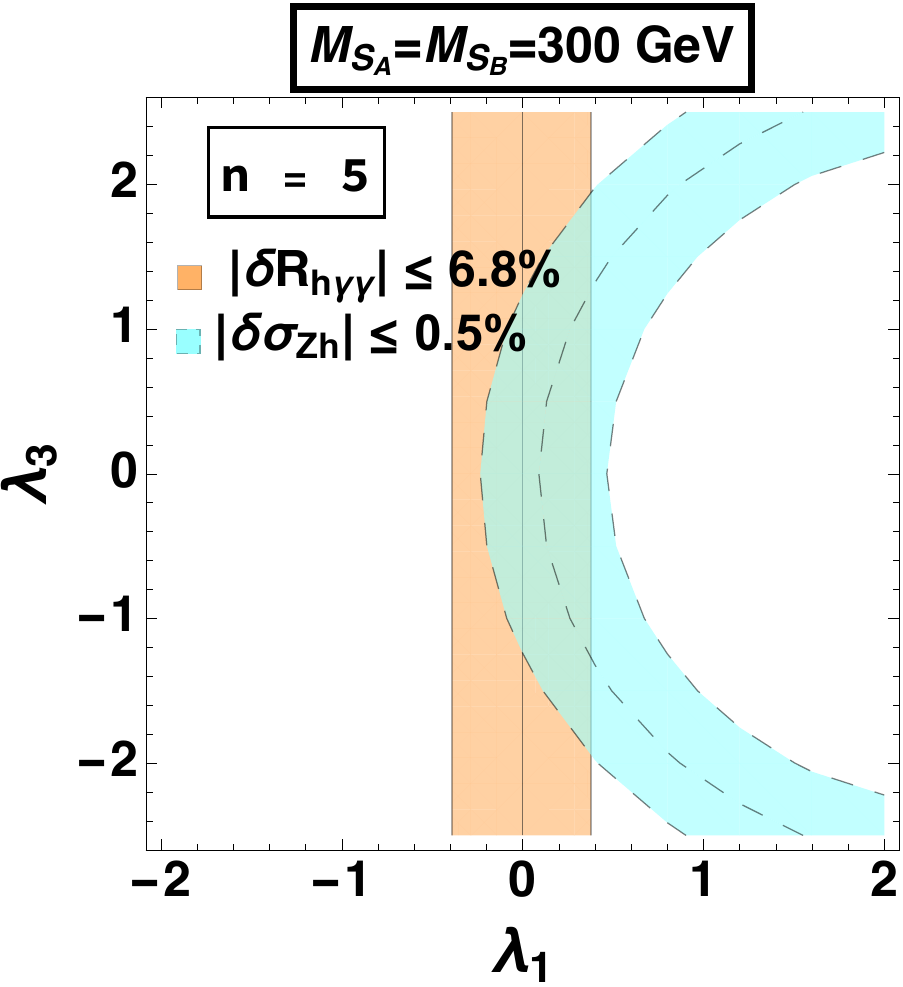}
    \includegraphics[scale=0.65]{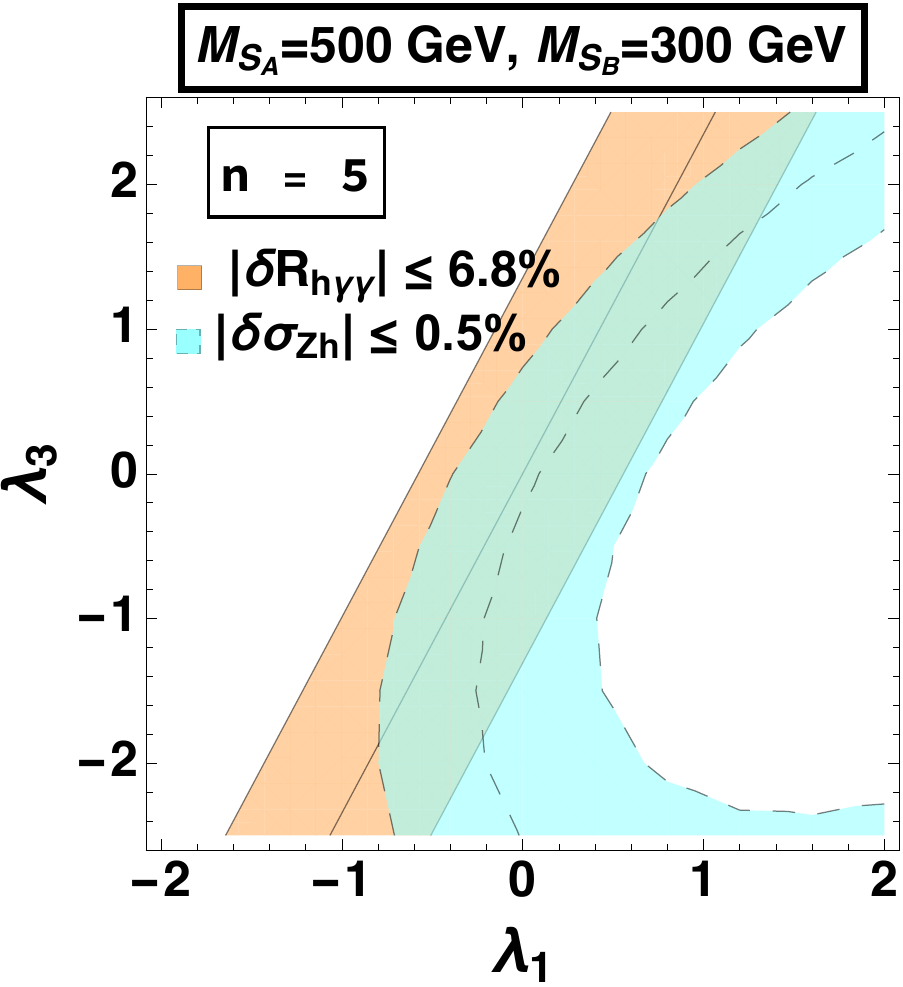}
    \includegraphics[scale=0.65]{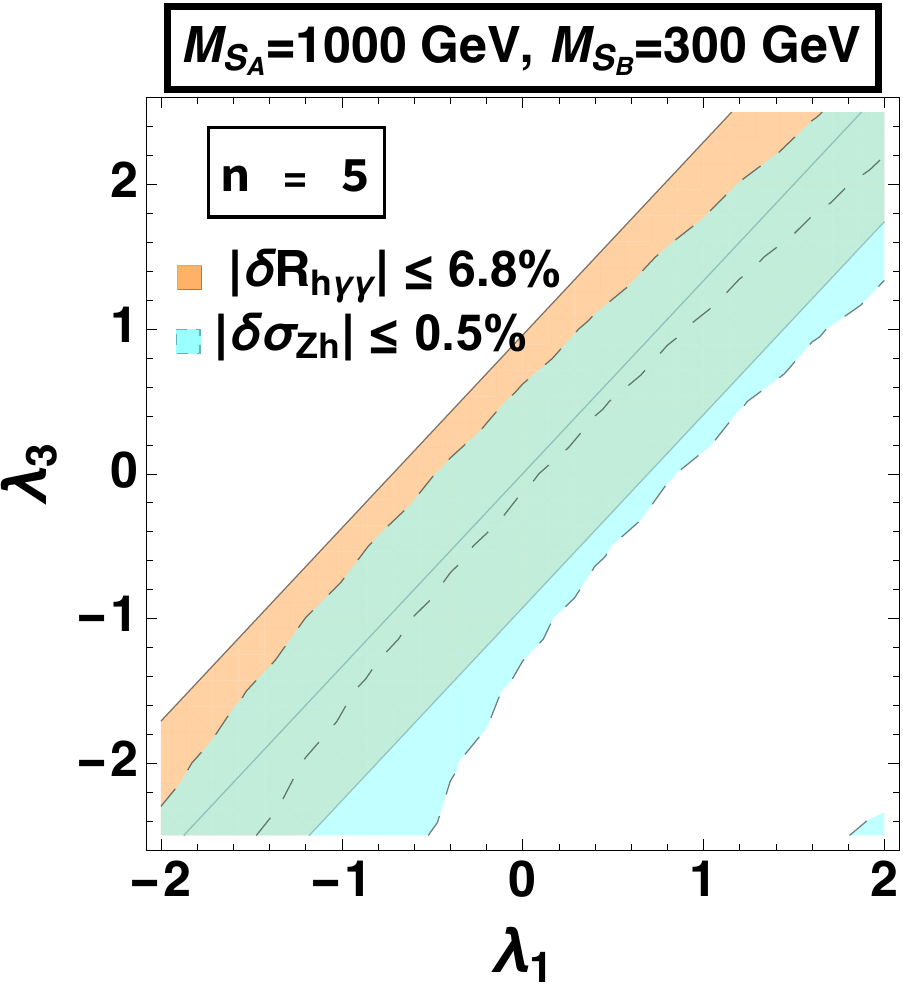}
    \caption{Quintuplet
    }
    \label{subfig:n5}
  \end{subfigure}
  \hfill
  \begin{subfigure}{0.45\textwidth}
    \centering
    \includegraphics[scale=0.65]{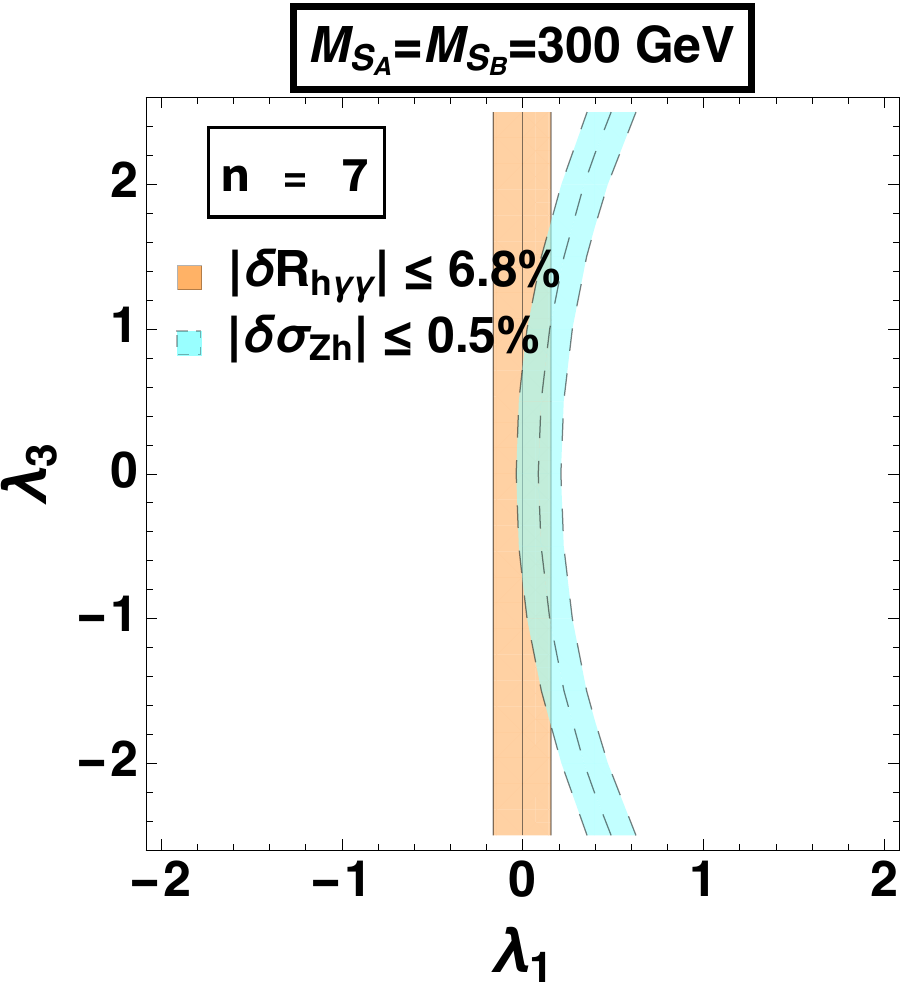}
    \includegraphics[scale=0.65]{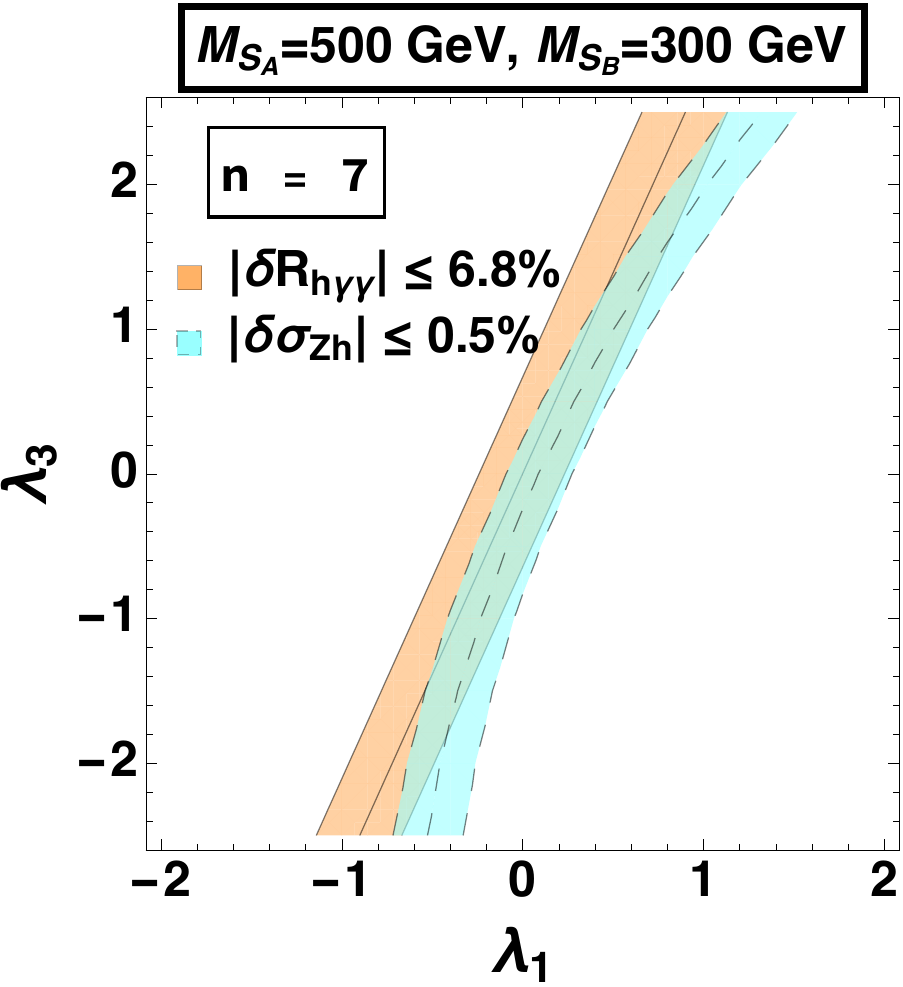}
    \includegraphics[scale=0.65]{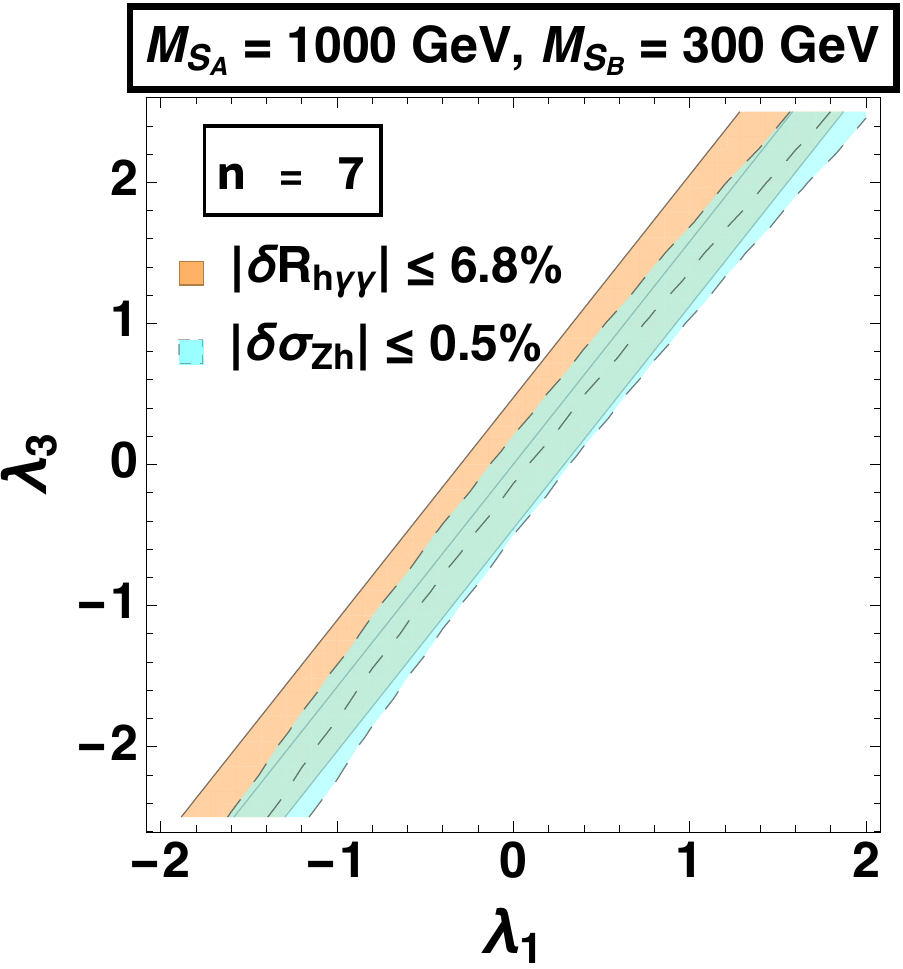}
    \caption{Septuplet
}
    \label{subfig:n7}
  \end{subfigure}
  \caption{Contour comparison for $\left|\delta\sigma_{Zh}\right|\le0.5\%$
  (cyan band outlined with dashed lines)
  and $\left|\delta R_{h\gamma\gamma}\right|\le6.8\%$ (orange band outlined with solid lines)
  for quintuplet and septuplet with fixed values of the physical masses
  $\brc{M_{S_A},~M_{S_B}}=\brc{300,300},~\brc{500,300},~\brc{1000,300}~\GeV$, respectively.}
  \label{fig:n5n7}
\end{figure}

We choose the physical masses $\brc{M_{S_A},M_{S_B}}$ to be fixed
and perform the contour comparison for the
two processes in the $n=5,~7$ scalar model. Fig. \ref{fig:n5n7} illustrates the comparisons with
three sets of fixed $\brc{M_{S_A},M_{S_B}}$. 
As we can see that the overlapping area varies with the setting of $\brc{M_{S_A},M_{S_B}}$,
and the two contours become more degenerate as increases of the mass difference
between the type A and type B scalars. 
When the two physical masses are set to be equal, one observes that the $h\gamma\gamma$ decay
rate constraint appears as a thin straight vertical band in the $\brc{\lambda_1,\lambda_3}$ plane,
meaning the result is independent of the parameter $\lambda_3$.
This is due to the natural of the tri-couplings $g_{s_i^ms_i^{-m}\gamma}$ and $g_{s_i^ms_i^{-m}h}$
($i=A,~B$, the tri-couplings correspond to notation of vertices
in Table \ref{tbl:feyn:gauge} and \ref{tbl:feyn:potential} in Appendix \ref{sec:feyn}),
which have the form
\begin{equation}
  g_{S_i^mS_i^{-m}\gamma}=-em,
  \quad
  g_{S_A^mS_A^{-m}h}=-\brc{\lambda_1+\frac{2}{\sqrt{n}}\lambda_3},
  \quad
  g_{S_B^mS_B^{-m}h}=-\brc{\lambda_1-\frac{2}{\sqrt{n}}\lambda_3}, 
\end{equation}
with $m=0,\ldots,\frac{n-1}{2}$. 
When $M_{S_A}=M_{S_B}$, the loop function $A_0^h\brc{\tau_s}$ in Eq. (\ref{eq:hgg})
and (\ref{eq:hgg:loopfunc}) returns the same value for both $S_A$ and $S_B$,
then the sum of the same charged components
makes the exact cancellation of $\lambda_3$ between type $A$ and $B$ scalars. 

\begin{figure}[thpb]
\begin{center}
\includegraphics[scale=0.75]{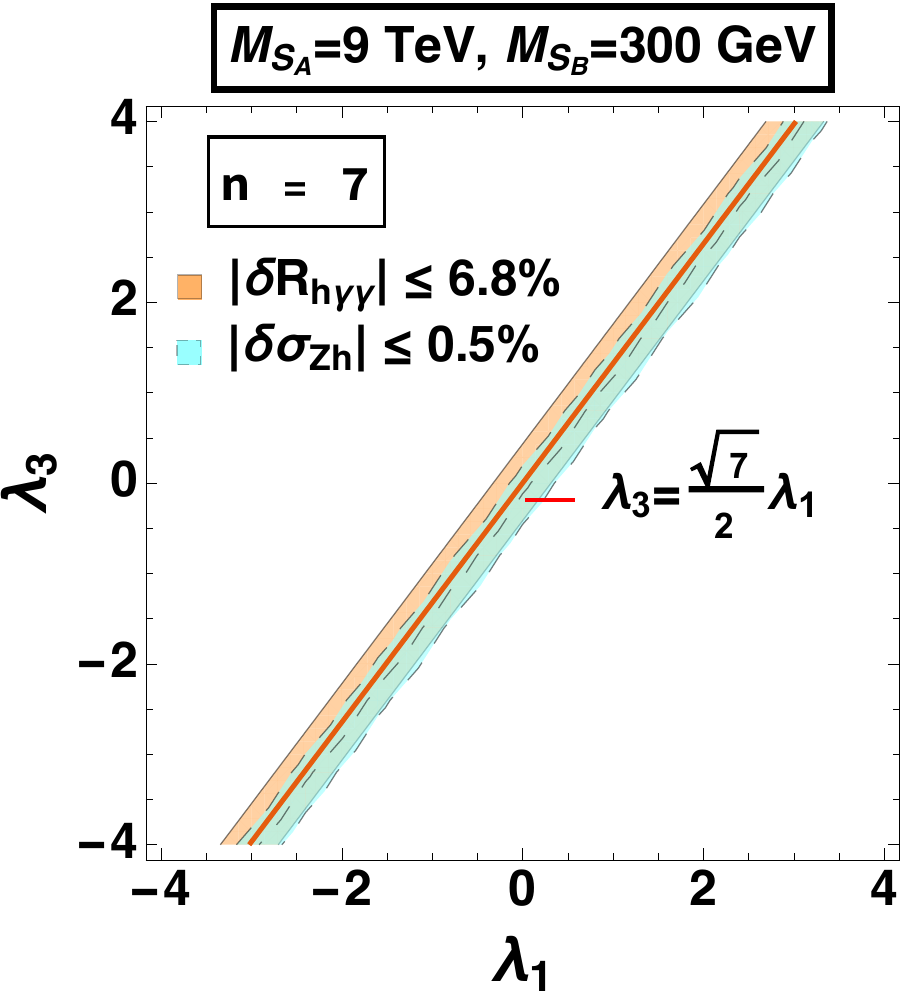}
\end{center}
\caption{Contours by $\left|\delta\sigma_{Zh}\right|\le0.5\%$ (cyan band outlined with dashed lines)
and $\left|\delta R_{h\gamma\gamma}\right|\le6.8\%$ (orange band outlined with solid lines) compared to
the line (red solid) for $\lambda_\text{eff}=0$ for septuplet with fixed values of $M_{S_A}=9$ TeV
and $M_{S_B}=300~\GeV$.}
\label{fig:n7:dm}
\end{figure}

In Ref. \cite{Chao:2018xwz} it was found that the effective coupling is rather small
$\left|\lambda_\text{eff}\right|\ll1$ when saturating the observed relic density and evading the
direct detection limits by LUX \cite{Akerib:2016vxi}, PandaX-II \cite{Cui:2017nnn}
and XENONT1 \cite{Aprile:2018dbl}.  
The effective coupling $\lambda_\text{eff}$ is a linear combination of parameter $\lambda_1$ and $\lambda_3$.
One could take the septuplet for illustration 
to determine the constraints on the parameter space in connection with the dark matter phenomenology. 
The effective coupling in accord with our definition for the septuplet reads
\begin{equation}
\lambda_\text{eff}=\lambda_1-\frac{2}{\sqrt{7}}\lambda_3. 
\label{eq:leff}
\end{equation}
The neutral component of $S_A$ is in general taken as the dark matter candidate.
According to the findings in \cite{Chao:2018xwz}, if we assume the septuplet is the only candidate of dark matter, 
then its mass should be $\sim9$ TeV in the one-species scenario for vanishing $\lambda_\text{eff}$.
In Fig. \ref{fig:n7:dm} it shows contours for $\left|\delta\sigma_{Zh}\right|\le0.5\%$ and
$\left|\delta R_{h\gamma\gamma}\right|\le6.8\%$ compared to the line for vanishing $\lambda_\text{eff}$
with the mass of the dark matter $M_{S_A}=9$ TeV and the choice of $M_{S_B}$ being $300~\GeV$
\footnote{$M_{S_B}$ is a free parameter with no constraint in dark matter phenomenology.}.
As can be seen that the line denoting a vanishing $\lambda_\text{eff}$
goes through the overlapping area of the two contours and 
almost aligns with the central line
of the contour area by $\left|\delta R_{h\gamma\gamma}\right|\le6.8\%$
(the contour with $\delta R_{h\gamma\gamma}\equiv0$).
This is not surprising when the dark matter mass is in multi-TeV regime and $M_{S_B}/M_{S_A}\ll1$.
Referring to Eq. (\ref{eq:hgg}) one of the solutions to minimize $\delta R_{h\gamma\gamma}$
in (\ref{eq:dr:hgg}) is a vanishing last term in (\ref{eq:hgg})
which is proportional to $g_{S_B^mS_B^{-m}h}$ since $\frac{M_W}{g_2M_{S_A}^2}\ll1$ when $M_{S_A}$ is
in multi-TeV regime. And one can see that $g_{S_B^mS_B^{-m}h}=-\lambda_\text{eff}$ for septuplet. 
Therefore, one can conclude that the constraint on parameter space in the dark matter phenomenology
is consistent with that by the precision measurement of $Zh$ production and $h\gamma\gamma$ decay rate.   

\section{Conclusion}
\label{sec:conclusion}


In this work we calculated the one-loop corrections on the Higgs-strahlung process $e^+ e^- \to Z h$ in the presence of an extended scalar sector. 
In the case of zero or tiny VEV for the neutral scalar components, the BSM contribution may be computed separately from the SM EW corrections, making the calculation simpler than one might otherwise expect.
We first classify the extended scalar sector according to their $\SUtwo$ representation.  We then exhibit universal analytic expressions for the one-loop corrections in the $\MSbar$ and on-shell schemes, and subsequently apply the results to representative scalar sector extensions.

Since the Higgs-strahlung process $e^+e^-\to Zh$ and the Higgs diphoton decay process
$h\to\gamma\gamma$ are relatively sensitive to the BSM,  
we investigated the BSM contributions running in loops to these two independent processes to obtain
the constraints on the scalar potentials in the extended scalar models.
The projected precision at the CEPC for the inclusive $\sigma\brc{Zh}$ and $h\to\gamma\gamma$ decay rate
were assumed for purposes of concrete illustration.
We note that the precision for the $Zh$ measurement with the Higgs decaying to bottom quark pair
further reduces compared to the inclusive Higgs production. For instance, the estimated precision of 
$\sigma\brc{Zh}\times\mathrm{BR}\brc{h\to b\bar{b}}$ is 0.27\% at the CEPC, which is approximately a half of
the precision for the inclusive one. One might use narrow width approximation (NWA) to estimate the BSM
contributions to the Higgs-strahlung process accompanying $h\to b\bar{b}$.
Provided the extended scalar sector does not interact with the bottom quark, the BSM contributions to the full
process in NWA would be simply broken down to the BSM loop effects to $\sigma\brc{Zh}$ and
BR$\brc{h\to b\bar{b}}$ at LO.  
In this scenario the contour area of $\left|\delta\sigma_{Zh}\right|$ in the Fig.~\ref{fig:ct} -
\ref{fig:n7:dm} constraining the parameter space in the extended scalar sector shrinks accordingly
with higher precision whereas the complementarity or degeneracy between $Zh$ production and
$h\to\gamma\gamma$ decay rate remains qualitatively.
Nevertheless, due to the NWA accuracy~\cite{Uhlemann:2008pm} and the Yukawa coupling between the extended
scalars and the bottom quark subject to $\mathcal{O}\brc{\frac{m_b}{M_W}}$, the resulting uncertainty
might offset the improvement of the precision by introducing the subsequent $h\to b\bar{b}$ channel.

Based on the numerical results, we found the following generic features in extended scalar models.
First, similar to the oblique $T$ parameter, the $Zh$ cross section rate is sensitive to 
the mass splitting between different components of the multiplet. 
In the real triplet, quintuplet, and septuplet models there is no mass splitting at tree level~\footnote{In principle, there can be mass splitting at tree level in the quintuplet and septuplet models, see Eq. (16) and (17) in Ref.~\cite{Chao:2018xwz} for generic definition of mass eigenvalues, where the term proportional to $\lambda_2$ leads to mass splitting between differently charged scalars. In our study, the $\lambda_2$ contribution is omitted, which is necessary for the neutral scalar to be the DM candidate in the models.}.
In these cases, $\sigma(Zh)$ and the $h\to\gamma\gamma$ signal rate are sensitive to similar regions of model parameter space. The situation differs for the complex triplet model, wherein the mass splitting arises at tree level and the two precision Higgs observables provide complementary parameter space probes~\footnote{
Refer to the explanation in the case of the complex triplet model in Section \ref{subsec:res:comp} for cases with mass splittings. 
When there is no mass splitting, the scalar mass can be a direct input without mass splitting parameter(s), so that the BSM contribution to the self energy in $\sigma\brc{Zh}$ depends on the scalar mass only. 
As both self energy and vertex corrections decrease with the scalar mass,
the vertex corrections to $\sigma\brc{Zh}$ and $h\to\gamma\gamma$
decay rate are sensitive to the triple Higgs couplings in a similar manner 
with relatively large scalar masses.}.
Second, if the mass spectrum in the multiplet has been identified at a hadron collider, one could further extract information on the new scalar-Higgs couplings from the precision Higgs observables analyzed here, because they depend on the new scalar masses, gauge couplings (fixed by gauge invariance), and extended scalar potential couplings. 
In principle, should the neutral component of the new scalar electroweak multiplet contribute  to the dark matter relic density, these precision Higgs observables may further constrain the mass and interactions of some portion of the dark matter.

More specifically,  
\bit
\item In one class of models, such as the real scalar models, both the $e^+ e^- \to Z h$  and diphoton processes probe similar parameter regions in the scalar potential. Thus, for example, should one observable yield a significant difference from SM expectations while the other fall in agreement with the SM, a particular model in this class may be disfavored.

\item In another class of models, such as the complex triplet model, both the $e^+ e^- \to Z h$ and $h\to\gamma\gamma$ processes probe complementary regions of parameter space. To again follow our example hypothetical experimental outcome, agreement with the SM in one observable, coupled with disagreement in the other measurement could be accommodated in this case. 
\eit

One may also perform the foregoing analysis of $\sigma(Zh)$ and the $h\to\gamma\gamma$ signal rate  for other versions of the two Higgs doublet model and singlet scalar models. In these extended scenarios, the new scalar may obtain a non-zero VEV, and thus the new scalar contribution cannot in general be extracted from the SM one. For $\sigma(Zh)$ in particular, one must calculate the full one-loop corrections, including both weak and QED corrections. Such a computation for $\sigma(Zh)$ in the Minimal Supersymmetric Standard Model~\cite{Heinemeyer:2015qbu} and Two Higgs Doublet Model~\cite{Xie:2018yiv} have recently appeared.
The radiative corrections for $\sigma\brc{Zh}$ in the Minimal Dilaton Model (MDM)~\footnote{The Minimal Dilaton Model is an extension of the SM by introducing one singlet scalar called dilaton.}
was considered in~\cite{Cao:2014ita}, where partial contributions were neglected due to their size in the MDM\footnote{The negligible contributions include: (a) the correction mediated by $t'$ quark; (b) loops involving the $sZZ$ interaction but no possible large self-couplings among the scalars.}.
Looking to the future, it would be interesting to compare the $Zh$ cross section in the other extended scalar models with and without vanishing  VEVs.

\acknowledgments
J.H.Y. was supported by the National Natural Science Foundation of China (NSFC) under Grants No. 12022514 and No. 11875003. J.H.Y. was also supported by National Key Research and Development Program of China Grant No. 2020YFC2201501 and the NSFC under Grants No. 12047503.
MJRM and JZ were supported in part under U.S. Department of Energy contract No. DE-SC0011095. MJRM was also supported in part under National Natural Science Foundation of China grant No. 19Z103010239. 

\appendix

\def\bmtx#1{\begin{bmatrix}#1\end{bmatrix}}
\allowdisplaybreaks
\section{Feynman Rules}
\label{sec:feyn}
The Feynman rules with regard to the scalar multiplets coupling to gauge bosons are introduced
via the kinetic term in the Lagrange
$\mathcal{L}_{\mathrm{kin}}^{\bphi}\brc{D_\mu\bphi}^\dagger\brc{D^\mu\bphi}$
with the covariant derivative defined in Eq. (\ref{eq:cov_deriv}), and the $\SUtwo$ generators $T^a$ in
representation $j$ have the following generic matrix form,
{\scriptsize
\begin{eqnarray}
&&T^{+}=\left(\begin{array}{ccccccc}
0  &  \sqrt{2j}  & 0  & 0 &    \ldots  &  0  &  0 \\
0  &  0  &  \sqrt{2(2j-1)}  &  0  &  \ldots  &  0  &  0  \\
0  & 0  &  0  & \sqrt{6(j-1)}   &  \ldots  &  0  &  0  \\
\vdots  &  \vdots  &  \vdots  &  \vdots   &  \vdots  &  \vdots  &  \vdots   \\
0  &  0 &  0  &  0  & \ldots & \sqrt{2(2j-1)}  & 0  \\
0  &  0 &  0  &  0 & \ldots  &  0  & \sqrt{2j} \\
0  &  0 &  0  &  0  & \ldots  &  0  & 0
\end{array}\right),\\
&&T^{-}=\left(\begin{array}{ccccccc}
0 &  0  &  0 & \ldots & 0 &  0  &  0 \\
\sqrt{2j}  & 0  &  0 &  \ldots & 0 &  0  &  0 \\
0  & \sqrt{2(2j-1)}  &   0  & \ldots & 0 &  0  &  0 \\
0  & 0  &  \sqrt{6(j-1)} &  \ldots & 0 &  0  &  0 \\
\vdots  & \vdots & \vdots & \vdots & \vdots & \vdots  &  \vdots \\
0  & 0  & 0 & \ldots &  \sqrt{2(2j-1)} & 0  & 0 \\
0  & 0  & 0 & \ldots &  0 & \sqrt{2j}  & 0
\end{array}
\right),\\
&&T^{3}=\left(\begin{array}{cccccc}
j  ~&~ 0 ~&~  0 ~&~ \ldots  ~&~ 0  ~&~ 0\\
0 ~&~  j-1 ~&~ 0 ~&~ \ldots ~&~ 0  ~&~ 0\\
0  ~&~  0 ~&~ j-2 ~&~ \ldots  ~&~ 0  ~&~ 0\\
\vdots ~&~  \vdots ~&~ \vdots ~&~ \vdots ~&~ \vdots  ~&~ \vdots \\
0  ~&~  0 ~&~ 0 ~&~ \ldots  ~&~ -j+1 ~&~ 0\\
0  ~&~  0 ~&~ 0 ~&~ \ldots  ~&~ 0 ~&~ -j
\end{array}
\right).
\end{eqnarray}
}
Expanding the kinetic Lagrange $\mathcal{L}_\mathrm{kin}^{\bphi}$
with the specific form of generator $T^a$ in representation $j$, it is easy to read off the trilinear
and quartic couplings. 
The generic gauge Feynman rules are listed in Table. \ref{tbl:feyn:gauge} below
for the scalar multiplets,
\begin{table}[thpb]
  \centering
  \scriptsize
  \caption{Gauge Feynman rules for the EW scalar multiplets}
  \label{tbl:feyn:gauge}
  \begin{tabular}{|c|c|}
    \hline\hline
    Vertices &Couplings \\
    \hline
    $S^{\mp Q}S^{\pm Q\mp 1}W_\mu^\pm$
    &$ig_2 N_{S^\mp Q}N_{S^{\pm Q\mp 1}}\sqrt{2}\sqrt{\brc{j+m}\brc{j-m+1}}\brc{k_{S^{\mp Q}}-k_{S^{\pm Q\mp 1}}}_\mu$, $(1\le Q\le j+\frac{Y}{2})$ \\
    \hline
    $S^{Q}S^{-Q}W_\mu^+W_\nu^-$
    &$\pm 2g_2^2\left[j\brc{j+1}-m^2\right]g_{\mu\nu}$ \\
    \hline
    $S_a^Q S_b^{-Q}Z_\mu$
    &$\brc{i}^{\delta_{ab}}\brc{g_2c_W m + \brc{-1}^{\delta_{ab}} g_1 s_W\frac{Y}{2}}\brc{k_{S^Q}-k_{S^{-Q}}}_\mu$ \\
    \hline
    $S_a^Q S_b^{-Q}\gamma_\mu$
    &$\brc{i}^{\delta_{ab}}\brc{g_2s_W m - \brc{-1}^{\delta_{ab}}g_1c_W\frac{Y}{2}}\brc{k_{S^Q}-k_{S^{-Q}}}_\mu$ \\
    \hline
    $S^Q S^{-Q}Z_\mu Z_\nu$
    &$2\brc{g_2c_W m - g_1 s_W\frac{Y}{2}}^2 g_{\mu\nu}$ \\
    \hline
    $S^Q S^{-Q}\gamma_\mu\gamma_\nu$
    &$2\brc{g_2s_W m + g_1c_W\frac{Y}{2}}^2 g_{\mu\nu}$ \\
    \hline
    $S^Q S^{-Q}Z_\mu\gamma_\nu$
    &$2\brc{g_2c_W m - g_1s_W\frac{Y}{2}}\brc{g_2s_W m + g_1c_W\frac{Y}{2}} g_{\mu\nu}$ \\
    \hline\hline
  \end{tabular}
\end{table}
where the normalized factor
$N_{S^\mp Q}N_{S^{\pm Q\mp 1}}=\braket{\Phi^{j\pm m}}{\Phi^{j\pm m\mp 1}}$,
$Q=m+\frac{Y}{2}$ is the charge of the scalar component and positive definite in the first vertex
that is indicated in parenthesis.
The Feynman rules in the scalar potential are more complicated as they are representation dependent.
In what follows, these Feynman rules are given in Table \ref{tbl:feyn:potential}
for each model under consideration.
\begin{table}[thpb]
  \centering
  \scriptsize
  \caption{Feynman rules in the scalar potential for Inert Doublet, Real/Complex Triplet,
    Quintuplet and Septuplet Models}
  \label{tbl:feyn:potential}
  \begin{tabular}{|c|c|c|}
    \hline\hline
    Models &vertices &couplings \\
    \hline
    \multirow{6}{*}{Inert Doublet:\quad
      $\bphi=\begin{bmatrix}H^+ \\ \frac{1}{\sqrt{2}}\brc{H^0+iA^0}\end{bmatrix}$}
    &$H^+H^-h$ &$-\lambda_3 v_\phi$ \\
    &$H^0H^0h$ &$-\brc{\lambda_3+\lambda_4+\lambda_5}v_\phi$ \\
    &$A^0A^0h$ &$-\brc{\lambda_3+\lambda_4-\lambda_5}v_\phi$ \\
    &$H^+H^-hh$ &$-\lambda_3$ \\
    &$H^0H^0hh$ &$-\brc{\lambda_3+\lambda_4+\lambda_5}$ \\
    &$A^0A^0hh$ &$-\brc{\lambda_3+\lambda_4-\lambda_5}$\\
    \hline
    \multirow{2}{*}{Real Triplet:\quad
      $\bphi=\bmtx{H^+, ~H^0, ~-H^-}^T$}
    &$H^+H^-h,~H^0H^0h$ &$-\lambda_3 v_\phi$ \\
    &$H^+H^-hh,~H^0H^0hh$ &$-\lambda_3$ \\
    \hline
    \multirow{6}{*}{Complex Triplet:\quad
      $\bphi=\bmtx{H^{++} \\H^+ \\\frac{1}{\sqrt{2}}\brc{H+iA}}$}
    &$H^{++}H^{--}h$ &$-\lambda_4 v_\phi$ \\
    &$H^+H^-h$ &$-\brc{\lambda_4+\frac{\lambda_5}{2}}v_\phi$ \\
    &$HHh,~AAh$ &$-\brc{\lambda_4+\lambda_5}v_\phi$ \\
    &$H^{++}H^{--}hh$ &$-\lambda_4$ \\
    &$H^+H^-hh$ &$-\brc{\lambda_4+\frac{\lambda_5}{2}}$ \\
    &$HHhh,~AAhh$ &$-\brc{\lambda_4+\lambda_5}$ \\
    \hline
    \multirow{4}{*}{Quintuplet ($n=5$) \& Septuplet ($n=7$): \quad
      $\bphi=\bmtx{S_A^j+iS_B^j \\iS_A^{j-1}-S_B^{j-1} \\\vdots \\S_A^{-j}+iS_B^{-j}}$}
    &$S_A^mS_A^{-m}h,~\brc{m=0,\ldots,j}$ &$-\brc{\lambda_1+\frac{2}{\sqrt{n}}\lambda_3}v_\phi$ \\
    &$S_B^mS_B^{-m}h,~\brc{m=0,\ldots,j}$ &$-\brc{\lambda_1-\frac{2}{\sqrt{n}}\lambda_3}v_\phi$ \\
    &$S_A^mS_A^{-m}hh,~\brc{m=0,\ldots,j}$ &$-\brc{\lambda_1+\frac{2}{\sqrt{n}}\lambda_3}$ \\
    &$S_B^mS_B^{-m}hh,~\brc{m=0,\ldots,j}$ &$-\brc{\lambda_1-\frac{2}{\sqrt{n}}\lambda_3}$ \\
    \hline\hline
  \end{tabular}
\end{table}

\section{Potentials in Specific Models}
\label{sec:potentials}
Let us discuss the models with the extended scalar sector of dimension $n$.
For the models under consideration in the context ($n=2,3,5,7$), 
there is an imposition of $\Ztwo$ symmetry to have
stable neutral component as DM candidate\footnote{$\Ztwo$ symmetry is protected in complex triplet model when the triplet has vanishing VEV.}.

\noindent\underline{1. Real and Complex Singlet model~\cite{Barger:2007im, Barger:2008jx}:}
In this case, usually $Y = 0$ is taken, and thus additional term exists:
\bea
 \left({\overline H} H\right)_0 \Phi, \quad \left({\overline \Phi} \Phi\right) \Phi, \quad  \Phi^3. 
\eea
The scalar potential is
\begin{itemize}
\item real scalar singlet
\begin{equation}
\begin{aligned}
V\brc{H,\bphi}=&\frac{m^2}{2}\bh^\dagger\bh + \frac{\lambda}{4}\brc{\bh^\dagger\bh}^2
+ \frac{\delta_1}{2}\bh^\dagger\bh\bphi + \frac{\delta_2}{2}\bh^\dagger\bh\bphi^2 \\
&+ \brc{\frac{\delta_1 m^2}{2\lambda}}\bphi + \frac{\kappa_2}{2}\bphi^2
+ \frac{\kappa_3}{3}\bphi^3 + \frac{\kappa_4}{4}\bphi^4.
\end{aligned}
\end{equation}
\item complex scalar singlet
\begin{equation}
\begin{aligned}
V\brc{H,\bphi}=&\frac{m^2}{2}\bh^\dagger\bh + \frac{\lambda}{4}\brc{\bh^\dagger\bh}^2
+\brc{\frac{\left|\delta_1\right|e^{i\phi_{\delta_1}}}{4}\bh^\dagger\bh\bphi+\mathrm{h.c.}}
+\frac{\delta_2}{2}\bh^\dagger\bh\left|\bphi\right|^2 \\
&+\brc{\frac{\left|\delta_3\right|e^{i\phi_{\delta_3}}}{4}\bh^\dagger\bh\bphi^2+\mathrm{h.c.}}
+\brc{\left|a_1\right|e^{i\phi_{a_1}}\bphi+\mathrm{h.c.}}
+\brc{\frac{\left|b_1\right|e^{i\phi_{b_1}}}{4}\bphi^2+\mathrm{h.c.}} \\
&+\frac{b_2}{2}\left|\bphi\right|^2
+\brc{\frac{\left|c_1\right|e^{i\phi_{c_1}}}{6}\bphi^3+\mathrm{h.c.}}
+\brc{\frac{\left|c_2\right|e^{i\phi_{c_2}}}{6}\bphi\left|\bphi\right|^2+\mathrm{h.c.}} \\
&+\brc{\frac{\left|d_1\right|e^{i\phi_{d_1}}}{8}\bphi^4+\mathrm{h.c.}}
+\brc{\frac{\left|d_3\right|e^{i\phi_{d_3}}}{8}\bphi^2\left|\bphi\right|^2+\mathrm{h.c.}}
+\frac{d_2}{4}\left|\bphi\right|^4
\end{aligned}
\end{equation}
\end{itemize}

\noindent\underline{2. Two Higgs Doublet model~\cite{Ma:2006km,Barbieri:2006dq,Branco:2011iw}:}
For the complex doublet $n = 2, Y = + 1$, additional terms appear
\bea
	\left[\left(H H\right)_1 \left({\overline \Phi}~{\overline\Phi}\right)_1 \right]_0, \quad \left[\left(H H\right)_1 \left({\overline H}~{\overline\Phi}\right)_1 \right]_0
	\quad \left[\left(H \Phi\right)_1 \left({\overline \Phi}~{\overline\Phi}\right)_1 \right]_0. 
\eea
If $n = 2, Y = - 1$, we have the above expression with $\Phi$ replaced by $\overline \Phi$. 
Thus the scalar potential in the two Higgs doublet model is written as
\begin{equation}
  \begin{aligned}
    V\brc{\bh,\bphi} = &~\mu_1^2~\bh^\dagger\bh + \mu_2^2~\bphi^\dagger\bphi
    + \mu_{12}^2\brc{\bphi^\dagger\bh+\mathrm{h.c.}}
    + \lambda_1\brc{\bh^\dagger\bh}^2
    + \lambda_2\brc{\bphi^\dagger\bphi}^2 \\
    &+ \lambda_3\brc{\bh^\dagger\bh}\brc{\bphi^\dagger\bphi}
    + \lambda_4\brc{\bh^\dagger\bphi}\brc{\bphi^\dagger\bh}
    + \left[
      \frac{\lambda_5}{2}\brc{\bh^\dagger\bphi}^2 + \mathrm{h.c.}
      \right], 
  \end{aligned}
\end{equation}
where the soft $\Ztwo$ breaking term (the one $\propto\mu_{12}^2$) is dropped in the inert doublet model.

\noindent\underline{3. Real Triplet model~\cite{Blank:1997qa, FileviezPerez:2008bj, Chen:2008jg}:}
Here we consider $Y = 0$. 
For real triplet $n = 3, Y = 0$, there are additional term
\bea
  \left[\left({\overline H} H\right)_1 \Phi\right]_0.
\eea
Thus the general potential is
\begin{equation}
  \begin{aligned}
    V\brc{\bh,\bphi} = &~\mu_1^2~\bh^\dagger\bh
    + \frac{\mu_2^2}{2}~\bphi^\dagger\bphi
    + \lambda_1\brc{\bh^\dagger\bh}^2 
    + \frac{\lambda_2}{4}\brc{\bphi^\dagger\bphi}^2
    + \frac{\lambda_3}{2}\brc{\bh^\dagger\bh}\brc{\bphi^\dagger\bphi}. 
  \end{aligned}
\end{equation}

\noindent\underline{4. Complex Triplet model~
\cite{AbdusSalam:2013eya,Konetschny:1977bn,Magg:1980ut,Schechter:1980gr,Cheng:1980qt}:}
For the complex triplet, there is another possibility with $n = 3, Y = \pm 2$, additional terms appear
\bea
\left[\left({\overline H} H\right)_1 \overline{\Phi}\right]_0, \quad \left[\left({\overline H} H\right)_1 \Phi\right]_0. 
\eea
The general potential is  
\begin{equation}
  \begin{aligned}
    V\brc{\bh,\bphi} = &~\mu_1^2~\bh^\dagger\bh + \mu_2^2~\bphi^\dagger\bphi
    + \lambda_1\brc{\bh^\dagger\bh}^2
    + \lambda_2\brc{\bphi^\dagger\bphi}^2
    + \lambda_3\left|\bphi^\dagger T^a\bphi\right|^2 \\
    & + \lambda_4\bh^\dagger\bh\bphi^\dagger\bphi
    + \lambda_5\bh^\dagger\tau^a\bh\bphi^\dagger T^a\bphi. 
\end{aligned}
\end{equation}
It is customary to write the complex scalar triplet fields in a $2\times2$ representation $\Delta$.
Following the notation of \cite{Du:2018eaw} with $Y=2$\footnote{$Y=-2$ gives the multiplet that is conjugate to the one with $Y=2$.}, the potential is written
\begin{equation}
\begin{aligned}
V\brc{\bh,\Delta} = &-m^2\bh^\dagger\bh + M^2\mathrm{Tr}\brc{\Delta^\dagger\Delta}
+\brc{\mu\bh^\mathrm{T}i\tau_2\Delta^\dagger\bh + \mathrm{h.c.}}
+\lambda_1\brc{\bh^\dagger\bh}^2 
+\lambda_2\mathrm{Tr}^2\brc{\Delta^\dagger\Delta} \\
&+\lambda_3\mathrm{Tr}\brc{\Delta^\dagger\Delta\Delta^\dagger\Delta}
+\lambda_4\brc{\bh^\dagger\bh}\mathrm{Tr}\brc{\Delta^\dagger\Delta}
+\lambda_5\bh^\dagger\Delta\Delta^\dagger\bh,
\label{eq:V:ct}
\end{aligned}
\end{equation}
with
\[
\Delta=\left(
\begin{array}{cc}
\frac{\Delta^+}{\sqrt{2}}\, &H^{++} \\
\frac{1}{\sqrt{2}}\brc{\delta+v_\Delta+i\eta}\, &-\frac{\Delta^+}{\sqrt{2}}
\end{array}
\right)
\]
The relation between the $3\times1$ and $2\times2$ representation is
\[
\Delta=\bphi\cdot\tau^a.
\]

\noindent\underline{5. Complex Quadruplet model~\cite{AbdusSalam:2013eya}:}
For complex triplet $n = 4, Y = \pm 1$, additional terms appear
\bea
	\left[\left(H H\right)_1 \left( {\overline H}~{\overline\Phi}\right)_1 \right]_0, \quad \left[\left(H H\right)_1 \left( {\overline H}  { \Phi}\right)_1 \right]_0, 
\eea
and for $Y = \pm 3$
\bea
	\left[\left(H H\right)_1 \left( H {\overline\Phi}\right)_1 \right]_0, \quad \left[\left(H H\right)_1 \left( H { \Phi}\right)_1 \right]_0. 
\eea
The scalar potential is
\begin{equation}
  \begin{aligned}
    V\brc{\bh,\bphi} = &~\mu_1^2~\bh^\dagger\bh + \mu_2^2~\bphi^\dagger\bphi
    + \lambda_1\brc{\bh^\dagger\bh}^2
    + \lambda_2\brc{\bphi^\dagger\bphi}^2
    + \lambda_3\left|\bphi^\dagger T^a\bphi\right|^2 \\
    & + \alpha\bh^\dagger\bh\bphi^\dagger\bphi
    + \beta\bh^\dagger\tau^a\bh\bphi^\dagger T^a\bphi 
    + \gamma\left[
    \brc{\bh^T\epsilon\tau^a\bh}\brc{\bphi^T C T^a\bphi}^\dagger
    + \text{h.c.}
    \right], 
\end{aligned}
\end{equation}
where $C$ is an antisymmetric matrix defined as
\[
CT^aC^{-1}=-T^{aT}, 
\]
and its explicit form as well as that of $\epsilon$ can be found in appendix (6.2) in \cite{AbdusSalam:2013eya}.

\noindent\underline{6. Complex and Real Quintuplet and Septuplet models~\cite{Chao:2018xwz}:} For complex $n = 5, 7$ and $Y = 0$, additional terms appear
\bea
 \left[\left(\Phi{\Phi}\right)_J   \Phi \right]_0, \quad \left[\left(\Phi{\Phi}\right)_J  {\overline\Phi} \right]_0.
\eea
The scalar potential is 
\begin{equation}
  \begin{aligned}
    V\brc{\bh,\bphi} = &~-\mu^2~\bh^\dagger\bh
    + M_A^2\brc{\bphi^\dagger\bphi}
    + \left[
      M_B^2\brc{\bphi\bphi}_0 + \mathrm{h.c.}
      \right]
    + \lambda\brc{\bh^\dagger\bh}^2 \\
    &+ \lambda_1\brc{\bh^\dagger\bh}\brc{\bphi^\dagger\bphi}
    + \lambda_2\left[\brc{\overline{\bh}\bh}_1\brc{\overline{\bphi}\bphi}_1\right]
    + \left[
      \lambda_3\brc{\overline{\bh}\bh}_0\brc{\overline{\bphi}\bphi}_0 + \mathrm{h.c.}
      \right]
  \end{aligned}
\end{equation}
It was, however, noted that for the quintuplet ($n=5$) the stability of the DM is ensured
by either the imposition of $\Ztwo$ symmetry or a heavy mass scale $\Lambda$
well above the Planck scale inversely in a non-renormalizable dimension five operator;
while for the septuplet ($n=7$) DM stability could be ensured by choosing $\Lambda$
below the Planck scale.

\noindent\underline{7. Generic  models~\cite{Chao:2018xwz}:} For generic multiplets other than special cases above, the general potential (except the mass terms) is written as
\begin{itemize}
\item Integer isospin $j$
\begin{equation}
\begin{aligned}
V\brc{\bh,\bphi}=&\sum_{J=0}^{2j}\lambda_J\left[\brc{\bphi{\bphi}}_J\brc{{\overline\bphi}\,{\overline\bphi}}_J\right]_0
+\alpha\brc{\bh^\dagger\bh}\brc{\bphi^\dagger\bphi}
+\beta\left[\brc{{\overline \bh}\bh}_1\brc{{\overline \bphi} \bphi}_1 \right]_0 \\
&+\delta_{0,Y}\Bigg\{
\sum_{\kappa=0}^{2j}\left\{\lambda_\kappa'\left[\brc{\bphi\bphi}_\kappa\brc{\bphi\bphi}_\kappa\right]_0
+\lambda_\kappa''\left[\brc{\overline{\bphi}\bphi}_\kappa\brc{\bphi\bphi}_\kappa\right]_0+\mathrm{h.c.}\right\}
+\left[\gamma\brc{\overline{\bh}\bh}_0\brc{\bphi\bphi}_0+\mathrm{h.c.}\right] \\
&+\delta_{0,j\brc{\text{mod 2}}}\left[\mu_1\brc{\bphi\brc{\bphi\bphi}_j}_0+\mu_2\brc{\overline{\bphi}\brc{\bphi\bphi}_j}_0+\text{h.c.}\right]
\Bigg\}
\end{aligned}
\end{equation}
\item Half integer isospin $j$
\begin{equation}
\begin{aligned}
V\brc{\bh,\bphi}=&\sum_{J=1}^{2j}\lambda_J\left[\brc{\bphi{\bphi}}_J\brc{{\overline\bphi}\,{\overline\Phi}}_J\right]_0
+\alpha\brc{\bh^\dagger\bh}\brc{\bphi^\dagger\bphi}
+\beta\left[\brc{{\overline \bh}\bh}_1\brc{{\overline \bphi} \bphi}_1 \right]_0 \\
&+\delta_{0,Y}\Bigg\{
\sum_{\kappa=1}^{2j}\left\{\lambda_\kappa'\left[\brc{\bphi\bphi}_\kappa\brc{\bphi\bphi}_\kappa\right]_0
+\lambda_\kappa''\left[\brc{\overline{\bphi}\bphi}_\kappa\brc{\bphi\bphi}_\kappa\right]_0+\mathrm{h.c.}\right\} \\
&+\left\{\gamma\left[\brc{\overline{\bh}\bh}_1\brc{\bphi\bphi}_1\right]_0+\mathrm{h.c.}\right\}
\Bigg\}
+\delta_{1,Y}\Bigg\{
\Big\{
\gamma_1\left[\brc{\bh\bphi}_{j+\frac{1}{2}}\brc{\overline{\bphi}~\overline{\bphi}}_{j+\frac{1}{2}}\right]_0\delta_{0,j-\frac{1}{2}~\brc{\text{mod 2}}} \\
&+\gamma_2\left[\brc{\bh\bphi}_{j-\frac{1}{2}}\brc{\overline{\bphi}~\overline{\bphi}}_{j-\frac{1}{2}}\right]_0\delta_{0,j+\frac{1}{2}~\brc{\text{mod 2}}}
+\text{h.c.}
\Big\} 
+\Big\{
\kappa_1\left[\brc{\bh\bh}_1\brc{\overline{\bphi}~\overline{\bphi}}_1\right]_0
+\text{h.c.}
\Big\}
\Bigg\} \\
&+\delta_{-1,Y}\Bigg\{
\Big\{
\gamma_1\left[\brc{\bh\overline{\bphi}}_{j+\frac{1}{2}}\brc{\bphi\bphi}_{j+\frac{1}{2}}\right]_0\delta_{0,j-\frac{1}{2}~\brc{\text{mod 2}}} 
+\gamma_2\left[\brc{\bh\overline{\bphi}}_{j-\frac{1}{2}}\brc{\bphi\bphi}_{j-\frac{1}{2}}\right]_0\delta_{0,j+\frac{1}{2}~\brc{\text{mod 2}}} \\
&+\text{h.c.}
\Big\} 
+\Big\{
\kappa_1\left[\brc{\bh\bh}_1\brc{\bphi\bphi}_1\right]_0
+\text{h.c.}
\Big\}
\Bigg\}
\end{aligned}
\end{equation}
\end{itemize}

\section{Self Energy and Vertex Corrections}
\label{sec:cor}
Since in the models under consideration the neutral components in the scalar multiplet sector
decouple to that in the SM, one can treat the scalar multiplet loop contributions independently. 
The self energy and vertex loop contributions are written in terms of
the one-, two- and three-point integrals.
In the following the relevant scalar and vector/tensor integrals in D-dimension ($D=4-2\epsilon$),
self energy and vertex corrections
in each model are given, respectively. 

\subsection{The N-point Integrals with $N\le 3$}\label{append:N-int}
The scalar one-point integral is
\begin{equation}
  \frac{i}{16\pi^2}A_0\brc{m}=\mu^{2\epsilon}\int\frac{d^Dl}{\brc{2\pi}^D}\frac{1}{l^2-m^2}, 
\end{equation}
and $A_0\brc{m}$ has simple form
\[
A_0\brc{m}=m^2\brc{\frac{1}{\epsilon}+1+\log\frac{\mu^2}{m^2}} + \m{O}\brc{\epsilon}. 
\]
The scalar and vector two-point integrals $B_{0,1}$ are defined as
\begin{equation}
  \frac{i}{16\pi^2}\brc{B_0,B_\mu}\brc{k,m_1,m_2}=\mu^{2\epsilon}\int\frac{d^Dl}{\brc{2\pi}^D}
  \frac{\brc{1,l_\mu}}{\brc{l^2-m_1^2}\left[\brc{l+k}^2-m_2^2\right]}, 
\end{equation}
with \(B_\mu=k_\mu B_1\) and 
\[
\begin{aligned}
  B_0\brc{k,m_1,m_2}&=\frac{1}{\epsilon}-\int_0^1dx\log\frac{x^2k^2-x\brc{k^2+m_1^2-m_2^2}+m_1^2-i\epsilon}{\mu^2}+\m{O}\brc{\epsilon}, \\
  B_1\brc{k,m_1,m_2}&=\frac{1}{2k^2}\left[
    A_0\brc{m_1}-A_0\brc{m_2}-B_0\brc{k,m_1,m_2}\brc{k^2+m_1^2-m_2^2}
    \right].
\end{aligned}
\]
The scalar and (rank 2) tensor three-point integrals are
\begin{equation}
  \frac{i}{16\pi^2}\brc{C_0,C_{\mu\nu}}\brc{k_1,k_2,m_1,m_2,m_3}
  =\mu^{2\epsilon}\int\frac{d^Dl}{\brc{2\pi}^D}
  \frac{\brc{1,l_\mu l_\nu}}
       {\brc{l^2-m_1^2}\left[\brc{l+k_1}^2-m_2^2\right]\left[\brc{l+k_1+k_2}^2-m_3^2\right]}, 
\end{equation}
with
\[
C_{\mu\nu}=k_{1\mu}k_{1\nu}C_{21}+k_{2\mu}k_{2\nu}C_{22}
+\brc{k_{1\mu}k_{2\nu}+k_{2\mu}k_{1\nu}}C_{23}
+g_{\mu\nu}C_{24}, 
\]
where $C_{22},~C_{23},~C_{24}$ are the relevant contributions entering our calculation
and their explicit expressions with specific arguments are given in \ref{append:NLO:vert}. 

\subsection{Gauge Boson Self energies}\label{append:NLO:self}  
In the following the gauge boson self energies involving the virtual scalar multiplet contributions
are given in the models below. 
\begin{enumerate}
\item Inert Double
  {\scriptsize
\begin{align}
  \Sigma_T^{ZZ}\brc{k^2}&=\frac{1}{16\pi^2}\frac{1}{3}\times\Bigg\{
  +\frac{e^2\brc{c_W^2-s_W^2}^2}{4c_W^2s_W^2}\left[
    4\Mp^2 B_0\brc{0,\Mp,\Mp} + \frac{2}{3}k^2 + \brc{k^2-4\Mp^2}B_0\brc{k,\Mp,\Mp}
    \right] \notag \\
  &+\frac{e^2}{4c_W^2s_W^2}\bigg[
    - \frac{\brc{\MHn^2-\MAn^2}^2}{k^2}
    - \frac{\MHn^2-\MAn^2}{k^2}\left[\MHn^2B_0\brc{0,\MHn,\MHn}-\MAn^2B_0\brc{0,\MAn,\MAn}\right] \notag \\
    &+ 2\left[\MHn^2B_0\brc{0,\MHn,\MHn}+\MAn^2B_0\brc{0,\MAn,\MAn}\right]
    + \frac{2}{3}k^2
    + \frac{\brc{\MHn^2-\MAn^2}^2}{k^2}B_0\brc{k,\MHn,\MAn} \notag \\
    &+ \brc{k^2-2\MHn^2-2\MAn^2}B_0\brc{k,\MHn,\MAn}
    \bigg]
  \Bigg\} \notag \\
  \Sigma_T^{\gamma Z}\brc{k^2}&=\frac{1}{16\pi^2}\frac{1}{3}\times\Bigg\{
  +\frac{e^2\brc{c_W^2-s_W^2}}{2c_Ws_W}\left[
    4\Mp^2B_0\brc{0,\Mp,\Mp} + \frac{2}{3}k^2 + \brc{k^2-4\Mp^2}B_0\brc{k,\Mp,\Mp}
    \right]
  \Bigg\} \notag \\
  \Sigma_T^{\gamma\gamma}\brc{k^2}&=\frac{1}{16\pi^2}\frac{1}{3}\times\Bigg\{
  +e^2\left[
    4\Mp^2B_0\brc{0,\Mp,\Mp} + \frac{2}{3}k^2 + \brc{k^2-4\Mp^2}B_0\brc{k,\Mp,\Mp}
    \right]
  \Bigg\} \notag \\
  \Sigma_T^{WW}\brc{k^2}&=\frac{1}{16\pi^2}\frac{1}{3}\times\frac{e^2}{4s_W^2}\Bigg\{
  -\frac{\brc{\Mp^2-\MHn^2}^2}{k^2}B_0\brc{0,\Mp,\MHn}
  - \frac{\brc{\Mp^2-\MAn^2}^2}{k^2}B_0\brc{0,\Mp,\MAn} \notag \\
  &+ 4\Mp^2B_0\brc{0,\Mp,\Mp} + 2\MHn^2B_0\brc{0,\MHn,\MHn} + 2\MAn^2B_0\brc{0,\MAn,\MAn}
  + \frac{4}{3}k^2 \notag \\
  &+ \frac{1}{k^2}\left[\brc{\Mp-\MHn}^2-k^2\right]\left[\brc{\Mp+\MHn}^2-k^2\right]B_0\brc{k,\Mp,\MHn} \notag \\
  &+ \frac{1}{k^2}\left[\brc{\Mp-\MAn}^2-k^2\right]\left[\brc{\Mp+\MAn}^2-k^2\right]B_0\brc{k,\Mp,\MAn} \notag \Bigg\} \notag \\
  \Sigma^{H}\brc{k^2} &= \frac{1}{16\pi^2}\times\Bigg\{
  +\lambda_3 A_0\brc{\Mp} + \lambda_3^2 v_\phi^2 B_0\brc{k,\Mp,\Mp} \notag \\
  &+\frac{1}{2}\left[
  \lambda_L A_0\brc{\MHn} + \lambda_L^2 v_\phi^2 B_0\brc{k,\MHn,\MHn}
  +\lambda_A A_0\brc{\MAn} + \lambda_A^2 v_\phi^2 B_0\brc{k,\MAn,\MAn}
  \right]
  \Bigg\}
\end{align}
}

\item Real Triplet ($V_1V_2=ZZ,\gamma Z,WW,\gamma\gamma$)
\begin{equation}\label{eq:slf:rtriplet}
  \small
  \begin{aligned}
    \Sigma_T^{V_1V_2}\brc{k^2} &= 
    \frac{1}{16\pi^2}~\frac{1}{3}~g_{ssV_1}g_{ssV_2}\left[4 M_S^2 B_0\brc{0,M_S,M_S} + \frac{2}{3}k^2 + \brc{k^2 - 4M_S^2}B_0\brc{k,M_S,M_S}\right], \\
    \Sigma^H\brc{k^2} &= \frac{1}{16\pi^2}\frac{3}{2}\left[\lambda_3A_0\brc{M_S} + \lambda_3^2 v^2 B_0\brc{k,M_S,M_S}\right].
  \end{aligned}
\end{equation}

\item Complex Triplet
  \begin{equation}\label{eq:self:ctriplet}
  \scriptsize
  \begin{aligned}
    &\begin{aligned}
      \Sigma_T^{ZZ}\brc{k^2} &= \frac{1}{16\pi^2}\frac{1}{3}\times\Bigg\{
      +\frac{e^2 s_W^2}{c_W^2}
      \left[
        4 \Mp^2 B_0\brc{0,\Mp,\Mp} + \frac{2}{3}k^2
        + \brc{k^2-4 \Mp^2}B_0\brc{k,\Mp,\Mp}
        \right] \\
      &+\frac{e^2\brc{c_W^2-s_W^2}^2}{c_W^2 s_W^2}
      \left[
        4 \Mpp^2 B_0\brc{0,\Mpp,\Mpp} + \frac{2}{3}k^2
        + \brc{k^2-4\Mpp^2}B_0\brc{k,\Mpp,\Mpp}
        \right] \\
      &+\frac{e^2}{c_W^2 s_W^2}
      \left[
        4 \Mn^2 B_0\brc{0,\Mn,\Mn} + \frac{2}{3}k^2
        + \brc{k^2-4\Mn^2}B_0\brc{k,\Mn,\Mn}
        \right]
      \Bigg\}
     \end{aligned} \\
    &\begin{aligned}
       \Sigma_T^{\gamma Z}\brc{k^2} &=\frac{1}{16\pi^2}\frac{1}{3}\times\Bigg\{
       -\frac{e^2 s_W}{c_W}
       \left[
         4\Mp^2 B_0\brc{0,\Mp,\Mp} + \frac{2}{3}k^2
      + \brc{k^2-4\Mp^2}B_0\brc{k,\Mp,\Mp}
      \right] \\
       &+\frac{2e^2\brc{c_W^2-s_W^2}}{c_W s_W}
       \left[
         4\Mpp^2 B_0\brc{0,\Mpp,\Mpp} + \frac{2}{3}k^2
        + \brc{k^2-4\Mpp^2}B_0\brc{k,\Mpp,\Mpp}
         \right]
       \Bigg\}
     \end{aligned} \\
    &\begin{aligned}
       \Sigma_T^{\gamma\gamma}\brc{k^2} &= \frac{1}{16\pi^2}\frac{1}{3}\times\Bigg\{
       +e^2
       \left[
         4\Mp^2 B_0\brc{0,\Mp,\Mp} + \frac{2}{3}k^2
      + \brc{k^2-4\Mp^2}B_0\brc{k,\Mp,\Mp}
      \right] \\
       &+4e^2
       \left[
         4\Mpp^2 B_0\brc{0,\Mpp,\Mpp} + \frac{2}{3}k^2
        + \brc{k^2-4\Mpp^2}B_0\brc{k,\Mpp,\Mpp}
         \right]
       \Bigg\}
     \end{aligned} \\
    &\begin{aligned}
       \Sigma_T^{WW}\brc{k^2} &= \frac{1}{16\pi^2}\frac{1}{3}\times\frac{e^2}{s_W^2}\Bigg\{
       -\frac{\brc{\Mp^2-\Mn^2}^2}{k^2}B_0\brc{0,\Mp,\Mn}
       -\frac{\brc{\Mpp^2-\Mp^2}^2}{k^2}B_0\brc{0,\Mpp,\Mp} \\
       &+2\Mn^2 B_0\brc{0,\Mn,\Mn}
       +4\Mp^2 B_0\brc{0,\Mp,\Mp}
       +2\Mpp^2 B_0\brc{0,\Mpp,\Mpp} + \frac{4}{3}k^2\\
       &+\frac{1}{k^2}\left[\brc{\Mp-\Mn}^2-k^2\right]\left[\brc{\Mp+\Mn}^2-k^2\right]B_0\brc{k,\Mp,\Mn} \\
       &+\frac{1}{k^2}\left[\brc{\Mpp-\Mp}^2-k^2\right]\left[\brc{\Mpp+\Mp}^2-k^2\right]B_0\brc{k,\Mpp,\Mp}
       \Bigg\}
     \end{aligned} \\
    &\begin{aligned}
       \Sigma^H\brc{k^2} &= \frac{1}{16\pi^2}\times
       \Bigg\{
         +\brc{\lambda_4+\frac{\lambda_5}{2}}A_0\brc{\Mp}
         +\brc{\lambda_4+\frac{\lambda_5}{2}}^2 v_\phi^2 B_0\brc{k,\Mp,\Mp} \\ 
         &+\lambda_4 A_0\brc{\Mpp} + \lambda_4^2 v_\phi^2 B_0\brc{k,\Mpp,\Mpp} \\
         &+\brc{\lambda_4+\lambda_5} A_0\brc{\Mn} + \brc{\lambda_4+\lambda_5}^2 v_\phi^2 B_0\brc{k,\Mn,\Mn}
         \Bigg\}
     \end{aligned}
  \end{aligned}
\end{equation}

\item Quintuplet ($n=5$) and Septuplet ($n=7$)
  {\scriptsize
  \begin{align}
  \Sigma_T^{V_1V_2}\brc{k^2} =& \frac{1}{16\pi^2}\frac{1}{3}\times\Bigg\{
  \sum_{i=A}^{B}\brc{\sum_{m=0}^{m=\frac{n-1}{2}}g_{S_i^mS_i^{-m+q}V_1}\cdot g_{S_i^mS_i^{-m+q}V_2}}
  \Big[4M_{S_i}^2 B_0\brc{0,M_{S_i},M_{S_i}}
    + \frac{2}{3}k^2 \notag \\ 
    &+ \brc{k^2-4M_{S_i}^2}B_0\brc{k,M_{S_i},M_{S_i}}
    \Big]
  \Bigg\}, \notag \\
  &\brc{q=1~\mathrm{for}~V_1V_2=WW,~q=0~\mathrm{for}~V_1V_2=ZZ,\gamma Z,\gamma\gamma}
  \notag \\
  \Sigma^h\brc{k^2} =& \frac{1}{16\pi^2}\frac{n}{2}\sum_{i=A}^{B}\Bigg[
    -g_{S_iS_ihh}A_0\brc{M_{S_i}}+g_{S_iS_ih}^2B_0\brc{k,M_{S_i},M_{S_i}}
    \Bigg], \notag \\
  &\brc{g_{S_iS_ih}=g_{S_iS_ihh}\cdot v, 
    ~g_{S_AS_Ahh}=-\lambda_1-\frac{2}{\sqrt{n}}\lambda_3,
    ~g_{S_BS_Bhh}=-\lambda_1+\frac{2}{\sqrt{n}}\lambda_3}
  \end{align}
  }

\end{enumerate}

\subsection{Three-point Tensor Integrals}\label{append:NLO:vert}
The vertex contribution is given in Eq. (\ref{eq:vert:m1m2}).
In addition to $B_0$ given in \ref{append:N-int},
the explicit expressions of $C_{22s},~C_{23s},~C_{24s}$ are given below.
The subscript $'s'$ and arguments $\brc{M_1,M_2}$ are short for the full arguments for the three-point
functions that $C_{..s}\brc{M_1,M_2}=C_{..}\brc{k_1,k_2,M_1,M_2,M_2}$,
and $k_1^2=M_Z^2,~k_2^2=M_h^2,~\brc{k_1+k_2}^2=s$.
The masses $\brc{M_1,M_2}$ represent the masses of the scalar components in each model,
where $M_1=M_2$ for each single vertex contribution except in the Inert Doublet when
two neutral components as virtual particles in the loop and one should take
$\brc{M_1,M_2}=\brc{M_{H^0},M_{A^0}},~\brc{M_{A^0},M_{H^0}}$ respectively for two types of vertex
correction.

{\tiny
\begin{align}\label{eq:cexp:m1m2}
  C_{22s}\brc{M_1,M_2} &= \frac{1}{2\left[M_h^4 + \brc{M_Z^2 - s}^2 - 2 M_h^2 \brc{M_Z^2 + s}\right]^2}
  \Bigg\{2 M_Z^2 \left[M_h^4 + \brc{M_Z^2 - s}^2 - 2 M_h^2 \brc{M_Z^2 + s}\right] \notag \\
  & - \frac{1}{s}\brc{M_h^2-M_Z^2-s}\left[M_h^4 + \brc{M_Z^2 - s}^2 - 2 M_h^2 \brc{M_Z^2 + s}\right]
  \left[A_0\brc{M_1} - A_0\brc{M_2}\right] \notag \\
  & - 6 M_Z^2 \left[M_Z^2\brc{M_h^2-M_Z^2+s} + \brc{M_1^2-M_2^2}\brc{M_h^2+M_Z^2-s}\right] B_0\brc{k_1,M_1,M_2} \notag \\
  & - \bigg[M_h^6 + \brc{M_Z^2 - s}^2 \brc{3 M_Z^2 - s} - M_h^4 \brc{5 M_Z^2 + 3 s} +
    M_h^2 \brc{M_Z^4 + 3 s^2} \notag \\
  & - 2\brc{M_1^2-M_2^2}\brc{M_h^4+M_h^2\brc{4M_Z^2-s}+\brc{M_Z^2-s}^2}\bigg] B_0\brc{k_2,M_2,M_2} \notag \\
  & + \bigg[
    \brc{M_h^2 - 3 M_Z^2} \brc{M_h^2 - M_Z^2}^2 - \brc{3 M_h^4 + M_Z^4} s + \brc{3 M_h^2 + 5 M_Z^2} s^2 - s^3 \notag \\
    & + \frac{1}{s}\brc{M_1^2-M_2^2}\brc{\brc{M_h^2-M_Z^2}^3 - 5\brc{M_h^4-M_Z^4}s
      + \brc{7M_h^2-M_Z^2}s^2 - 3s^3}
    \bigg] B_0\brc{k_1+k_2,M_1,M_2} \notag \\
  & + 2\bigg[
    M_Z^4\brc{M_h^4-2M_h^2\brc{M_Z^2-2s} + \brc{M_Z^2-s}}
    + \brc{M_1^4+M_2^4}\brc{M_h^4+2M_h^2\brc{2M_Z^2-s}+\brc{M_Z^2-s}^2} \notag \\
    & - 2M_Z^2\brc{M_h^4\brc{-2M_1^2+M_2^2}+\brc{M_Z^2-s}^2\brc{M_1^2-2M_2^2}
      + M_h^2\brc{M_Z^2+s}\brc{M_1^2+M_2^2}} \notag \\
    & - 2M_1^2 M_2^2\left[M_h^4+2M_h^2\brc{2M_Z^2-s}+\brc{M_Z^2-s}^2\right]
    \bigg] C_0\brc{k_1,k_2,M_1,M_2,M_2} \Bigg\} \notag \\
  C_{23s}\brc{M_1,M_2} & = \frac{1}{2\left[M_h^4 + \brc{M_Z^2 - s}^2 - 2 M_h^2 \brc{M_Z^2 + s}\right]^2}
  \Bigg\{\brc{M_h^2+M_Z^2-s}\left[M_h^4 + \brc{M_Z^2 - s}^2 - 2 M_h^2 \brc{M_Z^2 + s}\right] \notag \\
  & - \frac{1}{s}\brc{M_h^2-M_Z^2+s}\brc{M_h^4+\brc{M_Z^2-s}^2-2M_h^2\brc{M_Z^2+s}}\left[
    A_0\brc{M_1} - A_0\brc{M_2}
    \right] \notag \\
  & + \bigg[M_Z^2\brc{-M_h^4 + 5 \brc{M_Z^2 - s}^2 - 4 M_h^2 \brc{M_Z^2 + s}} \notag \\
    & - \brc{M_1^2-M_2^2}\brc{M_h^4+\brc{M_Z^2-s}^2+2M_h^2\brc{5M_Z^2-s}}
    \bigg] B_0\brc{k_1,M_1,M_2} \notag \\
  & - \bigg[
    M_h^6 + 2 \brc{M_Z^2 - s}^3 - 2 M_h^4 \brc{3 M_Z^2 + 2 s} + M_h^2 \brc{3 M_Z^4 - 8 M_Z^2 s + 5 s^2} \notag \\
    & - 6\brc{M_1^2-M_2^2}M_h^2\brc{M_h^2+M_Z^2-s}
    \bigg]
  B_0\brc{k_2,M_2,M_2} \notag \\
  & + \bigg[
    M_h^6 - \brc{M_Z^2 - s}^2 \brc{3 M_Z^2 + 2 s} - M_h^4 \brc{5 M_Z^2 + 4 s} + M_h^2\brc{7 M_Z^4 - 4 M_Z^2 s + 5 s^2} \notag \\
    & + \brc{M_1^2-M_2^2}\brc{\brc{M_h^2-M_Z^2}^2+2\brc{-3M_h^4+M_h^2M_Z^2+2M_Z^4}s+\brc{3M_h^2-5M_Z^2}s^2+2s^3}
    \bigg] B_0\brc{k_1+k_2,M_1,M_2} \notag \\
  & + 2\bigg[
    M_Z^2\brc{\brc{M_Z^2-s}^3 + M_h^4\brc{M_Z^2+2s}-M_h^2\brc{2M_Z^4-3M_Z^2s+2s^2}} \notag \\
    & + 3\brc{M_1^4+M_2^4}M_h^2\brc{M_h^2+M_Z^2-s}
    + M_1^2\brc{M_h^2-M_Z^2+s}\brc{M_h^4+2M_h^2\brc{2M_Z^2-s}+\brc{M_Z^2-s}^2} \notag \\
    & + 2M_2^2\brc{\brc{M_Z^2-s}^3-M_h^4\brc{2M_Z^2+s}+M_h^2\brc{M_Z^4-3M_Z^2 s+2s^2}} \notag \\
    & - 6M_1^2M_2^2M_h^2\brc{M_h^2+M_Z^2-s}
    \bigg] C_0\brc{k_1,k_2,M_1,M_2,M_2}
  \Bigg\} \notag \\
  C_{24s}\brc{M_1,M_2} & = \frac{1}{4\left[M_h^4 + \brc{M_Z^2 - s}^2 - 2 M_h^2 \brc{M_Z^2 + s}\right]}
  \Bigg\{M_h^4 + \brc{M_Z^2 - s}^2 - 2 M_h^2 \brc{M_Z^2 + s} \notag \\
  & - \left[M_Z^2 \brc{M_h^2-M_Z^2+s} + \brc{M_1^2-M_2^2}\brc{M_h^2+M_Z^2-s}\right] B_0\brc{k_1,M_1,M_2} \notag \\
  & + M_h^2 \brc{M_h^2-M_Z^2-s + 2M_1^2-2M_2^2} B_0\brc{k_2,M_2,M_2} \notag \\
  & + \left[s \brc{-Mh^2-Mz^2+s} - \brc{M_1^2-M_2^2}\brc{M_h^2-M_Z^2+s}\right] B_0\brc{k_1+k_2,M_1,M_2} \notag \\
  & + 2\bigg[
    M_h^2 M_Z^2 s + \brc{M_1^4+M_2^4}M_h^2 + M_1^2M_h^2\brc{M_h^2-M_Z^2-s}
    + M_2^2\brc{\brc{M_Z^2-s}^2 - M_h^2\brc{M_Z^2+s}} \notag \\
    & - 2M_1^2 M_2^2 M_h^2
    \bigg] C_0\brc{k_1,k_1,M_1,M_2,M_2}
  \Bigg\}
\end{align}
}

\section{On-shell Renormalization Scheme}
\label{sec:onshell}
The on-shell (OS) renormalization conditions consist of two parts: 
(a) the pole of the propagator defines the real part of the renormalized mass; 
(b) the real part of the residue of the propagator is equal to one.
The renormalized quantities and renormalization constants are in accordance with
the scheme of Denner \cite{Denner:1991kt}, and are defined as follows
\begin{equation}
  \begin{aligned}
    M_{W,0}^2 &= M_W^2+\delta M_W^2, \\
    M_{Z,0}^2 &= M_Z^2+\delta M_Z^2, \\
    M_{h,0}^2 &= M_h^2+\delta M_h^2, \\
    W_0^\pm &= Z_W^{1/2}W^\pm=\brc{1+\frac{1}{2}\delta Z_W}W^\pm, \\
    \begin{pmatrix}
       Z_0 \\
       A_0
    \end{pmatrix}
    &= \begin{pmatrix}
      Z_{ZZ}^{1/2} &Z_{Z\gamma}^{1/2} \\
      Z_{\gamma Z}^{1/2} &Z_{\gamma\gamma} 
    \end{pmatrix}
    \begin{pmatrix}
      Z \\
      A
    \end{pmatrix}
    =
    \begin{pmatrix}
      1+\frac{1}{2}\delta Z_{ZZ} &\frac{1}{2}\delta Z_{Z\gamma} \\
      \frac{1}{2}\delta Z_{\gamma Z} &1+\frac{1}{2}\delta Z_{\gamma\gamma}
    \end{pmatrix}
    \begin{pmatrix}
      Z \\
      A
    \end{pmatrix} \\
    h_0 &= Z_h^{1/2} h = \brc{1+\frac{1}{2}\delta Z_h} h, 
  \end{aligned}
\end{equation}
where quantities with subscript $0$ denote the bare parameters. 

In terms of the renormalized self-energy functions which are denoted with caret
(one should differentiate this with the caret notation in $\MSbar$ scheme),
the OS conditions are given
\begin{equation}
  \begin{aligned}
    &\Re~\hat{\Sigma}_T^{WW}\brc{M_W^2}=\delta M_W^2,
    \quad
    \Re\left.\frac{\partial\hat{\Sigma}_T^{WW}\brc{k^2}}{\partial k^2}\right|_{k^2=M_W^2}=0, \\
    &\Re~\hat{\Sigma}_T^{ZZ}\brc{M_Z^2}=\delta M_Z^2,
    \quad
    \Re\left.\frac{\partial\hat{\Sigma}_T^{ZZ}\brc{k^2}}{\partial k^2}\right|_{k^2=M_Z^2}=0, \\
    &\hat{\Sigma}_T^{\gamma\gamma}\brc{0}=0,
    \quad
    \Re\left.\frac{\partial\hat{\Sigma}_T^{\gamma\gamma}\brc{k^2}}{\partial k^2}\right|_{k^2=0}=0, \\
    &\Re~\hat{\Sigma}_T^{\gamma Z}\brc{M_Z^2}=\hat{\Sigma}_T^{\gamma Z}\brc{0}=0, \\
    &\Re~\hat{\Sigma}^{h}\brc{M_h^2}=\delta M_h^2,
    \quad
    \Re\left.\frac{\partial\hat{\Sigma}^{h}\brc{k^2}}{\partial k^2}\right|_{k^2=M_h^2}=0.  
  \end{aligned}
\end{equation}
The mass and wave function renormalization counterterms are then determined by the following relations
\begin{equation}
  \begin{aligned}
    &\delta M_W^2=\Re~\Sigma_T^{WW}\brc{M_W^2},
    \quad
    \delta Z_W=-\Re\left.\frac{\partial\Sigma_T^{WW}\brc{k^2}}{\partial k^2}\right|_{k^2=M_W^2}, \\
    &\delta M_Z^2=\Re~\Sigma_T^{ZZ}\brc{M_Z^2},
    \quad
    \delta Z_{ZZ}=-\Re\left.\frac{\partial\Sigma_T^{ZZ}\brc{k^2}}{\partial k^2}\right|_{k^2=M_Z^2}, \\
    &\delta Z_{\gamma\gamma}=
    -\Re\left.\frac{\partial\Sigma_T^{\gamma\gamma}\brc{k^2}}{\partial k^2}\right|_{k^2=0}, \\
    &\delta Z_{\gamma Z}=-2~\Re~\frac{\Sigma_T^{\gamma Z}\brc{M_Z^2}}{M_Z^2},
    \quad
    \delta Z_{Z\gamma}=2\frac{\Sigma_T^{\gamma Z}\brc{0}}{M_Z^2}, \\
    &\delta M_h^2=\Re~\Sigma^h\brc{M_h^2},
    \quad
    \delta Z_h=-\Re\left.\frac{\partial\Sigma^h\brc{k^2}}{\partial k^2}\right|_{k^2=M_h^2}, 
  \end{aligned}
  \label{eq:reno:wavefunc}
\end{equation}
where the expressions of the unrenormalized self energies for each scalar multiplet model can be found in
Appendix \ref{append:NLO:self}.

The counter terms of self-energy that are relevant to our calculation read
\begin{equation}
  \includegraphics[scale=0.6,valign=c]{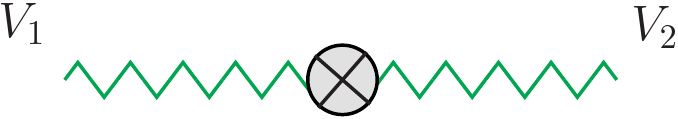} = \left\{
  \begin{tabular}{l}
    \(-i~\delta\Sigma_T^{WW} = -i\left[\brc{k^2-M_W^2}\delta Z_W - \delta M_W^2\right]\), \\
    \(-i~\delta\Sigma_T^{ZZ} = -i\left[\brc{k^2-M_Z^2}\delta Z_{ZZ} - \delta M_Z^2\right]\), \\
    \(-i~\delta\Sigma_T^{\gamma Z} = -\frac{i}{2}\left[\brc{k^2-M_Z^2}\delta Z_{Z\gamma} + k^2\delta Z_{\gamma Z}\right]\), 
  \end{tabular}
  \right.
\end{equation}
so that one can obtain the renormalized self-energies according to
\[
\hat{\Sigma}_T^{V_1V_2}\brc{k^2} = \Sigma_T^{V_1V_2}\brc{k^2} + \delta\Sigma_T^{V_1V_2}. 
\]

The self-energy contributions in terms of matrix element square factorize to Born, which read
\begin{equation}
  \begin{aligned}
    &2~\Re\left[\brc{\delta\MM_{\mathrm{self}}^{ZZ} + \delta\MM_{\mathrm{self}}^{\gamma Z}}\cdot\MM_{LO}^*\right] \\
    =&2\frac{2e^4}{s_W^2c_W^2}\left[-\frac{\brc{g_v^2+g_a^2}\brc{s-M_Z^2}}{\brc{s-M_Z^2}^2+M_Z^2\Gamma_Z^2}\cdot\hat{\Sigma}_T^{ZZ}\brc{s} - \frac{g_v}{s}\cdot\hat{\Sigma}_T^{\gamma Z}\brc{s}\right]\frac{t u + 2s M_Z^2 - M_h^2 M_Z^2}{\brc{s-M_Z^2}^2+M_Z^2\Gamma_Z^2}
    \end{aligned}
\end{equation}

It should be noted that the vertex contribution per se is UV finite.
Using OS renormalization scheme,
one has to cross check if the corresponding counter term contribution is also
UV finite. The vertex counter terms have the following contributions, 
\begin{equation}
  \includegraphics[scale=0.6,valign=c]{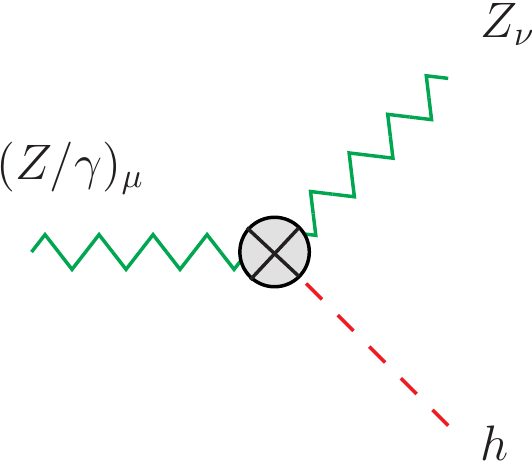} = \left\{
  \begin{tabular}{ll}
    $ieg_{\mu\nu}\cdot\delta_{\mathrm{ct}}^{ZZh}$: &\(\delta Z_e + \frac{2s_W^2-c_W^2}{c_W^2}\frac{\delta s_W}{s_W} + \frac{1}{2}\frac{\delta M_W^2}{M_W^2} + \frac{1}{2}\delta Z_h + \delta Z_{ZZ}\) \\
    $ieg_{\mu\nu}\cdot\delta_{\mathrm{ct}}^{\gamma Zh}$: &\(\frac{1}{2}\delta Z_{Z\gamma}\)
  \end{tabular}
  \right.
  \label{eq:vrt:cts}
\end{equation}
where $\delta Z_e$ and $\delta s_W^2/s_W^2$ can be written in terms of the mass and field renormalization constants
(FRCs) that
\begin{equation}
  \delta Z_e = -\frac{1}{2}\brc{\delta Z_{\gamma\gamma} + \frac{s_W}{c_W}\delta Z_{Z\gamma}}, \quad
  \frac{\delta s_W}{s_W} = -\frac{1}{2}\frac{c_W^2}{s_W^2}\brc{\frac{\delta M_W^2}{M_W^2} - \frac{\delta M_Z^2}{M_Z^2}}. 
\end{equation}
It is clear that both vertex counter terms in Eq.(\ref{eq:vrt:cts}) are UV finite.

In addition to the final vertex ($(Z/\gamma)Zh$) counter term,
one has to take the initial vertex ($e^-e^+Z$) counter term into account, 
even though there is no scalar loop corrections to the initial vertex.
The counter term of the initial vertex $e^-e^+Z$ has the following form. 
\begin{equation}
  \includegraphics[scale=0.6,valign=c]{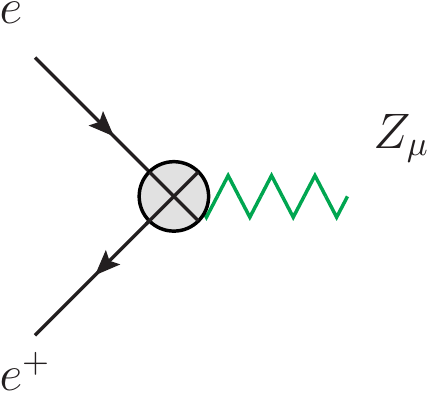} = 
  ie\gamma_\mu\brc{\delta c_v - \delta c_a\gamma^5}
  \label{eq:init:cts}
\end{equation}

\begin{equation*}
  \begin{aligned}
    &\delta c_v = \delta g_v + \frac{g_v}{2}\delta Z_{ZZ} - \frac{Q_e}{2}\delta Z_{\gamma Z}, \\
    &\delta c_a = \delta g_a + \frac{g_a}{2}\delta Z_{ZZ}, 
  \end{aligned}
\end{equation*}
where the variations of $\brc{g_v,~g_a}$ read
\[
\begin{aligned}
  &\delta g_v = \frac{I_{W,e}^3}{2s_Wc_W}\brc{\delta Z_e + \frac{s_W^2-c_W^2}{c_W^2}\frac{\delta s_W}{s_W}} - \frac{s_W}{c_W}Q_e\brc{\delta Z_e + \frac{1}{c_W^2}\frac{\delta s_W}{s_W}}, \\
  &\delta g_a = \frac{I_{W,e}^3}{2s_Wc_W}\brc{\delta Z_e + \frac{s_W^2-c_W^2}{c_W^2}\frac{\delta s_W}{s_W}}, 
\end{aligned}
\]
with the quantum numbers for electron taking $I_{W,e}^3 = -1/2, ~Q_e = -1$.
It is worth noting that the initial vertex counter term contribution is as well UV finite.

The vertex counter term stand alone contributions in terms of matrix element square take the form below,
\begin{equation}
  \begin{aligned}
    &2~\Re\left[\brc{\delta\MM_{\mathrm{vert,ct}}^{ZZh} + \delta\MM_{\mathrm{vert,ct}}^{\gamma Zh} + \delta\MM_{\mathrm{vert,ct}}^{e^-e^+Z}}\MM_{LO}^*\right] \\
    = &2\frac{2e^4}{s_W^2c_W^2}\brc{t u + 2 s M_Z^2 - M_h^2M_Z^2}\times \\
    &\left\{\frac{\brc{g_v^2+g_a^2}\delta_{\mathrm{ct}}^{ZZh}}{\brc{s-M_Z^2}^2+M_Z^2\Gamma_Z^2} + \frac{g_v\brc{s-M_Z^2}\delta_{\mathrm{ct}}^{\gamma Zh}}{s\left[\brc{s-M_Z^2}^2+M_Z^2\Gamma_Z^2\right]} + \frac{\delta c_v g_v + \delta c_a g_a}{\brc{s-M_Z^2}^2+M_Z^2\Gamma_Z^2}
    \right\}.
  \end{aligned}
\end{equation}

\bibliography{zhnlo}

\providecommand{\href}[2]{#2}\begingroup\raggedright\begin{thebibliography}{10}

\bibitem{ATLAS:2019slw}
{\bf ATLAS} Collaboration, {\it {Combined measurements of Higgs boson
  production and decay using up to $80$ fb$^{-1}$ of proton--proton collision
  data at $\sqrt{s}=$ 13 TeV collected with the ATLAS experiment}}, .

\bibitem{Sirunyan:2018koj}
{\bf CMS} Collaboration, A.~M. Sirunyan {\em et.~al.}, {\it {Combined
  measurements of Higgs boson couplings in proton\textendash{}proton collisions
  at $\sqrt{s}=13\,\text {Te}\text {V} $}},  {\em Eur. Phys. J. C} {\bf 79}
  (2019), no.~5 421 [\href{http://arXiv.org/abs/1809.10733}{{\tt 1809.10733}}].

\bibitem{Cepeda:2019klc}
M.~Cepeda {\em et.~al.}, {\it {Report from Working Group 2}: {Higgs Physics at
  the HL-LHC and HE-LHC}},  {\em CERN Yellow Rep. Monogr.} {\bf 7} (2019)
  221--584 [\href{http://arXiv.org/abs/1902.00134}{{\tt 1902.00134}}].

\bibitem{deBlas:2019rxi}
J.~de~Blas {\em et.~al.}, {\it {Higgs Boson Studies at Future Particle
  Colliders}},  {\em JHEP} {\bf 01} (2020) 139
  [\href{http://arXiv.org/abs/1905.03764}{{\tt 1905.03764}}].

\bibitem{Abada:2019lih}
{\bf FCC} Collaboration, A.~Abada {\em et.~al.}, {\it {FCC Physics
  Opportunities}: {Future Circular Collider Conceptual Design Report Volume
  1}},  {\em Eur. Phys. J. C} {\bf 79} (2019), no.~6 474.

\bibitem{CEPCStudyGroup:2018ghi}
{\bf CEPC Study Group} Collaboration, M.~Dong {\em et.~al.}, {\it {CEPC
  Conceptual Design Report: Volume 2 - Physics \& Detector}},
  \href{http://arXiv.org/abs/1811.10545}{{\tt 1811.10545}}.

\bibitem{Abada:2019zxq}
{\bf FCC} Collaboration, A.~Abada {\em et.~al.}, {\it {FCC-ee: The Lepton
  Collider}: {Future Circular Collider Conceptual Design Report Volume 2}},
  {\em Eur. Phys. J. ST} {\bf 228} (2019), no.~2 261--623.

\bibitem{Barklow:2015tja}
T.~Barklow, J.~Brau, K.~Fujii, J.~Gao, J.~List, N.~Walker and K.~Yokoya, {\it
  {ILC Operating Scenarios}},  \href{http://arXiv.org/abs/1506.07830}{{\tt
  1506.07830}}.

\bibitem{Fujii:2017vwa}
K.~Fujii {\em et.~al.}, {\it {Physics Case for the 250 GeV Stage of the
  International Linear Collider}},  \href{http://arXiv.org/abs/1710.07621}{{\tt
  1710.07621}}.

\bibitem{Fujii:2019zll}
{\bf LCC Physics Working Group} Collaboration, K.~Fujii {\em et.~al.}, {\it
  {Tests of the Standard Model at the International Linear Collider}},
  \href{http://arXiv.org/abs/1908.11299}{{\tt 1908.11299}}.

\bibitem{TheATLAScollaboration:2014ewu}
{\it {Projections for measurements of Higgs boson signal strengths and coupling
  parameters with the ATLAS detector at a HL-LHC}}, .

\bibitem{CMS:2013xfa}
{\bf CMS} Collaboration, {\it {Projected Performance of an Upgraded CMS
  Detector at the LHC and HL-LHC: Contribution to the Snowmass Process}},  in
  {\em {Community Summer Study 2013}: {Snowmass on the Mississippi}}, 7, 2013.
\newblock \href{http://arXiv.org/abs/1307.7135}{{\tt 1307.7135}}.

\bibitem{Profumo:2007wc}
S.~Profumo, M.~J. Ramsey-Musolf and G.~Shaughnessy, {\it {Singlet Higgs
  phenomenology and the electroweak phase transition}},  {\em JHEP} {\bf 08}
  (2007) 010 [\href{http://arXiv.org/abs/0705.2425}{{\tt 0705.2425}}].

\bibitem{Espinosa:2011ax}
J.~R. Espinosa, T.~Konstandin and F.~Riva, {\it {Strong Electroweak Phase
  Transitions in the Standard Model with a Singlet}},  {\em Nucl. Phys. B} {\bf
  854} (2012) 592--630 [\href{http://arXiv.org/abs/1107.5441}{{\tt
  1107.5441}}].

\bibitem{Niemi:2018asa}
L.~Niemi, H.~H. Patel, M.~J. Ramsey-Musolf, T.~V. Tenkanen and D.~J. Weir, {\it
  {Electroweak phase transition in the real triplet extension of the SM:
  Dimensional reduction}},  {\em Phys. Rev. D} {\bf 100} (2019), no.~3 035002
  [\href{http://arXiv.org/abs/1802.10500}{{\tt 1802.10500}}].

\bibitem{Niemi:2020hto}
L.~Niemi, M.~J. Ramsey-Musolf, T.~V.~I. Tenkanen and D.~J. Weir, {\it
  {Thermodynamics of a Two-Step Electroweak Phase Transition}},  {\em Phys.
  Rev. Lett.} {\bf 126} (2021), no.~17 171802
  [\href{http://arXiv.org/abs/2005.11332}{{\tt 2005.11332}}].

\bibitem{Barger:2007im}
V.~Barger, P.~Langacker, M.~McCaskey, M.~J. Ramsey-Musolf and G.~Shaughnessy,
  {\it {LHC Phenomenology of an Extended Standard Model with a Real Scalar
  Singlet}},  {\em Phys. Rev. D} {\bf 77} (2008) 035005
  [\href{http://arXiv.org/abs/0706.4311}{{\tt 0706.4311}}].

\bibitem{Barger:2008jx}
V.~Barger, P.~Langacker, M.~McCaskey, M.~Ramsey-Musolf and G.~Shaughnessy, {\it
  {Complex Singlet Extension of the Standard Model}},  {\em Phys. Rev. D} {\bf
  79} (2009) 015018 [\href{http://arXiv.org/abs/0811.0393}{{\tt 0811.0393}}].

\bibitem{Papaefstathiou:2020iag}
A.~Papaefstathiou and G.~White, {\it {The Electro-Weak Phase Transition at
  Colliders: Confronting Theoretical Uncertainties and Complementary
  Channels}},  \href{http://arXiv.org/abs/2010.00597}{{\tt 2010.00597}}.

\bibitem{Cirelli:2005uq}
M.~Cirelli, N.~Fornengo and A.~Strumia, {\it {Minimal dark matter}},  {\em
  Nucl. Phys. B} {\bf 753} (2006) 178--194
  [\href{http://arXiv.org/abs/hep-ph/0512090}{{\tt hep-ph/0512090}}].

\bibitem{McDonald:1993ex}
J.~McDonald, {\it {Gauge singlet scalars as cold dark matter}},  {\em Phys.
  Rev. D} {\bf 50} (1994) 3637--3649
  [\href{http://arXiv.org/abs/hep-ph/0702143}{{\tt hep-ph/0702143}}].

\bibitem{Burgess:2000yq}
C.~Burgess, M.~Pospelov and T.~ter Veldhuis, {\it {The Minimal model of
  nonbaryonic dark matter: A Singlet scalar}},  {\em Nucl. Phys. B} {\bf 619}
  (2001) 709--728 [\href{http://arXiv.org/abs/hep-ph/0011335}{{\tt
  hep-ph/0011335}}].

\bibitem{Diaz:2015pyv}
M.~A. D\'\i{}az, B.~Koch and S.~Urrutia-Quiroga, {\it {Constraints to Dark
  Matter from Inert Higgs Doublet Model}},  {\em Adv. High Energy Phys.} {\bf
  2016} (2016) 8278375 [\href{http://arXiv.org/abs/1511.04429}{{\tt
  1511.04429}}].

\bibitem{Ma:2006km}
E.~Ma, {\it {Verifiable radiative seesaw mechanism of neutrino mass and dark
  matter}},  {\em Phys. Rev. D} {\bf 73} (2006) 077301
  [\href{http://arXiv.org/abs/hep-ph/0601225}{{\tt hep-ph/0601225}}].

\bibitem{Barbieri:2006dq}
R.~Barbieri, L.~J. Hall and V.~S. Rychkov, {\it {Improved naturalness with a
  heavy Higgs: An Alternative road to LHC physics}},  {\em Phys. Rev. D} {\bf
  74} (2006) 015007 [\href{http://arXiv.org/abs/hep-ph/0603188}{{\tt
  hep-ph/0603188}}].

\bibitem{Banerjee:2016vrp}
S.~Banerjee and N.~Chakrabarty, {\it {A revisit to scalar dark matter with
  radiative corrections}},  {\em JHEP} {\bf 05} (2019) 150
  [\href{http://arXiv.org/abs/1612.01973}{{\tt 1612.01973}}].

\bibitem{FileviezPerez:2008bj}
P.~Fileviez~Perez, H.~H. Patel, M.~Ramsey-Musolf and K.~Wang, {\it {Triplet
  Scalars and Dark Matter at the LHC}},  {\em Phys. Rev. D} {\bf 79} (2009)
  055024 [\href{http://arXiv.org/abs/0811.3957}{{\tt 0811.3957}}].

\bibitem{Araki:2011hm}
T.~Araki, C.~Geng and K.~I. Nagao, {\it {Dark Matter in Inert Triplet Models}},
   {\em Phys. Rev. D} {\bf 83} (2011) 075014
  [\href{http://arXiv.org/abs/1102.4906}{{\tt 1102.4906}}].

\bibitem{Hambye:2009pw}
T.~Hambye, F.-S. Ling, L.~Lopez~Honorez and J.~Rocher, {\it {Scalar Multiplet
  Dark Matter}},  {\em JHEP} {\bf 07} (2009) 090
  [\href{http://arXiv.org/abs/0903.4010}{{\tt 0903.4010}}]. [Erratum: JHEP 05,
  066 (2010)].

\bibitem{AbdusSalam:2013eya}
S.~S. AbdusSalam and T.~A. Chowdhury, {\it {Scalar Representations in the Light
  of Electroweak Phase Transition and Cold Dark Matter Phenomenology}},  {\em
  JCAP} {\bf 05} (2014) 026 [\href{http://arXiv.org/abs/1310.8152}{{\tt
  1310.8152}}].

\bibitem{Chao:2018xwz}
W.~Chao, G.-J. Ding, X.-G. He and M.~Ramsey-Musolf, {\it {Scalar Electroweak
  Multiplet Dark Matter}},  {\em JHEP} {\bf 08} (2019) 058
  [\href{http://arXiv.org/abs/1812.07829}{{\tt 1812.07829}}].

\bibitem{Konetschny:1977bn}
W.~Konetschny and W.~Kummer, {\it {Nonconservation of Total Lepton Number with
  Scalar Bosons}},  {\em Phys. Lett. B} {\bf 70} (1977) 433--435.

\bibitem{Magg:1980ut}
M.~Magg and C.~Wetterich, {\it {Neutrino Mass Problem and Gauge Hierarchy}},
  {\em Phys. Lett. B} {\bf 94} (1980) 61--64.

\bibitem{Schechter:1980gr}
J.~Schechter and J.~Valle, {\it {Neutrino Masses in SU(2) x U(1) Theories}},
  {\em Phys. Rev. D} {\bf 22} (1980) 2227.

\bibitem{Cheng:1980qt}
T.~Cheng and L.-F. Li, {\it {Neutrino Masses, Mixings and Oscillations in SU(2)
  x U(1) Models of Electroweak Interactions}},  {\em Phys. Rev. D} {\bf 22}
  (1980) 2860.

\bibitem{Blank:1997qa}
T.~Blank and W.~Hollik, {\it {Precision observables in SU(2) x U(1) models with
  an additional Higgs triplet}},  {\em Nucl. Phys. B} {\bf 514} (1998) 113--134
  [\href{http://arXiv.org/abs/hep-ph/9703392}{{\tt hep-ph/9703392}}].

\bibitem{Chen:2008jg}
M.-C. Chen, S.~Dawson and C.~Jackson, {\it {Higgs Triplets, Decoupling, and
  Precision Measurements}},  {\em Phys. Rev. D} {\bf 78} (2008) 093001
  [\href{http://arXiv.org/abs/0809.4185}{{\tt 0809.4185}}].

\bibitem{Du:2018eaw}
Y.~Du, A.~Dunbrack, M.~J. Ramsey-Musolf and J.-H. Yu, {\it {Type-II Seesaw
  Scalar Triplet Model at a 100 TeV $pp$ Collider: Discovery and Higgs Portal
  Coupling Determination}},  {\em JHEP} {\bf 01} (2019) 101
  [\href{http://arXiv.org/abs/1810.09450}{{\tt 1810.09450}}].

\bibitem{Fleischer:1982af}
J.~Fleischer and F.~Jegerlehner, {\it {Radiative Corrections to Higgs
  Production by $e^+ e^- \to Z H$ in the {Weinberg-Salam} Model}},  {\em Nucl.
  Phys. B} {\bf 216} (1983) 469--492.

\bibitem{Kniehl:1991hk}
B.~A. Kniehl, {\it {Radiative corrections for associated $Z H$ production at
  future $e^{+} e^{-}$ colliders}},  {\em Z. Phys. C} {\bf 55} (1992) 605--618.

\bibitem{Denner:1992bc}
A.~Denner, J.~Kublbeck, R.~Mertig and M.~Bohm, {\it {Electroweak radiative
  corrections to e+ e- ---\ensuremath{>} H Z}},  {\em Z. Phys. C} {\bf 56}
  (1992) 261--272.

\bibitem{Sirlin:1980nh}
A.~Sirlin, {\it {Radiative Corrections in the SU(2)-L x U(1) Theory: A Simple
  Renormalization Framework}},  {\em Phys. Rev. D} {\bf 22} (1980) 971--981.

\bibitem{Abouabid:2020eik}
H.~Abouabid, A.~Arhrib, R.~Benbrik, J.~E. Falaki, B.~Gong, W.~Xie and Q.-S.
  Yan, {\it {One-loop radiative corrections to $e^+ e^-\to Zh^0/H^0A^0$ in the
  Inert Higgs Doublet Model}},  \href{http://arXiv.org/abs/2009.03250}{{\tt
  2009.03250}}.

\bibitem{Pilkington:2016erq}
T.~Pilkington, {\em {Dark Matter and Collider Phenomenology of Large
  Electroweak Scalar Multiplets}}.
\newblock PhD thesis, Carleton U., 2016.

\bibitem{Denner:2019vbn}
A.~Denner and S.~Dittmaier, {\it {Electroweak Radiative Corrections for
  Collider Physics}},  {\em Phys. Rept.} {\bf 864} (2020) 1--163
  [\href{http://arXiv.org/abs/1912.06823}{{\tt 1912.06823}}].

\bibitem{Djouadi:2005gj}
A.~Djouadi, {\it {The Anatomy of electro-weak symmetry breaking. II. The Higgs
  bosons in the minimal supersymmetric model}},  {\em Phys. Rept.} {\bf 459}
  (2008) 1--241 [\href{http://arXiv.org/abs/hep-ph/0503173}{{\tt
  hep-ph/0503173}}].

\bibitem{Gonderinger:2012rd}
M.~Gonderinger, H.~Lim and M.~J. Ramsey-Musolf, {\it {Complex Scalar Singlet
  Dark Matter: Vacuum Stability and Phenomenology}},  {\em Phys. Rev. D} {\bf
  86} (2012) 043511 [\href{http://arXiv.org/abs/1202.1316}{{\tt 1202.1316}}].

\bibitem{Riesselmann:1996is}
K.~Riesselmann and S.~Willenbrock, {\it {Ruling out a strongly interacting
  standard Higgs model}},  {\em Phys. Rev. D} {\bf 55} (1997) 311--321
  [\href{http://arXiv.org/abs/hep-ph/9608280}{{\tt hep-ph/9608280}}].

\bibitem{Durieux:2017rsg}
G.~Durieux, C.~Grojean, J.~Gu and K.~Wang, {\it {The leptonic future of the
  Higgs}},  {\em JHEP} {\bf 09} (2017) 014
  [\href{http://arXiv.org/abs/1704.02333}{{\tt 1704.02333}}].

\bibitem{Bambade:2019fyw}
P.~Bambade {\em et.~al.}, {\it {The International Linear Collider: A Global
  Project}},  \href{http://arXiv.org/abs/1903.01629}{{\tt 1903.01629}}.

\bibitem{Aghanim:2018eyx}
{\bf Planck} Collaboration, N.~Aghanim {\em et.~al.}, {\it {Planck 2018
  results. VI. Cosmological parameters}},  {\em Astron. Astrophys.} {\bf 641}
  (2020) A6 [\href{http://arXiv.org/abs/1807.06209}{{\tt 1807.06209}}].

\bibitem{Zyla:2020zbs}
{\bf Particle Data Group} Collaboration, P.~Zyla {\em et.~al.}, {\it {Review of
  Particle Physics}},  {\em PTEP} {\bf 2020} (2020), no.~8 083C01.

\bibitem{Borah:2012pu}
D.~Borah and J.~M. Cline, {\it {Inert Doublet Dark Matter with Strong
  Electroweak Phase Transition}},  {\em Phys. Rev. D} {\bf 86} (2012) 055001
  [\href{http://arXiv.org/abs/1204.4722}{{\tt 1204.4722}}].

\bibitem{Gil:2012ya}
G.~Gil, P.~Chankowski and M.~Krawczyk, {\it {Inert Dark Matter and Strong
  Electroweak Phase Transition}},  {\em Phys. Lett. B} {\bf 717} (2012)
  396--402 [\href{http://arXiv.org/abs/1207.0084}{{\tt 1207.0084}}].

\bibitem{Blinov:2015vma}
N.~Blinov, S.~Profumo and T.~Stefaniak, {\it {The Electroweak Phase Transition
  in the Inert Doublet Model}},  {\em JCAP} {\bf 07} (2015) 028
  [\href{http://arXiv.org/abs/1504.05949}{{\tt 1504.05949}}].

\bibitem{Chowdhury:2011ga}
T.~A. Chowdhury, M.~Nemevsek, G.~Senjanovic and Y.~Zhang, {\it {Dark Matter as
  the Trigger of Strong Electroweak Phase Transition}},  {\em JCAP} {\bf 02}
  (2012) 029 [\href{http://arXiv.org/abs/1110.5334}{{\tt 1110.5334}}].

\bibitem{Cline:2013bln}
J.~M. Cline and K.~Kainulainen, {\it {Improved Electroweak Phase Transition
  with Subdominant Inert Doublet Dark Matter}},  {\em Phys. Rev. D} {\bf 87}
  (2013), no.~7 071701 [\href{http://arXiv.org/abs/1302.2614}{{\tt
  1302.2614}}].

\bibitem{Hambye:1996wb}
T.~Hambye and K.~Riesselmann, {\it {Matching conditions and Higgs mass upper
  bounds revisited}},  {\em Phys. Rev. D} {\bf 55} (1997) 7255--7262
  [\href{http://arXiv.org/abs/hep-ph/9610272}{{\tt hep-ph/9610272}}].

\bibitem{Cirelli:2007xd}
M.~Cirelli, A.~Strumia and M.~Tamburini, {\it {Cosmology and Astrophysics of
  Minimal Dark Matter}},  {\em Nucl. Phys. B} {\bf 787} (2007) 152--175
  [\href{http://arXiv.org/abs/0706.4071}{{\tt 0706.4071}}].

\bibitem{Aad:2013yna}
{\bf ATLAS} Collaboration, G.~Aad {\em et.~al.}, {\it {Search for charginos
  nearly mass degenerate with the lightest neutralino based on a
  disappearing-track signature in pp collisions at $\sqrt(s)$=8 TeV with the
  ATLAS detector}},  {\em Phys. Rev. D} {\bf 88} (2013), no.~11 112006
  [\href{http://arXiv.org/abs/1310.3675}{{\tt 1310.3675}}].

\bibitem{CMS:2014gxa}
{\bf CMS} Collaboration, V.~Khachatryan {\em et.~al.}, {\it {Search for
  disappearing tracks in proton-proton collisions at $ \sqrt{s}=8 $ TeV}},
  {\em JHEP} {\bf 01} (2015) 096 [\href{http://arXiv.org/abs/1411.6006}{{\tt
  1411.6006}}].

\bibitem{Aaboud:2017mpt}
{\bf ATLAS} Collaboration, M.~Aaboud {\em et.~al.}, {\it {Search for long-lived
  charginos based on a disappearing-track signature in pp collisions at $
  \sqrt{s}=13 $ TeV with the ATLAS detector}},  {\em JHEP} {\bf 06} (2018) 022
  [\href{http://arXiv.org/abs/1712.02118}{{\tt 1712.02118}}].

\bibitem{Sirunyan:2018ldc}
{\bf CMS} Collaboration, A.~M. Sirunyan {\em et.~al.}, {\it {Search for
  disappearing tracks as a signature of new long-lived particles in
  proton-proton collisions at $\sqrt{s} =$ 13 TeV}},  {\em JHEP} {\bf 08}
  (2018) 016 [\href{http://arXiv.org/abs/1804.07321}{{\tt 1804.07321}}].

\bibitem{Chiang:2020rcv}
C.-W. Chiang, G.~Cottin, Y.~Du, K.~Fuyuto and M.~J. Ramsey-Musolf, {\it
  {Collider Probes of Real Triplet Scalar Dark Matter}},  {\em JHEP} {\bf 01}
  (2021) 198 [\href{http://arXiv.org/abs/2003.07867}{{\tt 2003.07867}}].

\bibitem{Akerib:2016vxi}
{\bf LUX} Collaboration, D.~Akerib {\em et.~al.}, {\it {Results from a search
  for dark matter in the complete LUX exposure}},  {\em Phys. Rev. Lett.} {\bf
  118} (2017), no.~2 021303 [\href{http://arXiv.org/abs/1608.07648}{{\tt
  1608.07648}}].

\bibitem{Cui:2017nnn}
{\bf PandaX-II} Collaboration, X.~Cui {\em et.~al.}, {\it {Dark Matter Results
  From 54-Ton-Day Exposure of PandaX-II Experiment}},  {\em Phys. Rev. Lett.}
  {\bf 119} (2017), no.~18 181302 [\href{http://arXiv.org/abs/1708.06917}{{\tt
  1708.06917}}].

\bibitem{Aprile:2018dbl}
{\bf XENON} Collaboration, E.~Aprile {\em et.~al.}, {\it {Dark Matter Search
  Results from a One Ton-Year Exposure of XENON1T}},  {\em Phys. Rev. Lett.}
  {\bf 121} (2018), no.~11 111302 [\href{http://arXiv.org/abs/1805.12562}{{\tt
  1805.12562}}].

\bibitem{Uhlemann:2008pm}
C.~F. Uhlemann and N.~Kauer, {\it {Narrow-width approximation accuracy}},  {\em
  Nucl. Phys. B} {\bf 814} (2009) 195--211
  [\href{http://arXiv.org/abs/0807.4112}{{\tt 0807.4112}}].

\bibitem{Heinemeyer:2015qbu}
S.~Heinemeyer and C.~Schappacher, {\it {Neutral Higgs boson production at
  $e^+e^-$ colliders in the complex MSSM: a full one-loop analysis}},  {\em
  Eur. Phys. J. C} {\bf 76} (2016), no.~4 220
  [\href{http://arXiv.org/abs/1511.06002}{{\tt 1511.06002}}].

\bibitem{Xie:2018yiv}
W.~Xie, R.~Benbrik, A.~Habjia, B.~Gong and Q.-S. Yan, {\it {Signature of 2HDM
  at Higgs Factories}},  \href{http://arXiv.org/abs/1812.02597}{{\tt
  1812.02597}}.

\bibitem{Cao:2014ita}
J.~Cao, Z.~Heng, D.~Li, L.~Shang and P.~Wu, {\it {Higgs-strahlung production
  process $e^{+} e^{-} \to Zh$ at the future Higgs factory in the Minimal
  Dilaton Model}},  {\em JHEP} {\bf 08} (2014) 138
  [\href{http://arXiv.org/abs/1405.4489}{{\tt 1405.4489}}].

\bibitem{Branco:2011iw}
G.~Branco, P.~Ferreira, L.~Lavoura, M.~Rebelo, M.~Sher and J.~P. Silva, {\it
  {Theory and phenomenology of two-Higgs-doublet models}},  {\em Phys. Rept.}
  {\bf 516} (2012) 1--102 [\href{http://arXiv.org/abs/1106.0034}{{\tt
  1106.0034}}].

\bibitem{Denner:1991kt}
A.~Denner, {\it {Techniques for calculation of electroweak radiative
  corrections at the one loop level and results for W physics at LEP-200}},
  {\em Fortsch. Phys.} {\bf 41} (1993) 307--420
  [\href{http://arXiv.org/abs/0709.1075}{{\tt 0709.1075}}].

\end{thebibliography}\endgroup

\end{document}